\documentclass[preprint]{elsarticle}
\usepackage[utf8]{inputenc}
\journal{Theoretical Population Biology}

\usepackage{amssymb, latexsym, amsthm, amsmath, lineno, epsfig, mathtools, hyperref, todonotes, booktabs, cite, graphicx,natbib}
\usepackage[singlelinecheck=false]{caption}
\modulolinenumbers[1]

\newcommand{\bs}[1]{\ensuremath{\boldsymbol{#1}}}
\newcommand\Var{\operatorname{Var}}
\newcommand\Cov{\operatorname{Cov}}
\newcommand\E{\operatorname{E}}

\newcommand\given{{\,|\,}}

\newcommand\eg{{\it e.g.,}}

\newcommand\ie{{\it i.e.,}}

\newcommand\etc{{\it etc}}

\begin{document}

\begin{frontmatter}

\title{The Expected Sample Allele Frequencies from Populations of Changing Size via Orthogonal Polynomials}

\author[address1]{Lynette Caitlin Mikula\corref{correspondingauthor}}
\cortext[correspondingauthor]{Corresponding author}
\ead{lcm29@st-andrews.ac.uk}
\author[address2,address3]{Claus Vogl}
\ead{claus.vogl@vetmeduni.ac.at}

\address[address1]{Centre for Biological Diversity, School of Biology, University of St. Andrews, St Andrews KY16 9TH, UK}
\address[address2]{Department of Biomedical Sciences, Vetmeduni Vienna, Veterin\"arplatz 1, A-1210 Wien, Austria}
\address[address3]{Vienna Graduate School of Population Genetics, Vetmeduni Vienna, Veterin\"arplatz 1,  A-1210 Wien, Austria}
\date{2022}

\begin{abstract}

In this article, discrete and stochastic changes in (effective) population size are incorporated into the spectral representation of a biallelic diffusion process for drift and small mutation rates. A forward algorithm inspired by Hidden-Markov-Model (HMM) literature is used to compute exact sample allele frequency spectra for three demographic scenarios: single changes in (effective) population size, boom-bust dynamics, and stochastic fluctuations in (effective) population size. An approach for fully agnostic demographic inference from these sample allele spectra is explored, and sufficient statistics for stepwise changes in population size are found. Further, convergence behaviours of the polymorphic sample spectra for population size changes on different time scales are examined and discussed within the context of inference of the effective population size. Joint visual assessment of the sample spectra and the temporal coefficients of the spectral decomposition of the forward diffusion process is found to be important in determining departure from equilibrium. Stochastic changes in (effective) population size are shown to shape sample spectra particularly strongly. 
\end{abstract}


\end{frontmatter}
\newpage


\section{Introduction}\label{sec:intro}

Population demography is difficult to infer reliably, even without considering its interplay with different modes of selection \citep{Johri22}. Here, we consider an analytically tractable framework for modelling and inferring demography that is efficient in terms of its use of the often limited information within site frequency data. Specifically, we utilise a previously proposed hidden Markov method (HMM) inspired approach \citep{Bergman18a} that allows for exact computation of the expected sample allele frequency spectrum of a single population across time, assuming this population evolves according to a diffusion process with a generator that permits a spectral decomposition. We will assume a Moran diffusion process for drift and small mutation rates \citep{Vogl16}. The HMM-inspired method, especially when applied to a diffusion with small mutation rates, is particularly efficient for modelling non-constant population size dynamics through approximation by sequences of epochs of piecewise constant population sizes, and this will form the backbone of our article. We develop, analyse, and discuss three specific demographic models: i) a single deterministic shift in population size, to model simple population growth or shrinkage, ii) series of alternating fixed population sizes, which mimic population boom-bust cycles, and iii) stochastic changes in population size that occur according to an autoregression model. The deterministic demographic models i) and ii) are hardly new within this modelling context \citep{Song12, Lukic11, Zivkovic11, Zivkovic15}, however, our assumption of small mutation rates further simplifies the exact calculation of expected sample allele frequencies in these cases. To the best of our knowledge, the stochastic demographic models iii) are novel. They utilise the theory of autoregression processes commonly used in econometrics \citep[see][Chapter 7]{Johnston60}. Further, we find sufficient statistics for inferring changing population sizes (under simplifying assumptions). This enables us to propose a small-scale but reliable inference framework and contribute to the discussion around the limitations of inference of demography from site frequency data, particularly using spectral methods \citep{Myers08, Bhaskar14}.

Both Kingman's $n$-coalescent \citep[][Chapter~3]{Wake09} and diffusion approaches have been used to generate the expected distribution of allele frequencies (\ie{} the site frequency spectrum) of population samples \citep[\eg{}][]{Sawyer92, Griffiths98, Fu95, Griffiths03, Evans07}. Coalescent approaches yield accurate closed form solutions for small to moderate sample sizes, but for large sample sizes numerical problems generally arise. Coalescent approaches involve multiple steps: Starting with all observed lineages at the present time and looking back, each expected coalescent time, \ie{} the duration for which the number of lineages remains constant, is determined. Conditional on the coalescent time and number of lineages present, the expected number of mutations on each thereby constructed branch of the genealogy is calculated. The expected sample allele frequencies follow from summing mutations along the bifurcating lineages. This process is cumbersome particularly with changing population sizes: To circumvent numerical pitfalls, computationally intense simulations are often employed \citep[\eg{}][]{Hudson02, Excoffier13}. Even computationally more efficient analytical solutions for simplified scenarios can require slow high-precision numerics \citep{Marth04}. Coalescent methods based on utilising the spectral decomposition of the transition matrix describing the decrease of distinct lineages across generations from the sample to the most recent common ancestor (\ie{} the ``matrix coalescent'', \citep{Wooding02}) for determining the expected coalescent times, do not require such libraries for stepwise constant population sizes \citep{Wooding02, Polanski03} and exponential growth scenarios \citep{Polanski03, Bhaskar15}. Other growth patterns have also been well-approximated analytically \citep{Chen19}. Aside from the incorporation of different growth patterns, the ``matrix coalescent'' approach has also been extended to the calculation of the expected site frequency spectra of general coalescent processes, which include coalescent models that explicitly include multiple and simultaneous coalescent events \citep{Spence16}. A more recent, compellingly straightforward but not yet popularised matrix method for determining expected sample spectra assuming either the classic or a more general coalescent model utilises phase-type theory \citep{Hobolth19}: So-called phase-type distributions are constructed via convolutions or mixtures of individual exponential distributions; since the individual branch lengths in coalescent trees are exponentially distributed, many measures of interest to population geneticists can easily be determined by application of results from phase-type theory.  

Incorporation of non-constant population sizes into diffusion equations is considered comparatively conceptually straightforward to the classic coalescent approaches; however, closed form solutions for expected sample allele spectra are available only when considering concrete sampling schemes in combination with the backward diffusion process \citep{Griffiths03}. Analytically tractable solutions to the diffusion density itself have long been studied in population genetics: Kimura \citep{Kimura55} showed that the biallelic Wright-Fisher diffusion for population allele frequencies can be re-parameterised to the so-called oblate spheroidal differential equation. The general solution to the system is an infinite sum over:
i) A spatial component constituted by the eigenvectors that depends solely on the allele frequencies. It is given by a class of weighted orthogonal polynomials with a normalisation constant that is determined by the initial population conditions. ii) A temporal component modulating the change of polymorphic allele frequencies over time by the corresponding eigenvalues.
Kimura often assumed the infinite sites mutation model \citep{Kimura69a}, where each novel mutation hits a previously unmutated site. Formal treatment of neutral parent-independent mutation diffusions by others followed \citep{Shimakura77, Griffiths79}, and a more direct mathematical approach was introduced by \citet{Song12}: If the generator associated with the biallelic diffusion process fulfills certain mathematical requirements, a spectral decomposition can be applied to find an analytical representation of the allele transition density (extension to the multiallelic case in \citep{Steinrucken13}). Applications to diffusion processes with recurrent mutations and selection of arbitrary strength have been demonstrated (which goes beyond Kimura, who models only weak selection). Note that the spectral decomposition of the diffusion generator for populations with deterministic, piecewise constant population sizes under selection and with mutations segregating as per the infinite sites model has also been derived \citep{Zivkovic15}. To obtain explicit results for the expected sample allele frequencies from the transition densities for populations obtained using either of the above methods, binomial sampling at the extant time is required \citep{Griffiths03}. This sampling, coupled with a system of ordinary differential equations on the moments of the sample spectrum derived from the time-reversed Wright-Fisher diffusion \citep{Evans07}, also provides analytical equations for the sample site frequency spectrum while by-passing exact calculation of the transition density itself \citep{Zivkovic11, Zivkovic15}.

In \citet{Bergman18a}, analytical formulae for the transition densities are used to obtain the sample allele spectra via a dynamic programming approach: The continuous allele proportions of biallelic Moran diffusion processes are treated as latent/hidden variables, and realisations of discrete sample allele frequencies at predetermined intervals across time are treated as emission/observed variables. The logic of the classic forward-backward algorithm from Hidden Markov Model (HMM) literature \citep{Rabiner86} can then be applied for inference: \eg{} the probability of observed sample allele frequencies at extant time given assumptions on the population parameters can be determined via the forward algorithm, and the combination of the forward and backward algorithms can be utilised to calculate the joint probabilities of population and sample allele proportions at any point in time. Recall that standard Moran/Wright-Fisher diffusions are considered dual processes to the coalescent in the sense that the expected allele proportions that these models yield are identical, once the state spaces are made comparable by an appropriate function. It can, \eg{} be shown that the neutral biallelic Moran  diffusion model with recurrent mutations (which is equivalent to parent-independent mutations) in the first half of \citep{Bergman18a} is dual to variants of Kingman's coalescent with respect to the binomial sampling scheme \citep{Tavare84,Etheridge09,Chaleyat-Maurel09,Papaspiliopoulos14}. From the HMM theory it follows that the hidden probabilities of population allele proportions can be computed at a cost polynomial in the number of observations via finite mixtures of distributions involving the coalescing dual process (note that this is for constant population sizes, \citep{Papaspiliopoulos14}). By comparison, the transition densities of the forward diffusion processes in \citet{Bergman18a} are already made tractable by expansion into infinite spectral sums involving polynomials, and representation of the binomial sampling scheme as a finite sum of corresponding orthogonal polynomials then enables all HMM tasks to be performed using finite spectral sums of the order of the sample size. This ties to the long-established duality between the spectral representation and the diffusion \citep{Griffiths83, Griffiths10}. However, the finite expansion up to the haploid sample size $K$ (rather than an infinite expansion) also reduces the computational burden of the system. Note that similar reasoning is implicit in the moment method of \citet{Zivkovic11, Zivkovic15}, who forgo exact calculation of the transition densities.  

In the second half of \citep{Bergman18a}, the same HMM approach is applied to the spectral representation of the diffusion limit of the so-called boundary mutation Moran model \citep{Vogl12,Vogl16}. It is assumed that the expected heterozygosity is less than about $2.5\cdot 10^{-2}$ \citep{Vogl12}, which is true for most eukaryotes \citep{Lynch16}. Through a first-order approximation in the overall scaled mutation rate to the general mutation Moran model, mutations segregating in a sample of moderate size can then be modelled to occur only at previously monomorphic sites and normalised so they enter the polymorphic region at a constant equilibrium rate in relation to drift. While ``back-mutations'' are generally allowed, they are excluded while the allele is polymorphic.

In this article, we extend the spectral representation of the boundary mutation Moran model for drift and small mutation rates \citep{Vogl16} (recap in Appendix Section ~\ref{sec:maths_intro_BM}) to both deterministic and stochastic piecewise constant (effective) population sizes (see Sections ~\ref{sec:model1}, ~\ref{sec:model2a}, ~\ref{sec:model3}, ~\ref{sec:model4}). At this point we would like to note that any appropriate measure of the census or effective population size could be assumed to vary within our models (hence our use of (effective) population sizes); we explicitly discuss estimation of the effective population size as it relates to our demographic models in our Conclusions (Section~\ref{sec:discussion}). In the main part of the article, we examine changes in size occurring on different time scales, we determine analytical expressions for the resulting sample site frequency spectra via the forward algorithm from the HMM-inspired method \citep{Vogl12,Vogl16} (recap in Appendix~\ref{sec:maths_intro_BM}). These enable us to assess the convergence behaviour of the sample spectra vs the expected convergence of the effective population size for these differing time scales, and discuss whether this matches expectations from past literature. Further, we propose an agnostic inference framework for inferring past (effective) population sizes from these sample spectra (see Section~\ref{sec:inference}). Throughout, we wish to emphasise that the use of the boundary mutation model simplifies the spectral representation of the transition densities (which we discuss in Section~\ref{sec:maths_intro_general_framework}): Changes in population size, and thus the overall scaled mutation rate, affect only the temporal component. Therefore, only a single full spectral decomposition is required rather than one per population size change as in the more general mutation models used both in the first half of \citet{Bergman18a} and in the moment methods of \citet{Zivkovic11, Zivkovic15}. 


\section{Methods: General Mathematical Framework}\label{sec:maths_intro}

\subsection{Outline}
To start off, we provide an introductory sketch of the methods utilised in this article, and their purpose within the context of this article in particular. We will assume $L$ genomic sites, which have evolved independently according to a biallelic mutation-drift model. Such biallelic systems are determined from the four letter DNA alphabet either by polarisation of alleles into ancestral and derived (as in the infinite sites model), or by intentional nucleotide grouping, \eg{} by contrasting the bases adenine and thymine vs the bases cytosine and guanine; here, we will assume reversible mutations and thus data in the form of the latter. At the extant time, which we designate as $t=0$, a sample of $K$ haploid individuals with these $L$ sites is taken and partitioned into an observed site frequency spectrum $\bs{y}=(y_0,\dots,y_k,\dots,y_{K})$ according to the number of focal alleles $k$ in the sample, which will generally be either the ancestral type or a chosen group of nucleotides, whereby $\sum \bs{y} =\sum_k y_k=L$. If we denote the probability of observing $k$ focal alleles as $p_k$  we can write the likelihood of observing the full site frequency spectrum as the following multinomial distribution:
\begin{equation}\label{eq:multinomial}
    \Pr(y_0,\dots,y_K\given L, p_0,\dots,p_K)\,.
\end{equation}
Note that the exact distribution the event probabilities $p_k$ follow will differ according to the underlying mutation model, and need not correspond to the stationary sampling distributions; when they do, we will denote them as $\bar p_k$. In any case, the shape of the resulting spectrum given the event probabilities is governed by the overall scaled mutation rate $\theta$ (which is a model-dependent multiple of the (effective) population size and the overall genomic point-wise mutation rate per generation $\mu$), as well as the mutation biases $\beta$ towards the focal allele and $\alpha= 1 - \beta$ away from the focal allele. Throughout this article, we will assume that the source population of our sample has reached an evolutionary equilibrium with respect to the mutation bias. However, the (effective) population size and thus $\theta$ are assumed to have changed over time, which runs from $-\infty$ to $0$, and consequently the shape of the polymorphic site frequency spectrum sampled at the extant time may differ from equilibrium expectations. Specifically, this means there is an underlying vector $\bs{\theta}$ of past overall scaled mutation rates, where the entries $\theta_j$ represent a constant value that the mutation rate takes within the epoch $j$ between the epoch break points at times $t_{j-1}$ and $t_{j}$. In the rest of this section, we will provide a mathematical introduction to the specific continuous (diffusion) model that describes the evolution of the source population, and then introduce the method we apply to analy\-tically determine the discrete event probabilities $p_k$ given the vector $\bs{\theta}$ and the corresponding epoch break points; the expected observed site frequency spectrum given these event probabilities then follows. The novelty of the current article and focus of the following sections is then to vary $\bs{\theta}$ and the epoch break points to model specific demographic scenarios, and then assess the effect on the expected sample spectrum.

\subsection{General Mutation Model}

Let us begin by considering the transition rate density $\phi(x,t)$ of population allele frequencies $x\in [0,1]$ along a continuous time axis. The standard forward Kolmogorov (Fokker-Planck) diffusion equation operating on this transition density for biallelic, reversible mutations and drift at each genomic site is given by:
\begin{equation}\label{eq:mutation-drift_diffusion}
\begin{split}
    \frac{\partial\phi(x,t)}{\partial t}&=-\frac{\partial}{\partial x}\bigg(\theta\beta(1-x)\phi(x,t)-\theta\alpha x\phi(x,t)\bigg)+\frac{\partial^2}{\partial x^2} x(1-x)\phi(x,t)\\
    &=-\frac{\partial}{\partial x}\theta(\beta-x)\phi(x,t)+\frac{\partial^2}{\partial x^2} x(1-x)\phi(x,t)\,,
\end{split}
\end{equation}
where the first term on the right hand side describes mutation events and the second corresponds to drift. The usual boundary conditions at $x=0$ and $x=1$, which we will not reproduce here, ensure that there is no flow of probability outside the unit interval (\ie{} that the initial probability mass is conserved at all times) \citep[see][]{McKane07}. It is well known that the stationary distribution $\phi(x,t=\infty)$ is the beta distribution $beta(\beta\theta,\alpha\theta)$ \citep{Wright31}. Binomially sampling $K$ haploid individuals results in the following beta-binomially distributed event probabilities:
 \begin{equation}\label{eq:betabin}
 \begin{split}
    \bar p_k&=\Pr(k \mid K,\beta,\theta)\\
     &=\int_0^1 \binom{K}{k} x^k(1-x)^{K-k}\,\frac{\Gamma(\theta)}{\Gamma(\beta\theta)\Gamma(\alpha\theta)}\,x^{\beta\theta-1}(1-x)^{\alpha\theta-1}\,dx\\
     &=\binom{K}{k}\frac{\Gamma(\theta)}{\Gamma(\beta\theta)\Gamma(\alpha\theta)}\frac{\Gamma(k+\beta\theta)\Gamma(K-k+\alpha\theta)}{K+\theta}\\
     &=betabin(k\given K, \beta, \theta)\,.
\end{split}
\end{equation} 
Substituting these $\bar p_k$ into Eq.~(\ref{eq:multinomial}) yields the likelihood of the observed sample site frequency spectrum in the general mutation-drift equilibrium. 

\subsection{Boundary Mutation Model}\label{boundary_model}
Note that if the scaled mutation rate $\theta$ is very low, most of the probability mass associated with the stationary distribution of population allele frequencies $x$ will be concentrated at the boundaries (compare to the visualisation of the stationary distribution of a Wright-Fisher process approaching its diffusion limit in Fig.~2B of \citep{Der14}). Then the same will be true for a sample taken from this stationary distribution. The first order Taylor series expansion in the overall scaled mutation rate $\theta$ of the beta-binomial stationary sampling distributions in Eq.~(\ref{eq:betabin}) results in the following event probabilities \citep{Vogl14b}:
\begin{equation}\label{eq:first_order_stationary}
\bar p_k=\Pr(k \mid K,\beta,\theta)=
\begin{cases}
    \displaystyle \alpha(1-\beta\theta H_{K-1})+\mathcal{O}(\theta^2) & k=0; \\
    \displaystyle \alpha\beta\theta\frac{K}{k(K-k)}+\mathcal{O}(\theta^2) & 1\leq k \leq K-1;\\
    \displaystyle \beta(1-\alpha\theta H_{K-1})+\mathcal{O}(\theta^2) & k=K\,,
\end{cases}
\end{equation}
where $H_{K-1}=\sum_k^{K-1}\tfrac{1}{k}$ is the harmonic number. Omitting the higher order terms, these correspond to the equilibrium distribution of the discrete boundary mutation Moran model with population size $N=K$ \citep{Vogl12, Vogl21}. We will refer to this distribution as the following boundary binomial distribution
\citep[][Eq.~13]{Vogl15}:
\begin{equation}\label{eq:boundary_stationary}
boundbin(k \mid K,\beta,\theta)=
\begin{cases}
    \displaystyle \alpha(1-\beta\theta H_{K-1}) & k=0; \\
    \displaystyle \alpha\beta\theta\frac{K}{k(K-k)} & 1\leq k \leq K-1;\\
    \displaystyle \beta(1-\alpha\theta H_{K-1}) & k=K\,,
\end{cases}
\end{equation}
due to its similarities with the beta-binomial distribution (see also Appendix Section~\ref{section:simpl_ass}). Note that such boundary mutation models restrict mutations entirely to the monomorphic boundaries; the mutation rate is then normalised so that the expected equilibrium heterozygosity is maintained regardless of the population size \citep{Vogl12, Vogl14b}. Simulations confirm that the boundary mutation Moran model approximates the general mutation Moran model well if the expected equilibrium heterozygosity of the population $\alpha\beta\theta< 0.025$ \citep{Vogl12}, which holds for the protein coding genes of most eukaryotes \citep{Lynch16}. Note that the boundary mutation model can be extended to multiple alleles \citep{Schrempf16,Burden18a,Burden18b,Vogl20}. Samples from such multiallelic diffusion models with similarly low scaled overall mutation rates are also well approximated by the boundary mutation Moran model because the probability of sites being more than biallelic is also very low \citep{Burden18a,Burden18b}.

Different to the case of general mutation models, the boundary mutation diffusion cannot be obtained directly by passing $N=K$ to infinity in a corresponding discrete system: The stationary probabilities of observing monomorphic sites in Eq.~(\ref{eq:first_order_stationary}) contains the harmonic number $H_{K-1}$, which approaches infinity logarithmically and leads to boundary singularities; thus, the sample size must be lower than $K_{max} \approx e^{min(\alpha,\beta)/(\alpha\beta\theta)}$ \citep{Bergman18a} for the model to remain mathematically tractable. In Appendix Section~\ref{sec:FromMoran2Diffusion}, we provide an alternative derivation of the boundary mutation diffusion model, the dynamics of which we explain here: Methodologically, the derivation reflects the logic of Kimura's infinite sites model \citep{Kimura69a}, where novel mutations occur at a steady but infinitesimally small genomic rate of $\theta$ and hit a previously unmutated site where ``the mutant form is represented only once at the moment of its occurrence, $p=1/(2N)$'' \citep[][immediately after Eq.~(16)]{Kimura69a}, and is subsequently spread by drift. Note that $p$ is the proportion of the new allele in the population consisting of $N$ diploid individuals. A proportion of exactly $p$ alleles in the population can be represented via a Dirac delta function.

In contrast to Kimura, who differentiates between ancestral and derived (novel) alleles in his infinite sites model, we differentiate between mutations hitting the non-focal and focal alleles in our boundary mutation model. Let $b_0(t)$ be the proportion of the non-focal allele in the population at time $t$ and $b_1(t)=1-b_0(t)$ the proportion of the focal allele in the population. The initial proportion of a new mutations is $\epsilon_x=1/N$ or $1-\epsilon_x=1-1/N$ and with rates $b_0(t)\beta\theta$ and $b_1(t)\alpha\theta$ respectively. Convergence towards the equilibrium values $b_0(\infty)=\alpha$ and $b_1(\infty)=\beta$ occurs exponentially with rate $\theta$. This rate is low compared to the rate of drift operating on the polymorphic region, which scales with $1$. While the monomorphic and polymorphic regions exchange probability mass, no probability fluxes are assumed to cross the closed unit interval. Thus probability mass is conserved within the closed unit interval. Due to the distinct spatial and temporal dynamics of the monomorphic vs the polymorphic regions, we specifically denote the transition rate density of the polymorphic interior (\ie{} the open interval $\epsilon_x\leq x\leq 1-\epsilon_x$) as $\phi_I(x,t)$. Compactly, the forward boundary mutation diffusion equation can be written as (\citep[][Eq.~41]{Vogl16}, and see Appendix~\ref{sec:FromMoran2Diffusion}):
\begin{equation}\label{eq:boundary-mutation-drift_diffusion}
\begin{split}
\frac{\partial\phi_I(x,t)}{\partial t}&=\lim_{\epsilon_x\to 0}\bigg((\beta\theta b_0(t)/\epsilon_x)\delta(x-\epsilon_x)+(\alpha\theta b_1(t)/\epsilon_x)\delta(x-1+\epsilon_x)\bigg)\\
    &\qquad+\frac{\partial^2}{\partial x^2} x(1-x)\phi_I(x,t)\,,
    \end{split}
\end{equation}
where $\delta(.)$ denotes the Dirac delta function. Together with the dynamics at the boundaries, this system has a solution in the form of the limiting distribution:
\begin{equation}\label{eq:boundary_solution}
\begin{split}
    \phi(x,t)&=\lim_{\epsilon_x\to 0}\bigg(\bigg(b_0(t)-\int_{0+\epsilon_x}^{1-\epsilon_x} (1-x) \phi_I(x,t)\,dx\bigg)\delta(x)\\
    &\qquad+\bigg(b_1(t)-\int_{0+\epsilon_x}^{1-\epsilon_x} x \phi_I(x,t)\,dx\bigg) \delta(x-1)\bigg)+\phi_I(x,t)\,,
\end{split}
\end{equation}
which is a variant of the distribution\citep[][Eq.~7]{McKane07} (noting that $\delta(x-1)=\delta(1-x)$): 
\begin{equation}
\begin{split}
    \phi(x,t)&= \Pi_0(t)\delta(x)+\Pi_1(t)\delta(1-x)+\phi_I(x,t)\,.
\end{split}
\end{equation}

The form of $\phi(x,t)$ (Eq.~\ref{eq:boundary_solution}) ensures that the integral $\int_{0}^{1}\phi(x,t)\,dx=1$ for all possible values of $t$ (\ie{} that $\phi(x,t)$ is normed to one at all times). While probabilistic interpretation of the transition rates is not immediate because of the singularities at the boundaries, binomially sampling $K$ haploid individuals results in regular, well defined event probabilities (which we discuss in detail in Appendix~\ref{section:simpl_ass}):
\begin{equation}
\begin{split}
     p_k=\Pr(k\given K,\beta,\theta, t)&=\int_0^1 \Pr(k\given K, x)\phi(x,t)\,dx\\
        &=\int_0^1 \binom{K}{k}x^k(1-x)^{K-k}\phi(x,t)\,dx\,.
\end{split}
\end{equation}
Indeed, binomially sampling from the stationary limiting distribution $\phi(x)=\phi(x,t=\infty)$ yields stationary event probabilities $\bar p_k$ of the boundary-binomial distribution (Eq.~\ref{eq:boundary_stationary}). This validates our representation of the boundary mutation drift diffusion as the limiting approximation $\theta\to 0$ of the general mutation drift diffusion model.

\subsection{Spectral Representation of Diffusion Equations}\label{sec:maths_intro_BM}
\subsubsection{General Mutation Model}

In order to make the transition densities of diffusion equations analytically tractable, they can be represented as spectral sums; these sums are constituted by the weighted eigenfunctions and eigenvalues of the associated diffusion generator. For diffusion models with pure drift \citep{Kimura55a} and with parent-independent mutation \citep{Shimakura77, Griffiths79}, the eigendecomposition was found as a solution to the respective differential equations themselves. Much more recently, \citet{Song12} introduced a method for directly determining the spectral decomposition of the diffusion generator for models with recurrent mutation and selection. Their result for the transition density of biallelic reversible mutation models is utilised by \citet{Bergman18a}, who construct a forward-backward algorithm for calculating expected allele proportions across time. Differently to \citet{Song12}, they must therefore model an explicit ancestral population distribution of allele proportions \citep[compare][]{Lukic11} as well as an explicit sampling distribution (rather than simply starting from a Dirac delta function and waiting for an infinite amount of time), and furthermore must use both the forward and backward diffusion equations (rather than just the latter). We recapitulate the required results here \citep[details in Section 3.4 of][ for the remainder of this section, unless otherwise cited]{Bergman18a}:

The spectral expansion of the biallelic reversible mutation model clearly depends on both the mutation bias $\beta$ as well as the scaled mutation rate $\theta$. We will assume that at a time $t_s<0$ in the past, the distribution of ancestral population allele proportions can be represented by a density $\rho(x)$. This density must be defined on the unit interval and assumed to integrate to unity, but is otherwise arbitrary i.e. can be chosen according to the context. It can be expanded into a series of (modified) Jacobi polynomials, which are defined by the recursion (compare Eq.~22.3.2 in \citet{Abramowitz70})
\begin{equation}\label{eq:Jacobi_modified}
  R_n^{(\beta,\theta)}(x)=\sum_{l=0}^m(-1)^l\frac{\Gamma(m-1+l+\theta)\Gamma(m+\beta\theta)}{\Gamma(m-1+\theta)\Gamma(l+\beta\theta)l!(m-l)!}x^l\,.
\end{equation}
These polynomials satisfy the orthogonality relation:
\begin{equation}
    \int_0^1 R_m^{(\beta,\theta)}(x) R_n^{(\beta,\theta)}(x) w^{(\beta,\theta)}(x)\,dx=\delta_{n,m} \Delta_{n}^{(G;\beta,\theta)}\,,
\end{equation}
with respect to the beta-distributed weight function $w^{(\beta,\theta)}(x)=beta(\beta\theta,\alpha\theta)$, whereby $\delta_{n,m}$ is Kronecker's delta, implying that the constant $$
\Delta_{n}^{(G;\beta,\theta)}=\frac{\Gamma(n+\beta\theta)\Gamma(n+\alpha\theta)}{(2n+\theta-1)\Gamma(n+\theta-1)\Gamma(n+1)}\,,
$$ 
is non-zero only for $m=n$. Knowing this, the ancestral distribution can now be written as:
$$
\rho(x)=\phi(x,t=t_s)=\sum_{n=0}^{\infty}\rho_n w^{(\beta,\theta)}(x) R_n^{(\beta,\theta)}(x)\,,
$$
where the coefficients $\rho_n$ can be determined via
$$
    \rho_n= \frac{1}{\Delta_n^{(G;\beta,\theta)}} \int_{0}^1 R_n^{(\beta,\theta)}(x){\rho}(x)\,dx\,.
$$

Thus, the forward diffusion equation representing the trajectory of the population allele proportions for $t_s\leq t \leq 0$, \ie{} starting from this ancestral distribution and extending forwards in time, can be represented uniquely as the following spectral sum:
$$
\phi(x,t)=\sum_{n=0}^{\infty}\rho_n w^{(\beta,\theta)}(x) R_n^{(\beta,\theta)}(x)e^{-\lambda_n (t-t_s)}\,,
$$ 
where $\lambda_0=\lambda_1=0$ and $\lambda_n=n(n+\theta-1)$ for $n\geq2$ are the corresponding eigenvalues. Note that this assumes that the mutational parameters $\beta$ and $\theta$ remain constant over time.

As before, we wish to draw a binomially distributed sample of size $K$ from this population at the extant time $t=0$. In order to do so, we must first more closely consider this binomial sampling scheme, \ie{} the distribution: 
\begin{equation}\label{eq:binomial_AppBnew}
    \Pr(k\given K,x)=\binom{K}{k}\,x^k(1-x)^{K-k}\,.
\end{equation} 
 It can equivalently be seen as a polynomial of order $K$ with coefficients $a_{k+i}=\binom{K}{k}\binom{K-k}{i}(-1)^{i}$. Let $\mathbf{a}(k,K)$ then be the vector of coefficients $a_{n}(k,K)$, and let $\mathbf{R}^{(\beta,\theta)}$ be the lower triangular matrix of polynomial coefficients of the Jacobi polynomials $R_{n}^{(\beta,\theta)}(x)$. The binomial distribution can then be expressed in matrix form via the following linear algebraic equation:
\begin{equation}\label{eq:matr}
    \mathbf{d}^{(\beta,\theta)}(K,k)=\mathbf{a}(k,K)\mathbf{R}^{(\beta,\theta)}
\end{equation}
Note that the triangular structure of $\mathbf{R}^{(\beta,\theta)}$ obviates matrix inversion. The entries of the vector on the left hand side, \ie{} $\mathbf{d}^{(\beta,\theta)}(K,y)$, can also be obtained via:
\begin{equation}\label{eq:int_d}
    d_n(K,k)=\int_0^1\binom{K}{k}\,x^k(1-x)^{K-k}\mathbf{R}^{(\beta,\theta)}w(x)^{(\beta,\theta)}\,dx\,.
\end{equation}
It follows that the binomial sampling distribution from Eq.~(\ref{eq:binomial_AppBnew}) can be uniquely expanded into Jacobi polynomials:
\begin{equation}\label{eq:binom2jacobi}
   \Pr(k\given K,x)=\binom{K}{k}x^k(1-x)^{K-k}=\sum_{n=0}^K d_n^{(\beta,\theta)}(K,y) R_n^{(\beta,\theta)}(x)\,.
\end{equation}

Utilising the previous results, the marginal likelihood of the sample, \ie{} the event probability $p_k$ of observing precisely $k$ focal alleles within our binomial sample of size $K$ drawn at time $t=0$, can be evaluated via the following (wherein we recall the orthogonality relationship to simplify calculations):
\begin{equation}\label{eq:marginal_LH_general}
\begin{split}
 p_k=\Pr(k\given K,\beta,\theta, t=0)
    &=\int_0^1 \binom{K}{k}x^k(1-x)^{K-k}\phi(x,t=0)\,dx\\
    &=\int_0^1 \bigg(\sum_{n=0}^K d_n(k,K) U_n^{(\beta,\theta)}(x)\bigg)\\
    &\qquad\times\bigg(\sum_{m=0}^\infty \rho_m e^{\lambda_m t_s} w^{(\beta,\theta)}(x)U_m^{(\beta,\theta)}(x)\bigg)\,dx\\
    &=\sum_{n=0}^K d_n(k,K) \rho_n^{(\beta,\theta)}\Delta_n^{(G;\beta)} e^{\lambda_n t_s}\,.
\end{split}
\end{equation}
Note that this is technically a forward pass of the forward-backward algorithm, which we discuss in more detail in Appendix Section~\ref{sec:appendixB.1.new}). Substituting these $p_k$ into Eq.~(\ref{eq:multinomial}) again yields the likelihood of the observed sample site frequency spectrum.

\subsubsection{Boundary Mutation Model}\label{sec:boundary_spectral}

As previously mentioned, the spectral decomposition of the transition density of the pure drift diffusion equation has been known for the good part of a century \citep{Kimura55}, and the eigenvectors and eigenvalues can be recovered via the formalised methodology presented by \citet{Song12}. However, both boundaries in these approaches are considered exit boundaries, and only the polymorphic region is thus explicitly modelled by the spectral representation. To include specific boundary terms, the spectral representation must be augmented by appropriate boundary conditions related to mutation and/or fixation rates \citep{McKane07, Tran14a, Tran14b, Vogl16}. In Appendix Section~\ref{sec:Appendix_spectral_decomposition}, we provide a full derivation of the spectral representation of the forward and backward boundary mutation diffusion equations that follows the unified approach to solutions of differential equations proposed by \citet{McKane07}, and thus clarify the reasoning in the derivation originally provided by \citep{Vogl16}. We will cover the main results here.  

Traditionally, the eigenfunctions of the diffusion generators of pure drift models (considering only the polymorphic region) are constituted by Gegenbauer polynomials; we here use the modified Gegenbauer polynomomials $U_{n}(x)$ (see \citep[][Eq.~18.5.77]{DLMF}):
\begin{equation}\label{eq:Gegenbauer_modified}
\begin{split}
  U_{n}(x)&=\sum_{l=0}^{n-2}(-1)^{-1-l}\binom{n+l}{l}\binom{n-1}{l+1}x^l \text{ for $n\geq 2$}\,.
\end{split}
\end{equation}
These are orthogonal with respect to the weight function $w(x)=x(1-x)$ for all orders of expansion $n,m>2$:
\begin{equation}\label{eq:ortho_Gegen}
    \int_0^1 U_n(x) U_m(x) w(x)\,dx=\delta_{n,m} \Delta_n^{(B)}\,,
\end{equation}
where $\delta_{n,m}$ is Kronecker's delta, and 
\begin{equation}\label{eq:Delta_boundary}
    \Delta_n^{(B)}=\frac{(n-1)}{n(2n-1)}
\end{equation} 
with $n\geq 2$ is the proportionality constant. The corresponding eigenvalues determining the rate of drift are $\lambda_{n\geq 2}=n(n-1)$. The augmentation of this system required to include fixation at the boundaries yields the following system of forward eigenfunctions with the corresponding eigenvalues $\lambda_0=0$, $\lambda_1=0$ and $\lambda_{n\geq 2}=n(n-1)$ \citep{McKane07} (also used by \citep{Tran14b,Bergman18a}):
\begin{equation}\label{eq:forw_Us}
\begin{cases}
    \mathcal{F}_0^{(\beta)}(x)&=\alpha\delta(x)+\beta\delta(x-1)\\
    \mathcal{F}_1^{(\beta)}(x)&=-\delta(x)+\delta(x-1)\\
    \mathcal{F}_{n\geq2}^{(\beta)}(x)&=-\frac{(-1)^n}n\delta(x)+U_n(x)-\frac{1}n\delta(x-1)\,,
\end{cases}
\end{equation}
where $\delta(.)$ is Dirac's delta. Observe that the point masses at the boundaries of the eigenfunctions $n\geq 2$ balance the probability weight in the polymorphic interior with the probability mass point masses at the boundaries. Importantly for this article, these forwards eigenfunctions are identical to those of the diffusion generator of the boundary mutation model (compare \citep[][Eqs.~32, 37]{Vogl16}, and see the derivation in Appendix Section~\ref{sec:Appendix_spectral_decomposition}). However, the first-order eigenvalue $\lambda_1$ in the boundary mutation diffusion model is given by $\theta$ so as to explicitly model the influx of new mutations occurring at either boundary (with the other eigenvalues remaining the same). As covered in Section~\ref{boundary_model}, $\theta$ also determines the rate of convergence to the equilibrium boundary values associated with the eigenvalue $\lambda_0=0$ (see also Eq.~\ref{eq:bs_2_F0_F1}), which is slow compared to the rate of drift.

A set of backward eigenfunctions can be constructed to be orthogonal to the forward eigenfunctions above; these are the following polynomials of ascending order \citep{McKane07,Tran14a, Tran14b, Bergman18a} (see also Appendix Section~\ref{sec:Appendix_spectral_decomposition}):
\begin{equation}\label{eq:back_Bergman18}
\begin{cases}
    \mathcal{B}_0^{(\beta)}(x)&=1\\
    \mathcal{B}_1^{(\beta)}(x)&=x-\beta\\
    \mathcal{B}_{n\geq2}^{(p)}(x)&=x(1-x)U_{n}(x)\,.
\end{cases}
\end{equation}
This construction can be confirmed by setting $\Delta_0^{(B)}=\Delta_1^{(B)}=1$, and checking that an analogous orthogonality relation to before holds:
\begin{equation*}
    \int_0^1 \mathcal{B}_n^{(\beta)}(x)\mathcal{F}_m^{(\beta)}(x)\,dx=\delta_{nm}\Delta_n^{(B)}
\end{equation*}
for a nonzero $\Delta_n^{(B)}=\frac{n-1}{(2n-1)n}$ only if $m=n$. 

As with the general mutation model in the previous subsection, the spectral representation of the boundary mutation diffusion can be used within the context of a forward-backward algorithm \citep[details in Section 4.2]{Bergman18a}. This again requires the specification of a distribution of ancestral allele proportions $\rho(x)$ at time $t_s<0$. As before, this is an arbitrary distribution that we here obtain a spectral representation for by expanding the limiting distribution that represents the solution of the boundary mutation diffusion model (Eq.~\ref{eq:boundary_solution}) into orthogonal polynomials. Specifically:
\begin{equation}\label{eq:rho_def}
\rho(x)= \phi(x,t_s)=\tau_0(t_s)\mathcal{F}_0^{\beta}(x) + \tau_1(t_s)\mathcal{F}_1^{\beta}(x) + \sum_{n=2}^\infty \tau_n(t_s)\mathcal{F}_n^{\beta}(x)\,,
\end{equation}
where $\tau_0(t_s)=1$, and $\tau_1(t_s)=\beta-b_1(t_s)$ represents the proportion of the focal allele in the population at time $t_s$. To correctly model the temporal change in focal allele proportions due to mutational input between $t_s<t\leq 0$, we must set $\tau_0(t)=1$ and $\tau_1(t)=(b_1(t_s)-\beta)e^{-\lambda_1(t-t_s)}$ (compare Eq.~\ref{eq:DE_inhomogeneous_Fs_2}). For the polymorphic interior, the changes in population allele proportions over time $t_s<t\leq 0$ due to drift can be found by substituting the spectral sum for the polymorphic transition rate density $\phi_I(x,t)=\sum_{n=2}^\infty\tau_n(t)F_n^{\beta}(x)$ into the forward diffusion equation Eq.~(\ref{eq:boundary-mutation-drift_diffusion}). This induces the following system of inhomogeneous differential equations for the temporal coefficients \citep[][Eq.~43]{Vogl16} (see Appendix~\ref{section:eigensystem_boundary_mutation}, Eq.~\ref{eq:DE_inhomogeneous_Fs_1}):
\begin{equation}\label{eq:DE_inhomogeneous_main}
\begin{split}
    \frac{d}{dt}\tau_n(t) &= -\lambda_n\bigg(\tau_n(t)-\alpha\beta\theta\frac{E_n}{\lambda_n} \tau_0(t)-\theta \frac{O_n}{\lambda_n} \tau_1(t)\bigg)\,.
\end{split}
\end{equation}
for $n\geq 2$, with: 
\begin{equation*}
    E_n=\frac{U_n(0)+U_n(1)}{\Delta_n^{(B)}} =-(2n- 1)n\frac{(-1)^n+1}{\Delta_n^{(B)}}
\end{equation*}
and 
\begin{equation*}
    O_n=\frac{U_n(0)\beta+U_n(1)\alpha}{\Delta_n^{(B)}}=-(2n- 1)n\frac{(-1)^n\beta-\alpha}{\Delta_n^{(B)}}\,.
\end{equation*}
Importantly, this eigensystem can be diagonalised to a time-homogeneous form (which we show in Appendix Section~\ref{seq:diagonalisation}). However, we will proceed with the above system for this article (details follow in Section~\ref{sec:maths_intro_general_framework}).

At the extant time $t=0$, we once again draw a binomial sample of size $K$ from the population that has evolved according to the above. The binomial likelihood itself can, similarly to the previous subsection, be uniquely expanded into a series of appropriate backward polynomials at $t=0$:
\begin{equation}\label{eq:dens_geg}
\begin{split}
    \Pr(k\given x,K)&=\binom{K}{k}\,x^k(1-x)^{K-k}\\
    &=\sum_{n=0}^K  d_n(k,K) \mathcal{B}_n^{(\beta)}(x)\,,
\end{split}
\end{equation}
where $d_n(k,K)$ are again coefficients depending on $k$ and $K$ that can be obtained analogously to either Eq.~(\ref{eq:matr}) via a matrix multiplication or to Eq.~(\ref{eq:int_d}) via integration:
\begin{equation}
    d_n(K,k)=\int_0^1\binom{K}{k}\,x^k(1-x)^{K-k}\mathcal{F}_n^{\beta}(x)\,dx\,.
\end{equation}
Note that the coefficients $d_n(k,K)$ of the boundary mutation expansion are of course different from those of the general mutation expansion, but we choose to not distinguish them in their notation since they are always associated with the appropriate eigenfunctions $R_n^{(\beta,\theta)}$ vs $\mathcal{B}_n^{(\beta)}(x)$ or with the constants $\Delta_n^{(G;\dots)}$ or $\Delta_n^{(B)}$, from which context it is obvious which model is assumed. Either way, Eq.~(\ref{eq:dens_geg}) represents a straightforward linear transformation, since a binomial sample can be represented as polynomial of order $K$, just like the backwards eigenfunction $\mathcal{B}_n^{(\beta)}(x)$.

Overall, e can thus determine the likelihood of the sample, \ie{} the event probability $p_k$ of observing precisely $k$ focal alleles within our binomial sample of size $K$ drawn at time $t=0$, via the following (wherein we again use the orthogonality relationship of the orthogonal polynomials to simplify calculations):
\begin{equation}\label{eq:marginal_LH_boundary}
\begin{split}
     p_k=\Pr(k\given K,\beta,\theta, t=0)&=\int_0^1 \Pr(k\given K, x)\phi(x,0)\,dx\\
        &=\int_0^1 \binom{K}{k}x^k(1-x)^{K-k}\phi(x,0)\,dx\\
    &=\int_0^1 \bigg(\sum_{n=0}^K d_n(k,K) \mathcal{B}_n^{(\beta)}(x)\bigg)\\
    &\qquad\times\bigg(\sum_{m=0}^\infty \tau_m(0)\mathcal{F}_m^{(\beta)}(x)\bigg)\,dx\\
    &=\sum_{n=0}^K d_n(k,K) \tau_n(0)\Delta_n^{(B)}\,.
\end{split}
\end{equation}
Clearly, this is again a forward pass of the forward-backward algorithm (see Appendix Section~\ref{sec:appendixB.1.new}), and substituting these $p_k$ into Eq.~(\ref{eq:multinomial}) yields the likelihood of the observed sample site frequency spectrum.

\paragraph{Remark} In the formulae for the event probabilities $p_k$ assuming either the general or boundary mutation diffusion for the population allele frequencies (Eq.~(\ref{eq:marginal_LH_general} and Eq.~(\ref{eq:marginal_LH_boundary} respectively), polynomial expansions only up to the sample size $K$ rather than to infinity are required. This is due to our application of the HMM forward-backward scheme to a system of orthogonal polynomials: the sample size $K$ at time $t=0$ determines (backwards in time) the order of the expansion needed for the forward eigensystem. In principle, this approach to determining expected sample site frequency spectra therefore has a low computational burden. Directly implemented as above, the event probabilities $p_k$ can be evaluated in a matter of seconds in the open source statistics software R \citep{RCoreTeam17}; however, exceeding a sample size of $K=24$ incurs numerical instability. With the high precision floating point library MPFR \citep{FouLHLPZ-2007}, reliable results can be obtained up to $K=37$. Clearly, specialised programs are still lacking for application to larger samples. However, direct implementation of these formulae suffices for model-validating simulations and application to small or down-sampled data sets as required for this article.

\paragraph{Remark} 
In Appendix Section~\ref{boundary_model}, we show that the spectral representation of the backwards diffusion rate densities of both the general mutation and boundary mutation models has a dual representation as a time dependent pure-death jump process on the space of binomial distributions with respect to their respective transition rate distributions $\psi(x,t)$. Specifically, at each time point $t_s < t \leq 0$ and each order of expansion $0<n<K$, the expected event probabilities $p_k$ can be evaluated to the same result. In the case of the boundary mutation diffusion model, $\psi(x,t)$ is the limiting stationary distribution but not a density in the traditional sense. The application of the HMM framework and the interpretation of the dual process are still possible, however, precisely because the sampling distribution at each time point $t_s < t \leq 0$ and each order of expansion $0<n<K$ is a regular discrete probability distribution even in non-equilibrium (which can be seen in Appendix Section~\ref{boundary_model}, compare Section 4.2 in \citep{Papaspiliopoulos14}).

\paragraph{Remark} Recall that the forward eigenfunctions of the boundary mutation diffusion are given by $\mathcal{F}_{n\geq2}^{(\beta)}(x)$. Interestingly, these can also be considered a zeroth-order Taylor series expansion in the scaled mutation rate $\theta$ of the weighted Jacobi polynomials $w^{(\beta,\theta)} R_{n\geq2}^{(\beta,\theta)}(x)$ \citep[][Appendix A.1.]{Vogl16}. This is of note because analogously, the discrete boundary mutation Moran model \citep{Vogl12} was first derived as a first-order expansion in the overall scaled mutation rate $\theta$ of the general biallelic Moran model with separately parameterised drift and mutation terms \citep{Baake08}.

\subsection{Piecewise Constant (Effective) Population Size}\label{sec:maths_intro_general_framework}

Recall that our main aim is to analytically determine the expected site frequency spectra under a variety of demographic scenarios. So far, we have determined that utilising the spectral representation of diffusion models within a forward pass of a HMM forward-backward algorithm yields a computationally efficient framework for analytically calculating expected site frequency spectra, but have not explicitly introduced demographic changes into these models. In this section, however, we will introduce a time axis that is partitioned into a sequence of epochs between which the population undergoes changes in (effective) sizes; this will be reflected in a change of the overall scaled mutation rate $\theta$ and re-scaling of time between individual epochs. Incorporating this into the general mutation model requires that any change in the scaled mutation rate $\theta$ or the mutation bias $\beta$ must be matched by a change in the base of the Jacobi polynomials $R_n^{(\beta,\theta)}$. This is cumbersome and complicates the treatment of the demographic models. We will therefore proceed with the spectral representation of the boundary mutation diffusion model for the remainder of this article. As long as we use the un-diagonalised spectral representation of this model, demographic changes do not affect the base of the Gegenbauer polynomials within the eigensystem. These can thus be considered the purely spatial component of the spectral representation of the transition rate density, depending on the allele frequencies but not on the overall scaled mutation rate or mutation bias. Changes in the mutation parameters only require appropriate modification of the temporal coefficients that solve the time-inhomogeneous system of equations from Eq.~(\ref{eq:DE_inhomogeneous_main}).

\paragraph{A Simplification} As a further simplification within our modelling approach, we will assume that the mutation bias $\beta$ is evolutionarily stable, \ie{} does not change between epochs. This enables us to examine the effect of demography on the allele frequencies via the overall scaled mutation rate in isolation from any other evolutionary force. We can then assume that the proportion of the focal allele within the population has already converged to $\beta$, \ie{} that $\tau_1(t)$ has converged to $0$. Indeed $\tau_1(t)$ converges to $0$ at rate $\theta$ even with changing population sizes as long as the mutation bias $\beta$ remains constant (recall that the diffusion time is scaled by the population sizes). Thus, the only requirement for convergence of $\tau_1$ to zero is that our process starts at $t=-\infty$. 

Note that only a further change in $\beta$, which we do not permit, but not a change in $\theta$ can then induce $\tau_1(t)\neq 0$. Hence, we set $\tau_1(t)=0$ and thus $b_0(t)=\alpha$ and $b_1(t)=\beta$. As a result, the time-inhomogeneous system for the temporal coefficients from Eq.~(\ref{eq:DE_inhomogeneous_main}) can be simplified to (see Appendix Section \ref{section:simpl_ass}):
\begin{equation}\label{eq:DE_inhomogeneous_simple}
\begin{split}
    \frac{d}{dt}\tau_n(t) &= -\lambda_n\tau_n(t) + \alpha\beta\theta(2n-1)n((-1)^n + 1)
\end{split}
\end{equation}
for $n\geq 2$.
Since $\tau_1(t)=0$ and $\lambda_{n\geq 2}=n(n-1)$ are much greater than $\lambda_1=\theta$, we deduce that all odd $\tau_{n\geq 2}(t)$ converge to zero very rapidly. This implies that we need only consider the even temporal coefficients in our modelling approach, \ie{} that the polymorphic spectrum remains symmetric regardless of potential demographic changes because the mutation bias undergoes no further shifts. This symmetry does not pertain to the monomorphic boundaries, which converge to a symmetric equilibrium only if $\beta=\alpha=\tfrac{1}{2}$. 

\paragraph{Introducing Demography}
From now on, we will assume that time is segmented into epochs indexed by $j\in \{1,..,J\}$. Each epoch starts with time point $t_{j-1}$ and ends with time point $t_{j}$. We will usually end the last epoch at the extant time $t_{J}=0$, since our samples are assumed to come from the present. Let us define the epoch lengths as $s_j=t_j-t_{j-1}$. We will assume that the (effective) population size remains constant at $N_j$ within epoch $j$ and then instantaneously switches to $N_{j+1}$ at the onset of epoch $j+1$. This translates to corresponding changes in what we will call the bias-complemented overall scaled mutation rate, defined per epoch $j$ as $\theta_j^{*}= \alpha\beta\theta_j$. Note that $\theta_j^{*}$ also corresponds to the equilibrium solution of the boundary mutation diffusion model. Since we only consider the polymorphic part of the sample spectrum in our later demographic models, we will specify only the values of $\theta_j^{*}$ in our examples/simulations rather than any component values. 

Some caution is required around our treatment of the first epoch: In our model where we assume a single change in (effective) population size (Section~\ref{sec:model1}), as well as in our inference approach (Section~\ref{sec:inference}), we set the beginning of the first epoch to $t_{j=0}=-\infty$, and thus the initial interval $s_1$ is assumed to be infinitely long. In the boom-bust (Section~\ref{sec:model2a}) and stochastic fluctuation models (Sections~\ref{sec:model3},~\ref{sec:model4}), we obtain convergence results for an infinite number of epochs $J$, which means that the initial epoch has no real impact on the outcome. Therefore, we can readily assume that temporal functions are equal to their equilibrium values at the end of the first epoch, so at $t_1$ (having either converged there, or being generally negligible). These values are:

\begin{equation}\label{eq:TauExpansion_eqi}
\begin{cases}
    \tau_0=1\\ 
    \tau_1=0\\ 
    \tau_{n}(t)=\Xi_{n\geq 2}\theta_1^{*}\text{, for $t=t_1$,}
\end{cases}
\end{equation}
with $\Xi_n=-\tfrac{4n-2}{n-1}=\frac{E_n}{\lambda_n}$. 

Moving forwards in time, the bias-complemented overall scaled mutation rate $\theta_j^{*}$ changes from epoch to epoch; within each, the population allele frequencies converge towards their current expected equilibrium. This convergence is governed by the time-inhomogeneous equation for the odd temporal coefficients (Eq.~\ref{eq:DE_inhomogeneous_simple}); which can now be re-parameterised again. Specifically, for even $n$, \ie{} $n\in 2\mathbb{N}$, and $j\geq 2$, the system within each epoch is given by:
\begin{equation}\label{eq:TauExpansion_modified_poly}
\begin{split}
    \frac{d}{dt} \tau_{n}(t)\Xi_n^{-1}&=-\lambda_n\bigg(\tau_{n}(t)\Xi_n^{-1}+\theta_j^{*}\bigg)\text{, for $t_{j-1}<t\leq t_{j}$.}
\end{split}    
\end{equation}
This applies to all models within this paper. Setting the starting condition specifically as the epoch start point $\tau_{n}(t_{j-1})$, the solution of this differential equation is:
\begin{equation}\label{eq:TauExpansion_solution_poly}
\begin{split}
   \tau_n(t) \Xi_n ^{-1} &=\theta_{j}^*-
   \bigg(\theta_{j}^*-\tau_{n}(t_{j-1})\Xi_n ^{-1}\bigg)e^{-\lambda_n (t-t_{j-1})}\text{, for $t_{j-1}<t\leq t_{j}$.}
\end{split}    
\end{equation}
Note that each epoch with $j\geq 2$ now technically has an ancestral allele distribution $\rho(x)=\phi(x,t=t_{j-1})$ that is precisely the final allele configuration of the previous epoch; these are determined by substituting the temporal coefficients of the two equation systems immediately above into the representation of $\rho(x)$ as a spectral sum using Gegenbauer polynomials given in Eq.~(\ref{eq:rho_def}). Explicit modelling of the system therefore actually begins at the end of the first epoch, which represents an equilibrium state.

At the extant time, which is also the end of the last epoch, we will as always assume that a binomial sample of haploid size $K$ is drawn; more explicitly, we assume that the binomial sampling probabilities are expanded into a series of backward polynomials as in Eq.~(\ref{eq:dens_geg}). Then the event probability $p_k$ of observing precisely $k$ focal alleles within this sample, after the population has undergone a series of piecewise constant demographic scenarios, can be determined via Eq.~(\ref{eq:marginal_LH_boundary}):
\begin{equation}\label{eq:marg_like_general_geg_2a}
\begin{split}
p_{k}&=\Pr(k\given K,\beta,\theta, t=0)\\
 &=-\sum_{n=2}^{K}\frac{2}{n} d_n(k,K) \theta_1^{*}\qquad\text{for $k\in \{1,...,K-1 \}$}\,,
\end{split}
\end{equation}
for the polymorphic spectrum. The expected event probabilities at the boundaries follow accordingly:
\begin{equation}\label{eq:marg_like_general_geg_2b}
\begin{split}
&p_0=\Pr(k=0\given K,\beta,\theta, t=0) = \alpha - \sum_{l=1}^{K-1}\tfrac{1}{2}p_l\\
&p_K=\Pr(k=K\given K,\beta,\theta, t=0) = \beta - \sum_{l=1}^{K-1}\tfrac{1}{2}p_l\,.
\end{split}
\end{equation}
Importantly, recall that only an expansion of the temporal coefficients up to the sample size $K$ is required in order to determine the event probabilities given the sample at the extant time. Further, note that although we have assumed that $b_0(t)$ and $b_1(t)$ remain constant at $\alpha$ and $\beta$ respectively, the frequencies of the monomorphic classes in the sample may nevertheless vary over time with the changes in the mutation rate. 

It should be noted that Eqs.~(\ref{eq:TauExpansion_solution_poly}-\ref{eq:marg_like_general_geg_2a}) will form the basis for the analyses of specific demographic models in Section~\ref{sec:models}, where we look at the effect of these models on the polymorphic sample spectrum.

\section{Theory: Demographic Models}\label{sec:models}

\subsection{Model 1: Single Shift in Population Size}\label{sec:model1}

Now consider a population that undergoes a single change in (effective) population size, meaning there are only two epochs: The ancestral epoch runs between $t_0=-\infty$ and $t_1<0$ with the overall bias-complemented ancestral mutation rate of $\theta_1^{*}$; the overall bias-complemented mutation rate switches instantaneously to $\theta_2^{*}$ at $t_1$ and remains constant until the sample is taken at $t_2=0$. The population allele frequencies thus evolve from the old equilibrium attained before $t_1$ towards a new one according to Eq.~(\ref{eq:TauExpansion_solution_poly}) after this point, with $\tau_n(t_1)\Xi_n ^{-1}=\theta_1^{*}$ for $n\in 2\mathbb{N}$:
\begin{equation}\label{eq:TauExpansion_AC_solution}
\begin{split}
\tau_n(t_1<t\leq 0) \Xi_n ^{-1} &= \theta_2^{*} - \bigg( \theta_2^{*} - \theta_1^{*}\bigg)e^{-\lambda_n \frac{\theta_1^{*}}{\theta_2^{*}} (t-t_1)}\,.
\end{split}
\end{equation}

The marginal probabilities of sample allele frequencies can then be determined as outlined in Section~\ref{sec:maths_intro_general_framework}. Note that if $\theta_1^{*}\ll \theta_2^{*}$, the solution above can be approximated by the solution of a simple one-parameter model (see Appendix Section~\ref{sec:PopExpl}).

\subsubsection{Results and Discussion for Model~1}

\begin{figure}[!ht]
    \centering
    \includegraphics[width = 12cm, height=14cm]{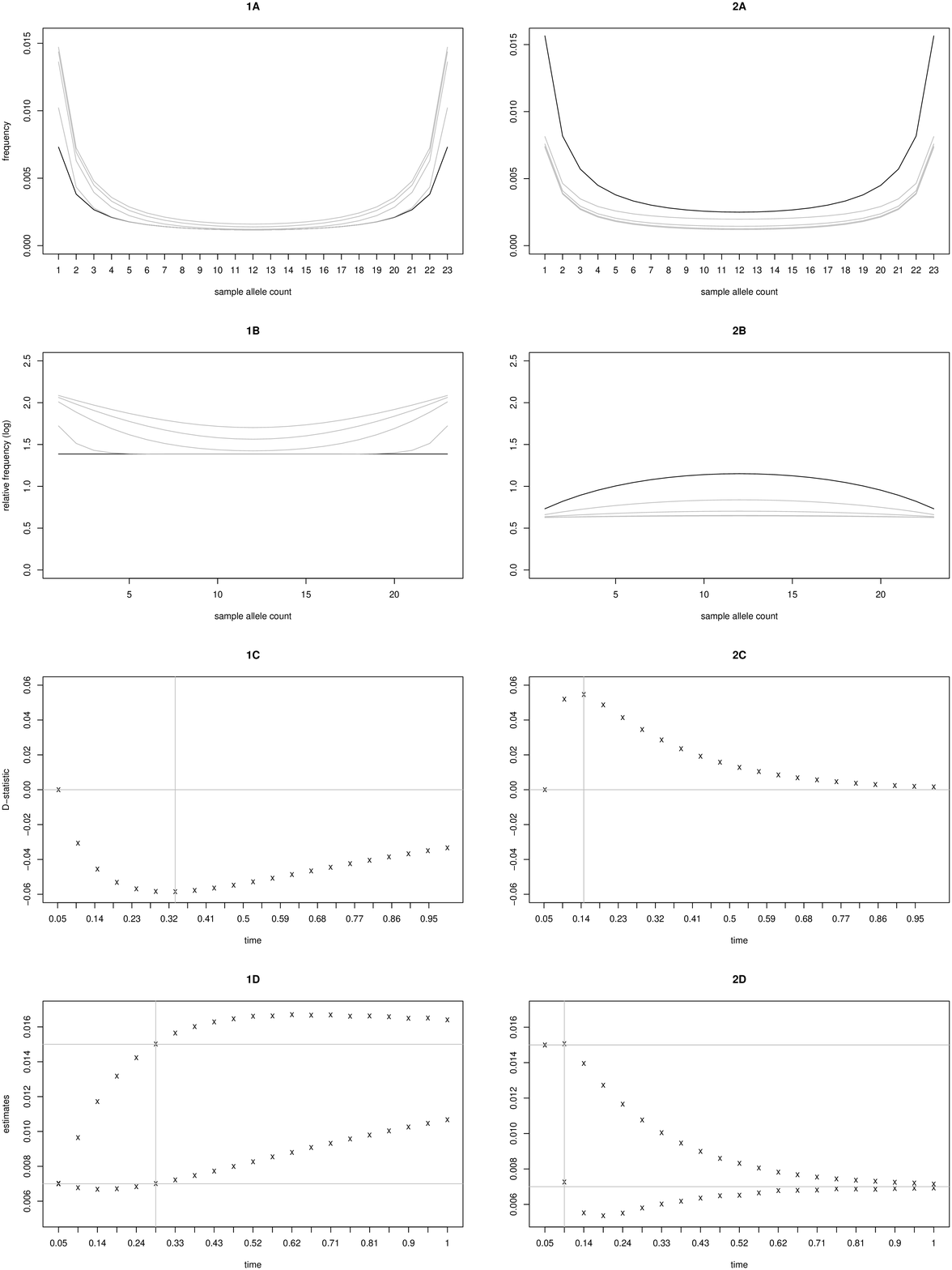}
    \caption{These figures examine the effect of an increase (1A-1D) and a decrease (2A-2D) in the overall bias-complemented scaled mutation rate of the polymorphic site frequency spectrum. Specifically, we examine samples of size $K=24$ from a population whose mutation rate has increased from $\theta_1^{*}=0.007$ to $\theta_2^{*}=0.015$ (1A-1D) or decreased in the reverse (2A-2D) at times $t_1=c(-0.083,-0.292,-0.500,-0.708)$ back from the time of sampling (see Section~\ref{sec:model1}). In (A), the sampled polymorphic site frequency spectra are shown with the first sample in black and the succeeding in grey; in (B) the log ratio between the sampled polymorphic spectra and the equilibrium distribution of the boundary mutation Moran model with the ancestral overall bias-complemented scaled mutation rate is shown in a similar way. The figures (C) show the value of the D-statistic inferred from each sample at the times since the shift, with the vertical grey line marking the most extreme value. The plots in (D) show our inferred values of both the ancestral and the current bias-complemented overall scaled mutation rate from each sample across time, with the true values indicated by the horizontal grey lines.}
   \label{fig:SingleShift}
\end{figure}

By calculating the marginal probabilities for each possible segregating value of the focal allele, an expected biallelic polymorphic sample site frequency spectrum can be constructed and assessed for departure from equilibrium. Visual representations that contrast an observed or simulated distribution of segregating sites with a neutral spectrum are simple and informative \citep{Nawa08, Achaz09}: Fig.~(\ref{fig:SingleShift}) shows a series of expected polymorphic sample spectra drawn from a growing (panel 1A) and a shrinking (panel 2A) population at successive time points after the shift in (effective) population size. The log ratio of these expected sample spectra vs a sample assumed to be in equilibrium at the ancestral overall-biased scaled mutation rate (see \citep[][Eq.~13]{Vogl12}) are depicted in Fig.~(\ref{fig:SingleShift}) panels 1B and 2B; these second plots emphasise the distinct demographic signatures. In Fig.~(\ref{fig:SingleShift}) panels 1A and 1B, which show the samples from a growing population, the presence of excess rare alleles is recognisable. Conversely, Fig.~(\ref{fig:SingleShift}) panels 2A and 2B show an increased proportion of intermediate frequency alleles for the samples from a shrinking population. These are well-established phenomena in population genetics and a battery of neutrality tests have been constructed to detect them \citep{Tajima83,Fu93,Fu95,Fay00, Korneliussen13}, among them Tajima's D. For such tests, the site frequency spectrum is construed as the minor allele frequency distribution and is given by either the number of polymorphic sites $\zeta_k$ at frequency $\frac{k}{K}$, where $k \in [1,K-1]$ when an outgroup is available, or by the number of polymorphic sites  $\zeta_{k,K-k}$ at frequencies $\frac{k}{K}$ and $\frac{K-k}{K}$ when ancestral and derived alleles cannot be distinguished. The classic neutrality tests contrast two estimators for the overall scaled mutation rate $\theta$ that are sensitive to deviations from neutrality in different regions of the allele frequency spectrum. These estimators are constructed by differently weighted linear combinations of $\hat{\theta}_k=k \zeta_k$ \citep{Achaz09}. In Fig.~(\ref{fig:SingleShift}) panels 1C and 2C, a version of Tajima's~D modified for the boundary mutation model, which we will call the D-statistic (see Appendix~\ref{sec:TajMahal} for its derivation), is inferred for the series of samples: This clearly shows how the signal of the demographic event first becomes clearer in the samples over time, and then fades to undetectable as a new equilibrium is approached (compare also the change in deviation from equilibrium over time in Fig.~(\ref{fig:SingleShift}) panels 1A, 1B and 2A, 2B. It is also apparent that populations lose polymorphism much faster than they accrue it (compare \citep{Nei75}), because the drift time scale is much faster than the mutation time scale.

\newpage

\newpage

\subsection{An Agnostic Inference Approach}\label{sec:inference}

Let us now assume a population that undergoes multiple shifts in (effective) population size over time, yielding a history of $J$ demographic epochs, indexed with $1\leq j\leq J$, and let us assume the time points $t_j$ as given, with $t_0=-\infty$, $t_J=0$ and $t_{j-1}<t_j$. The general solution of the time-inhomogeneous equations at $t=t_J$ is then for $n\in 2\mathbb{N}$:
\begin{equation}\label{eq:TauExpansion_inf}
\begin{split}
   \tau_n({t}_{J}) \Xi_n ^{-1} &= \theta_J^{*}\bigg( 1 - e^{-\lambda_n r_{J}}\bigg)
   +\sum_{j=2}^{J-1} \theta_{j}^{*}\bigg(e^{-\lambda_n r_J} - e^{-\lambda_n \sum_{j=1}^{J} r_{j}}\bigg)  \\
   &\qquad \times\theta_{1}^{*}\bigg(e^{-\lambda_n \sum_{j=1}^{J} r_{j}}\bigg)\,,
\end{split}
\end{equation}
where $r_j$ is shorthand for the scaled epoch length $r_j=\tfrac{\theta_{j-1}^{*}}{\theta_j^{*}} s_j$.

It has previously been noted that sampled spectra are not fully informative of a population's demographic history \citep{Myers08}. With orthogonal polynomial expansions, the demographic history is mapped onto a function space spanned by $e^{-\lambda_n r_{j}}$ with $1\leq j\leq J$, which implies several things: Because the exponential function decays rapidly, increasing orders of expansion have a decreasing influence on the shape of the spectrum. Furthermore, demographic histories orthogonal to the function space remain ``hidden'' - in effect, this means past events can cancel each other out and produce a `simplified' historical trajectory. In other words, an infinite number of demographic histories can actually produce the same set of spatial and temporal coefficients and therefore the same observed spectrum. However, it was later shown that the expected sample spectra results uniquely from demographic models as long as these models are defined piecewise using population size functions that do not oscillate/switch signs too often, and as long as the samples are sufficiently large \citep{Bhaskar14}; note, however, that fixed epoch start and end times were assumed. Generally, these past discussions have also always assumed an infinite-sites mutation scheme and corresponding either polarised or unpolarised (\ie{} folded) sample site frequency spectra. Here, we motivate a new analytic inference approach for piecewise constant demographic histories with unknown epoch break points, where the starting point is a biallelic sample spectrum. 

To start, let us return to Eq.~(\ref{eq:TauExpansion_inf}). In order to infer all $\theta_j$ for $1\leq j\leq J$ and therefore all the past (effective) population sizes attained at the end of each epoch, (i) the left hand side of the equation, \ie{} the temporal dynamics, must be determined from the sampled spectrum, and (ii) the $r_{j}$, \ie{} the scaled time points of demographic events, must also be specified in some way. Consider a sampled spectrum of size $K$ comprised of a total of $L$ loci, with $L_{(0)}$ and $L_{(1)}$ loci fixed for the respective alleles as well as $L_{(0,1)}$ polymorphic loci, so that $L=\sum (L_{(0)} + L_{(0,1)} + L_{(1)})$. In the context of the biallelic boundary mutation Moran model, these observed counts of loci are sufficient statistics for the probabilities of the respective monomorphic and polymorphic events, meaning that they contain all the information from the data that can be used to construct estimators of these events \citep{Vogl20}. It follows that $\hat{\alpha}=\tfrac{L_1+(L_{(0,1)}/2)}{L}$ is the minimum variance, unbiased maximum likelihood estimator for the mutation bias \citep{Vogl14a, Vogl20}. Hence, the mutation bias can be immediately estimated from the observed spectrum. Then the polymorphic spectrum can be symmetrised, yielding unbiased sample allele occupancy probabilities for $k=1,...,K-1$. These can be plugged into the left side of the polymorphic equation for expected sample allele frequencies given the past population history from Eq.~(\ref{eq:marg_like_general_geg_2a}). The purely spatial binomial coefficients on the right hand side of this equation can also be readily determined, so only the temporal coefficients $\tau_n(t_J)$ remain unknown. Seen for all the inferred polymorphic allele frequencies simultaneously, Eq.~(\ref{eq:marg_like_general_geg_2a}) clearly describes a consistent system of $K-1$ equations that can be uniquely solved for the $K-1$ temporal coefficients. These are required for the left hand side of the system of equations in Eq.~(\ref{eq:TauExpansion_inf}).

\begin{figure}[!ht]
    \centering
    \includegraphics[width = 8cm, height=8cm]{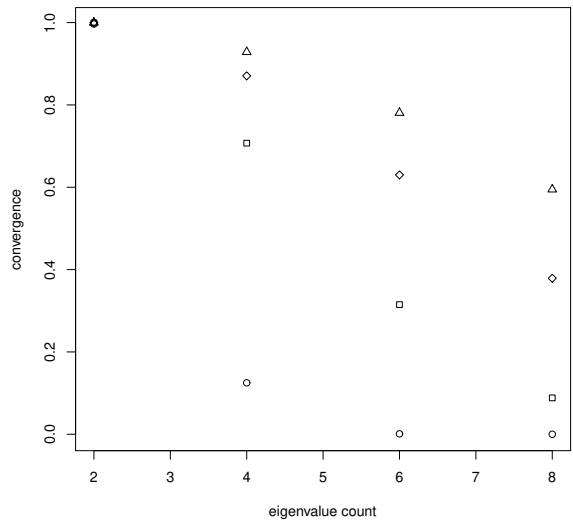}
    \caption{Plotted here are the values of the exponential $e^{-\lambda_n \tfrac{\log(2)}{\lambda_{2j+1}}}$ (on the y-axis) against the orders of expansion $n=2,4,6,8$ (on the x-axis). For each order of expansion, epochs $j=1,2,3,4$ are represented by circles, squares, diamonds, and triangles respectively.}
   \label{fig:Inf}
\end{figure}

 As the number and duration of the past population epochs is generally unknown, specification of the scaled epoch lengths $r_{j}$ must be \emph{ad hoc}. Recall that only the even-order solutions of the temporal equations are non-zero since the mutation bias is assumed constant. For example, a sample spectrum of size $K=6$ is shaped by the polynomials of degree $n=\{2,4,6\}$. Furthermore, for Eq.~(\ref{eq:TauExpansion_inf}) to constitute a unique solution, the number of epochs $J$ plus one (the additional one being for the ancestral state) must be equal to the number of informative polynomial coefficients. In our example with $K=6$, this means we can uniquely infer the current as well as two past overall bias-complemented overall mutation rates. A convenient agnostic placement of the time points at which to make these inferences is inspired by the half-life of exponential decay: Setting $\tilde{r_{j}}=\tfrac{\log(2)}{\lambda_{2j+1}}$, each epoch is placed where half the change in overall-scaled mutation compared to the preceding epoch is expected to have occurred based on the eigenvalues of the observed spectrum. And within epoch, the amount of change captured by each order of expansion is proportional to the corresponding eigenvalue (see Fig.~\ref{fig:Inf}). There is no guarantee that these $r_{j}$ are the true scaled time points at which the demographic events occurred; they are simply ``well placed'' for characterising the effect of exponential change. Note that only the lowest (or to a lesser extent the second lowest) order of expansion from the first epoch significantly influences the shape of the spectrum at the extant time, whereby all (or almost all) orders of expansion from later epochs influence the shape of the spectrum at the extant time, albeit less drastically. 
 
Altogether, we can conclude that \emph{if} we assume the same number of (effective) population sizes - current and historical - as there are even coefficients in the polynomial expansion of the observed site frequency spectrum, \emph{and if} the demographic events that caused the changes in size are assumed to be well placed in the above sense, unique solutions for all current and historical bias-complemented overall mutation rates $\theta_j$ can be found from the system of equations in Eq.~(\ref{eq:TauExpansion_inf}). These resulting estimators for all the $\theta_j$ are derived from sufficient statistics via two consistent systems of equations with unique solutions (one-to-one mappings of parameter spaces); therefore the resulting estimators for the $\theta_j$ are unique minimum unbiased estimators (see \citep[][Chapter 10]{Hogg95}).  

\begin{figure}[!ht]
    \centering
    \includegraphics[width = 5.5cm, height=12.5cm]{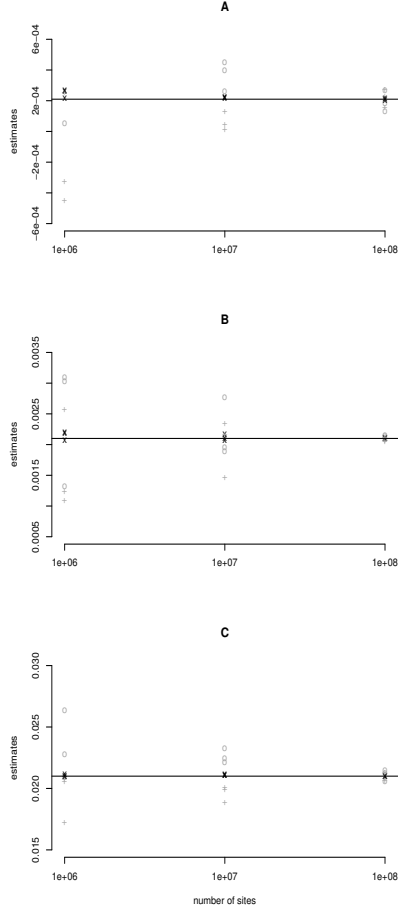}
    \caption{The above figures assess the numerical accuracy of the proposed inference framework by attempting to infer historic overall bias-complemented scaled mutation rates from equilibrium distributions; any deviation in the estimated rates from their equilibrium values must therefore be due to computational inaccuracy. We vary the magnitude of the overall scaled mutation rate and the number of sampled loci to assess which parameter ranges yield the most accurate results. To this end, the polymorphic equilibrium distribution of the discrete boundary-mutation Moran model of size $N=6$ is calculated (using Eq.~\ref{eq:first_order_stationary}) assuming $\theta^{*}=0.0021$ for panel (A), $\theta^{*}=0.021$ for panel (B), and $\theta^{*}=0.21$ for panel (C). Three corresponding sample spectra comprised of $1e^6$, $1e^7$, and $1e^8$ loci respectively are then simulated according to the allele occupancy probabilities $\bar{p_k}$ of each equilibrium distribution from (A)-(C). Then, the current as well as two past overall bias-complemented scaled mutation rates (which should be $\theta_1^{*}=\theta_2^{*}=\theta_3^{*}=\theta^{*}$) are inferred (via the above described method) from each of these three sample spectra for every equilibrium value of the overall scaled mutation rate (A)-(C). This procedure is repeated three times. For each value of the overall scaled mutation rate in (A)-(C), the estimators of the historical overall scaled mutation rates (y-axis) are represented in dependence of the number of loci (x-axis), where we use the symbols ``+'', ``x'', and ``o'' to denote estimates of $\theta_1^{*}$, $\theta_2^{*}$, and $\theta_3^{*}$ respectively. The horizontal lines in (A)-(C) mark the true/equilibrium values of the overall scaled bias-complemented mutation rates.}
   \label{fig:InfAccuracy}
\end{figure}

Both the accuracy and potential for wider application of this method are, as yet, limited by numerical instability: For our previous example with sample size $K=6$, a total number of loci in the order of $10^{8}$ is required for the accuracy of all three estimators to be adequate (see Fig.~\ref{fig:InfAccuracy}). This is, however, the upper limit of loci in a site frequency spectrum that can be simulated using the open software program R. It is clear that the feasibility of large-scale use of this method hinges not only on specialised implementation being developed in the future, but also on the availability of sets with a large enough number of loci. 

In Fig.~(\ref{fig:SingleShift} panels 1D and 2D), the sample allele occupancy probabilities from the single-shift examples of the previous subsection are hypergeometrically downsampled to $K=4$ and used to generate spectra of $2\cdot 10^8$ independent loci, from which both the extant and the ancestral overall bias-complemented scaled mutation rate are inferred. This novel inference approach infers the true values most accurately a touch sooner after the change in (effective) population size than the D-statistic detects the greatest signal of departure from equilibrium (compare Fig.~\ref{fig:SingleShift} panels 1C and 2C). Note the pattern of inferring slightly more extreme values for the ancestral mutation rate before and for the current mutation rate after this point to compensate for the positioning of the hypothetical epoch split time. 

\newpage

\newpage

\newpage

\subsection{Model 2: Deterministically Alternating Population Sizes}\label{sec:model2a}

In this subsection, we construct a caricature model of boom-bust population size dynamics: To do so, we assume a population that has haploid (effective) size $N_u$ for an epoch of length $s_u$ (where $u$ stands for the b$u$st epoch), and then switches instantaneously to having haploid (effective) size $N_o$ for an epoch of length $s_o$ (where $o$ stands for the b$o$om epoch). Together, these two epochs form a cycle that repeats indefinitely. 

Classic population genetic results apply to the overall (effective) size of this population: If the epochs are short compared to generation length (which is the inverse population size in the diffusion setting), the long-term effective size can be approximated by the harmonic mean across the two epochs \citep{Wright38b, Kimura63}. The heterozygosity of populations undergoing deterministic, cyclical changes in (effective) size has been studied extensively for various trajectories of rapid population growth followed by instantaneous collapse; the harmonic mean approximation holds both without mutations and when novel mutations enter at a low rate per locus per generation \citep{Nei75, Motro82}. Both the ratio between the maximum and minimum heterozygosity within each cycle as well as the ratio between the maximum heterozygosity and the harmonic average heterozygosity within each cycle increase with the severity and duration of the population collapse \citep{Nei75, Motro82}. We will examine both the expected sample polymorphic site frequency spectrum of populations undergoing boom-bust life cycles as well as the temporal part of their spectral decomposition.  

The harmonic mean of the (effective) population size across each cycle is: 
$$
    N_h=\frac{s_u+s_o}{\tfrac{s_u}{N_u}+\tfrac{s_o}{N_o}}\,.
$$ 
Diffusion time within each epoch can be scaled relative to this harmonic mean so that the scaled duration of the epochs become $s_u\tfrac{N_h}{N_u}$ and $s_o\tfrac{N_h}{N_o}$, \ie{} in this subsection only, time is scaled in units of the harmonic mean effective population size. Similarly, the harmonic mean of the overall bias-complemented scaled mutation rates is defined as: 
$$
\theta_h^{*}=\mu N_h=\frac{s_u+s_o}{\frac{s_u}{\theta_{s_u,n}^{*}}+\frac{s_o}{\theta_{s_o,n}^{*}}}\,.
$$ 

Then the solutions to the linear differential equations for the temporal dynamics of the boom and bust phases can be written as the following for $j\in 2\mathbb{N}$:
\begin{equation}\label{eq:TauExpansion_bb_solution}
\begin{split}
   \tau_n(t)\Xi_n^{-1}&= \theta_u^{*} - \bigg(\theta_u^{*}-\tau_n(t = t_{j-1})\Xi_n^{-1}\bigg)e^{-\lambda_n s_u\frac{\theta_h^{*}}{\theta_u^{*}} (t_{j-1})}\\ 
   \tau_n(t)\Xi_n^{-1}&= \theta_o^{*} - \bigg(\theta_o^{*}-\tau_n(t_j)\Xi_n^{-1}\bigg)e^{-\lambda_n s_o\frac{\theta_h^{*}}{\theta_o^{*}} (t_j)}\text{, for $t_{j} <t \leq t_{j+1}$.}
\end{split}
\end{equation}

As the number of cyclical iterations increases with $j\rightarrow \infty$, the temporal dynamics of the even and odd epochs evolve towards two separate solutions at $t_{j}$ and $t_{j+1}$, which we will denote as $\vartheta_{s_u,n}=\lim_{j\to\infty}\tau_n(t_j)\Xi_n^{-1}$ and  $\vartheta_{s_o,n}=\lim_{j\to\infty}\tau_n(t_{j-1})\Xi_n^{-1}$, respectively. These solutions follow the system of equations:
%
\begin{equation}\label{eq:bb_lindiffeqs3}
    \begin{split}
         \vartheta_{s_u,n}&=\theta_u^{*}-\bigg(\theta_u^{*}-\vartheta_{s_o,n}\bigg) e^{-\lambda_n\tfrac{\theta_h^{*}}{\theta_u^{*}}s_u}\\
        \vartheta_{s_o,n}&=\theta_o^{*}-\bigg(\theta_o^{*}-\vartheta_{s_u,n}\bigg)e^{\lambda_n\tfrac{\theta_h^{*}}{\theta_o^{*}}s_o}\,.
    \end{split}
\end{equation}

This system is easily solved: 
\begin{equation}\label{eq:system_lindiffeqs4}
    \begin{split}
        (\theta_u^{*}- \theta_o^{*})&=-\bigg(\vartheta_{s_o,n}- \theta_u^{*}\bigg) e^{-\lambda_n \tfrac{\theta_h^{*}}{\theta_u^{*}}s_u}+\bigg(\vartheta_{s_o,n}-\theta_o^{*}\bigg)e^{\lambda_n \tfrac{\theta_h^{*}}{\theta_o^{*}}s_o}\\
       \vartheta_{s_o,n}&=\frac{ \theta_u^{*}(1-e^{-\lambda_n \tfrac{\theta_h^{*}}{\theta_u^{*}}s_u})+ \theta_o^{*} (e^{\lambda_n \tfrac{\theta_h^{*}}{\theta_o^{*}}s_o}-1)}{(1-e^{-\lambda_n \tfrac{\theta_h^{*}}{\theta_u^{*}}s_u})+(e^{\lambda_n \tfrac{\theta_h^{*}}{\theta_o^{*}}s_o}-1)}\\
        \vartheta_{s_o,n}&= \frac{ \theta_u^{*}(1-e^{-\lambda_n \tfrac{\theta_h^{*}}{\theta_u^{*}}s_u})- \theta_o^{*}(1-e^{\lambda_n \tfrac{\theta_h^{*}}{\theta_o^{*}}s_o})}{(1-e^{-\lambda_n \tfrac{\theta_h^{*}}{\theta_u^{*}}s_u})-(1-e^{\lambda_n \tfrac{\theta_h^{*}}{\theta_o^{*}}s_o})}\,,
    \end{split}
\end{equation}
and 
\begin{equation}\label{eq:system_lindiffeqs5}
    \begin{split}
        \vartheta_{s_u,n}=\frac{ \theta_o^{*}(1-e^{-\lambda_n \tfrac{\theta_h^{*}}{\theta_o^{*}}s_o})- \theta_u^{*}(1-e^{\lambda_n \tfrac{\theta_h^{*}}{\theta_u^{*}}s_u})}{(1-e^{-\lambda_n \tfrac{\theta_h^{*}}{\theta_o^{*}}s_o})-(1-e^{\lambda_n \tfrac{\theta_h^{*}}{\theta_u^{*}}s_u})}\,.
    \end{split}
\end{equation}

Thus, analytical expressions for the long-term solutions of the temporal dynamics of populations of deterministically alternating (effective) population size have been found. When the epoch lengths are relatively short, so that $s_{max}\lambda_n\ll 1$ with $s_{max}=\operatorname{max}(s_u\tfrac{\theta_u^{*}}{\theta_h^{*}},s_o\tfrac{\theta_o^{*}}{\theta_h^{*}})$, the exponential decay of relative mutation effects in each epoch can be approximated by first order Taylor series expansions:
\begin{equation}\label{eq:bb_taylor1}
\begin{split}
    \vartheta_{s_o,n}
    &= \frac{\theta_u^{*}\lambda_n\tfrac{\theta_h^{*}}{\theta_u^{*}}s_u+
    \theta_o^{*}\lambda_n\tfrac{\theta_h^{*}}{\theta_o^{*}}s_o}
    {\lambda_n \tfrac{\theta_h^{*}}{\theta_u^{*}}s_u+\lambda_n\tfrac{\theta_h^{*}}{\theta_o^{*}}s_o}+\mathcal{O}(s_{max}^2)\\
    &=\frac{s_u+s_o}{\frac{s_u}{\theta_{s_u,n}^{*}}+\frac{s_o}{\theta_{s_o,n}^{*}}} +\mathcal{O}(s_{max}^2)\\
    &=\theta_h^{*} +\mathcal{O}(s_{max}^2)\,,
    \end{split}
\end{equation}
and equivalently:
\begin{equation}\label{eq:bb_taylor2}
    \begin{split}
       \vartheta_{s_u,n}&=\theta_h^{*}+\mathcal{O}(s_{max}^2)\,.
    \end{split}
\end{equation}
Essentially, rapidly alternating boom and bust phases ultimately result in the temporal coefficients of both epochs converging towards the harmonic mean of the individual overall bias-complemented scaled mutation rates. The effective population size is then accurately captured by $N_h$, as anticipated.

Conversely, it follows immediately from Eq.~(\ref{eq:bb_lindiffeqs3}) that for long epoch lengths $s_{min}\lambda_n\gg 1$ where $s_{min}=\operatorname{min}(s_u\tfrac{\theta_u^{*}}{\theta_h^{*}},s_o\tfrac{\theta_o^{*}}{\theta_h^{*}})$, the temporal coefficients of the epochs converge to their individual equilibrium solutions:
\begin{equation}\label{eq:bb_arith}
\begin{split}
    \vartheta_{s_o,n}&= \theta_o^{*}\\
    \vartheta_{s_u,n}&= \theta_u^{*}\,.
\end{split}
\end{equation}
Then the arithmetic mean of $\vartheta_{s_u,n}$ and $\vartheta_{s_o,n}$ is the appropriate approximation of the expected long-term effective population size, again as anticipated.

\subsubsection{Results and Discussion for Model~2}

\begin{figure}[!ht]
    \centering
    \includegraphics[width = 11.5cm]{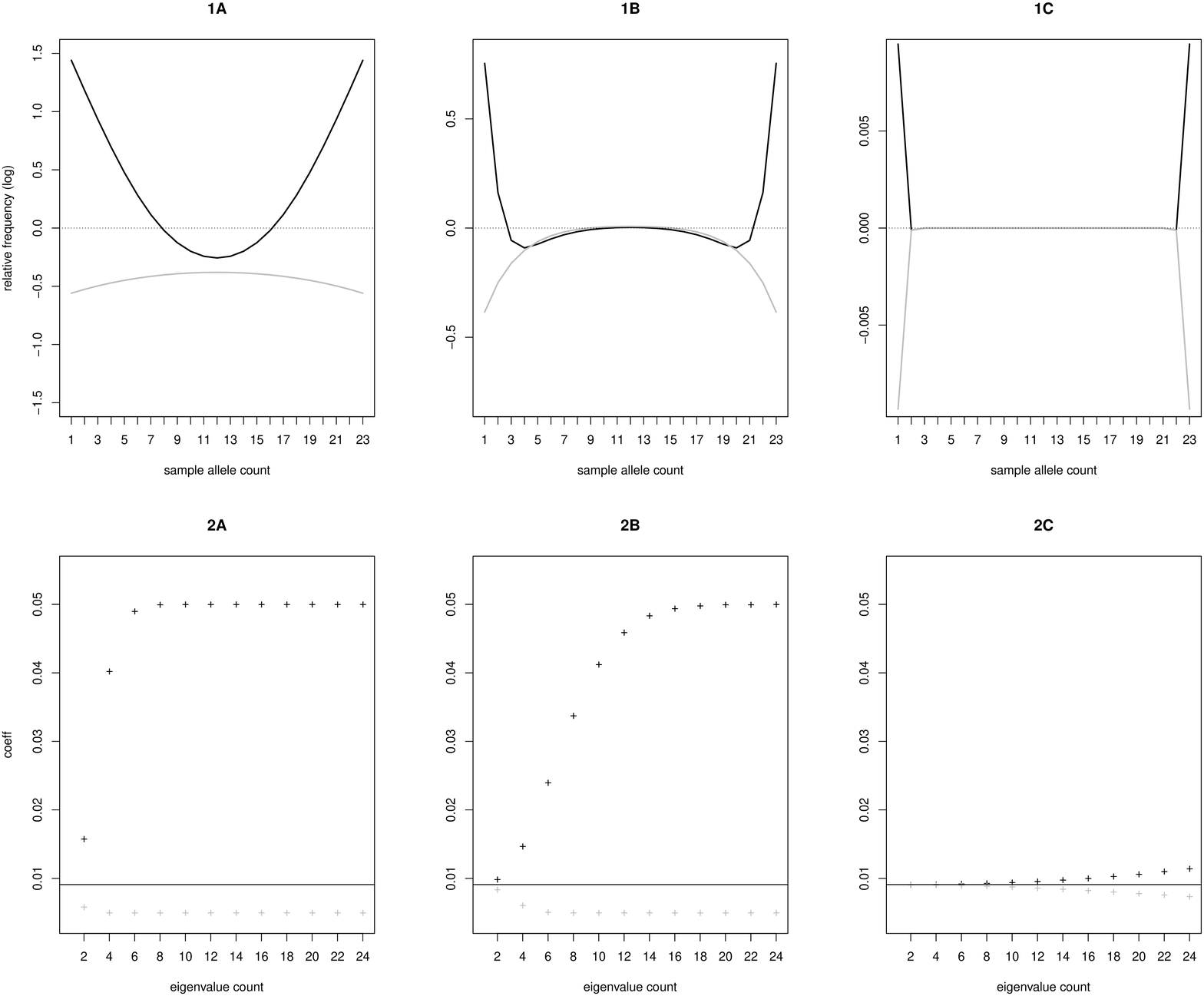}
    \caption{Samples of size $K=24$ drawn from a population evolving according to the caricature boom-bust model with $\theta^{*}_{o}=0.05$ in the boom epoch and $\theta^{*}_{u}=0.005$ in the bust epoch. In (1), the log-ratio of the expected sample site frequency spectra during the boom and bust epochs vs the equilibrium distribution at the harmonic mean of the overall bias-complemented scaled mutation rates are shown in black and grey respectively. The epoch lengths are set to  $s_u=s_o=7\cdot 10^{-1}$ (A),  $s_u=s_o=1\cdot 10^{-1}$ (B), and  $s_u=s_o=1\cdot 10^{-3}$ (C). In (2), the temporal coefficients are plotted for both boom (Eq.~(\ref{eq:system_lindiffeqs4}); in black) and bust (Eq.~(\ref{eq:system_lindiffeqs5}); in grey) epochs for each order of expansion.}
    \label{fig:BoomBust}
\end{figure}

In Fig.~\ref{fig:BoomBust} panels 1A-1C, the expected polymorphic sample spectra from boom and bust epochs are plotted relative to the harmonic mean for relatively long (A), intermediate (B), and short (C) epoch lengths compared to the average overall bias-complemented scaled mutation rate. Recall that this is easily achieved by plugging the long-term solution of the temporal dynamics of the boom (Eq.~\ref{eq:system_lindiffeqs4}) and bust epochs (Eq.~\ref{eq:system_lindiffeqs5}) into Eq.~(\ref{eq:marg_like_general_geg_2a}). Technically, according to the aforementioned critera, both scenarios A and B qualify as intermediate: In A, only the expansion order $n=2$ can be approximated by the harmonic mean with the remainder likely well-approximated by the arithmetic mean. Meanwhile in B, only the expansion order $n=2$ fulfills the approximation criteria of the harmonic mean but several higher order expansion are no longer accurately represented by the arithmetic mean. In panel 1A, the sample spectra from the boom epochs show the u-shape characteristic of population growth with an excess of low and dearth of high frequency alleles; similarly, the bust epochs show a flat, inverted u-shape that consistently lies below the harmonic equilibrium, characteristic of population contraction. Only the lowest order temporal coefficients have started converging towards the harmonic equilibrium (see panel 2A). The result of the D-statistics reflect these patterns (see Table.~\ref{tab:BoomBust_TajMahal}). 
In our intermediate scenario B, the sample spectrum of the boom epoch has a clear w-shape; the sample spectrum of the bust epoch simply shows a more pronounced inverse u-shape with the peak hitting the harmonic mean. For these epoch lengths, the lowest order temporal coefficients are almost equal to the harmonic mean, and some but not all the higher order coefficients have started converging (see panel 2B). In \citet{Nawa08}, visual assessment of the spectrum of segregating sites reveals w-shapes for populations recovering from a bottleneck: the number of intermediate alleles increase faster than the number of minor alleles, leading to confused signals in the D-statistic (see \citep[][Table 2]{Nawa08} alongside our Table~\ref{tab:BoomBust_TajMahal}). This w-shape becomes more pronounced with bottlenecks that are further back in the population history, are comparatively short, or are less extreme in terms of collapse in population size. Bottlenecks with the opposite characteristics lead to such an extreme w-shape that it becomes an inverted u-shape. The bust epoch can therefore be considered a mild, recurrent bottleneck. In  scenario C, where the epoch lengths are relatively short, all but the highest orders of expansion can be approximated by the harmonic mean. Both sample spectra deviate from the harmonic mean only by a small enrichment or diminution in the proportion of singletons respectively (see Table~\ref{tab:BoomBust_TajMahal}).

In Table~\ref{tab:BoomBust_Inf}, the results of applying our agnostic inference approach from Section~\ref{sec:inference}) to sample spectra drawn from both the boom and the bust phases are presented for each of the previous epoch length scenarios. In order to do this, the spectra are hypergeometrically downsampled to size $K=4$ so that the current and one past overall-biased mutation rate can be inferred. The results gained from the sample spectra drawn from the bust phase are reasonably informative of the underlying boom-bust population cycle: The results for long epoch lengths are near the individual equilibrium overall bias-complemented epoch lengths, and convergence towards the harmonic mean is noticeable as epoch lengths decrease. The results from the bust phase are less informative: Because the collapse in (effective) population size is so rapid (with $\exp\big(-\lambda_n \tfrac{\log(2)}{\lambda_{2j+1}}\big) \ll \exp\big(-\lambda_n s_u\frac{\theta_h^{*}}{\theta_u^{*}} t\big)$ for all $j$ when $n>2$), the spectrum contains little to no historical information. So while the estimators for the current overall bias-complemented scaled mutation rates are reasonable, the past estimates (inferred from not very far back in time) are no more than an indication that the (effective) population size was formerly larger.        

\begin{table}[!ht]
\centering
\begin{tabular}{l|lll}
\hline
 & boom epoch ($\theta^{*}_{1}=0.05$) & bust epoch ($\theta^{*}_{0}=0.005$) &  \\ \hline
 (A) $s_u=s_o=7e^{-1}$ & -0.2140 & 0.01351 \\
 (B) $s_u=s_o=1e^{-1}$& 0.0008& 0.03521 \\
 (C) $s_u=s_o=1e^{-3}$& -0.0008 & 0.0008 \\ \hline
\end{tabular}
\caption[]{D-statistic for sampled spectra from Fig.~(\ref{fig:BoomBust})}
\label{tab:BoomBust_TajMahal}
\end{table}

\begin{table}[!ht]
\centering
\begin{tabular}{l|lll}
\hline
 sample from boom epoch & $\widehat{\theta^{*}}_{1}$ & $\widehat{\theta^{*}}_{0}$ & \\ \hline
 (A) $s_u=s_o=7e^{-1}$& 0.00648 & 0.05151 \\
 (B) $s_u=s_o=1e^{-1}$& 0.00805 & 0.01692\\
 (C) $s_u=s_o=1e^{-3}$& 0.00908 & 0.00918 \\ \hline
  sample from bust epoch & $\widehat{\theta^{*}}_{0}$ & $\widehat{\theta^{*}}_{1}$ & \\ \hline
  (A) $s_u=s_o=7e^{-1}$ & 0.00463 & 0.00617 \\
 (B) $s_u=s_o=1e^{-1}$& 0.00503 & 0.00926 \\
 (C) $s_u=s_o=1e^{-3}$& 0.00898 & 0.00911 \\ \hline\end{tabular}
 \caption[]{Inference of temporal coefficients for sampled spectra from Fig.~(\ref{fig:BoomBust})}
\label{tab:BoomBust_Inf}
\end{table}

\newpage

\newpage

Generally, deterministic boom-bust type models are applicable to organisms that experience major, semi-regular environmental events on a time scale that is long compared to their generation lengths. For example, the El Nino currents in the equatorial pacific occur roughly every four years and are known to impact heterozygosity in some insect populations \citep{Franca20}. With the annual growth and subsequent collapse of naturally occurring populations of \textit{Drosophila} in temperate climates, however, generation times are short compared to the scale of seasonal variation in the environment, making seasonal adaptation possible \citep{Machado21}. This comparatively rapid turnover in generations implies that the difference between the (effective) sizes of boom and bust epochs will often be of several orders of magnitude, so $N_u/N_o\ll 1$. Realistic values may be $N_u/N_o\leq 10^{-4}$ with $\theta^{*}_{h}=0.01$ and epoch lengths of roughly $\tfrac{10}{N_h}$.
From such large ratios between $N_u:N_o$, it follows that $\frac{\theta_h^{*}}{\theta_o^{*}} \approx 0$. Note that then the long-term temporal coefficients of the boom and bust epochs (Eqs.~\ref{eq:system_lindiffeqs4},~\ref{eq:system_lindiffeqs5}) can be approximated by:
\begin{equation}\label{eq:system_lindiffeqs5_approx1a}
    \begin{split}
        \vartheta_{s_o,n}= \frac{\lambda_n \theta_h^{*} s_o + \theta_u^{*} \bigg(1-e^{-\lambda_n \tfrac{\theta_h^{*}}{\theta_u^{*}}s_u}\bigg)}{\bigg(1-e^{-\lambda_n \tfrac{\theta_h^{*}}{\theta_u^{*}}s_u}\bigg)}+\mathcal{O}\bigg(\frac{\theta_h^{*}}{\theta_o^{*}}\bigg)\\
    \end{split}
\end{equation}
and
\begin{equation}\label{eq:system_lindiffeqs5_approx1b}
    \begin{split}
        \vartheta_{s_u,n}= \frac{\lambda_n \theta_h^{*} s_o - \theta_u^{*}\bigg(1-e^{\lambda_n \tfrac{\theta_h^{*}}{\theta_u^{*}}s_u}\bigg)}{-\bigg(1-e^{\lambda_n \tfrac{\theta_h^{*}}{\theta_u^{*}}s_u}\bigg)}+\mathcal{O}\bigg(\frac{\theta_h^{*}}{\theta_o^{*}}\bigg)\,.
    \end{split}
\end{equation}
Furthermore, observe that $\theta_h^{*} =\theta^{*}_{s_u,n}\frac{s_u+s_o}{s_u}+\mathcal{O}(\frac{\theta_h^{*}}{\theta_o^{*}})$. Substituting this, the difference between the long-term dynamics of the epochs becomes:
\begin{equation}\label{eq:system_lindiffeqs5_approxdiff}
    \begin{split}
        \vartheta_{s_o,n} - \vartheta_{s_u,n}&=\frac{\lambda_n \theta_u^{*}\frac{s_u+s_o}{s_u}s_o + \theta_u^{*}\bigg(1-e^{-\lambda_n (s_u+s_o)}\bigg)}{\bigg(1-e^{-\lambda_n (s_u+s_o)}\bigg)}\\ 
        &\qquad-\frac{-\lambda_n \theta_u^{*}  \frac{s_u+s_o}{s_u}s_o + \theta_u^{*}\bigg(1-e^{\lambda_n (s_u+s_o)}\bigg)}{\bigg(1-e^{\lambda_n (s_u+s_o)}\bigg)}+\mathcal{O}\bigg(\frac{\theta_h^{*}}{\theta_o^{*}}\bigg)\\
    &=\lambda_n \theta_u^{*} \frac{s_u+s_o}{s_u}s_o \Bigg(\frac{1}{1-e^{-\lambda_n  (s_u+s_o)}}+\frac{1}{1-e^{\lambda_n (s_u+s_o)}}\Bigg)+\mathcal{O}\bigg(\frac{\theta_h^{*}}{\theta_o^{*}}\bigg)\\
    &=\lambda_n \theta_u^{*} \frac{s_u+s_o}{s_u}s_o+\mathcal{O}\bigg(\frac{\theta_h^{*}}{\theta_o^{*}}\bigg)\,.
    \end{split}
\end{equation}


The difference between the long-term temporal mutation dynamics of the boom and bust epochs is therefore independent of the overall bias-complemented scaled mutation rate of the boom epoch. Rather, for each order $n$ of the temporal expansion, the corresponding eigenvalue is scaled by a term comprised of the overall bias-complemented scaled mutation rate of the bust epoch, the cycle length, and the duration of the boom epoch relative to the bust epoch. While this effect was apparent from past simulations \citep{Motro82}, we have now shown it analytically. The accuracy of this final expression depends on the conditions ahead of Eq.~(\ref{eq:system_lindiffeqs5_approx1a}) and Eq.~(\ref{eq:system_lindiffeqs5_approx1b}). If necessary, upper bounds for accuracy can be determined via the error expansion of the Taylor series. Note that one can see from these last calculations that unrealistically large sample sizes may be necessary to be able to detect the signatures of the boom-bust dynamics in the spectra of \textit{Drosophila}, since the number of generations per year will be very much lower than $N_h$.

\subsection{Model 3: Serially Correlated Population Sizes}\label{sec:model3}

\subsubsection{Autoregression Model for Temporal Dynamics}

In Sections~\ref{sec:model1},\ref{sec:model2a}, populations are subject to major and predictable changes in (effective) population size. However, most fluctuations in population size are responses to demographic \citep{Lee11} or environmental \citep{Lande14} changes that are not necessarily regular over long time periods. While these fluctuations can be crucial to the fate of the populations, they are generally not easily quantifiable. In the following, stochastic variation in the overall bias-complemented scaled mutation rate is incorporated into the system of temporal equations of the forwards population process from Eq.~(\ref{eq:TauExpansion_eqi}). 

Let us denote the long-term solution for the temporal coefficient of $\tau_n(t)$ for each epoch $t_{j-1} < t \leq t_{j}$ as $\vartheta_{j,n}\Xi_n$. Furthermore, let us fix the relative scaled epoch length $c=s_{j}4N_j\alpha\beta\mu$. This relative scaled epoch length is independent of $j$ because the epoch lengths $s_{j}$ scale with $1/N_j$ on the continuum time scale and we assume that the number of generations (and thus the mutation parameters) is the same in each epoch. The resulting long-term temporal coefficients solve the following equation for $n\in 2\mathbb{N}$ and $j>0$ (compare Eqs.~\ref{eq:TauExpansion_eqi},\ref{eq:bb_lindiffeqs3}):
\begin{equation}\label{eq:system_detAR}
\begin{split}
   &\vartheta_{j,n} = \vartheta_{j-1,n} e^{-\lambda_n \frac{c}{ \theta_j^{*}}} + \theta_j^{*} \bigg(1-e^{-\lambda_n \frac{c}{ \theta_j^{*}}} \bigg)\,.
\end{split}
\end{equation}
This equation has the general form $\gamma_j=\gamma_{j-1}m_j+\epsilon_j(1-m_j)$ and is therefore a first order autoregression model. Because $|m_j|<1$, the process is also stationary, \ie{} its moments are independent of $j$ \citep[][Chapter 7]{Johnston60}. In the classic case, $\epsilon_j$ enters the process without a coefficient and is a serially uncorrelated, zero-mean and constant variance stochastic process (\ie{} white noise). Here, it corresponds to $\theta_j^{*}$, and we must characterise its distribution in order to make statements about $\vartheta_{j,n}$. Note that the occurrence of the overall bias-complemented scaled mutation rates $\theta_j^{*}$ in Eq.~(\ref{eq:system_detAR}) is reminiscent of an inverse gamma distributed variable. Therefore, we assume for $j>0$:
\begin{equation}\label{eq:IG}
\begin{split}
     &\theta_j^{*} \overset{iid}\sim invgamma(a,b) \text{\small{ with $a>2,b>0$}}\\
     &p(\theta^{*})=\Pr(\theta^{*}_j=\theta^{*} \given a,b)=\frac{b^a}{\Gamma(a)}\bigg(\frac{1}{ \theta^{*}}\bigg)^{a+1}e^{\frac{-b}{ \theta^{*}}}
    \end{split}
\end{equation}
The range of the hyperparameter $a$ is restricted to ensure existence of the mean and the variance.

\subsubsection{Characterising the Solutions to the Temporal Dynamics}
Next, we will use the distribution of the overall scaled mutation parameter $\theta_j^{*}$ together with the properties of the first-order autoregression processes to determine the mean, variances, and covariances of the solutions of the long-term temporal coefficients $\vartheta_{j,n}$. 

\paragraph{Mean}
Returning to the autoregression process in Eq.~(\ref{eq:system_detAR}) and invoking stationarity, we have for $n\in 2\mathbb{N}$ and $j>0$:
\begin{equation}\label{eq:system_detAR_mean1}
    \begin{split}
    \E_\vartheta(\vartheta_{j,n})-\E_\vartheta(\vartheta_{j-1,n})&=0\\
     \E_\vartheta\E_{\theta^{*}}(\vartheta_{j,n})-\E_\vartheta
     E_{\theta^{*}}(\vartheta_{j-1,n})&=\E_\vartheta\E_{\theta^{*}}\bigg((\theta^{*} - \vartheta_{j,n})(1-e^{-\lambda_n \frac{c}{\theta^{*}}})\bigg)=0\\
     \E_\vartheta(\vartheta_{j,n})\E_{\theta^{*}}\bigg(1-e^{-\lambda_n \frac{c}{\theta^{*}}}\bigg) &= \E_{\theta^{*}}(\theta^{*}) \E_{\theta^{*}}\bigg(1-e^{- \lambda_n \frac{c}{\theta^{*}´}}\bigg)
    \end{split}
\end{equation}
Recall that $\E_{\theta^{*}}=\int_0^{\infty}p(\theta^{*})d\theta^{*}$, where $p(\theta^{*})$ is the density of the inverse gamma distribution from Eq.~(\ref{eq:IG}). By straightforward application of the substitution and integration by parts rules for evaluation of each integral, we obtain for $n\in 2\mathbb{N}$ and $j>0$:
\begin{equation}\label{eq:detAR_mean2}
    \begin{split}
       \E_\vartheta(\vartheta_{j,n})\bigg(1-\frac{b^a}{(b+\lambda_n c)^a}\bigg)&= \frac{b}{(a-1)}- \frac{1}{(a-1)}\frac{b^a}{(b+\lambda_n c)^{a-1}}\\
       \E_\vartheta(\vartheta_{j,n})\frac{(b+\lambda_n c)^{a}-b^{a}}{(b+\lambda_n c)^{a}}&= \frac{b}{(a-1)}\bigg( \frac{(b+\lambda_n c)^{a-1}-b^{a-1}}{(b+\lambda_n c)^{a-1}} \bigg)\\
        \E_\vartheta(\vartheta_{j,n})&= \frac{b}{(a-1)}\bigg( \frac{(b+\lambda_n c)^{a}-(b+\lambda_n c)b^{a-1}}{(b+\lambda_n c)^{a}-b^{a}}\bigg)\\
        &= \frac{b}{(a-1)}\bigg(1- \frac{\lambda_n cb^{a-1}}{(b+\lambda_n c)^{a}-b^{a}}\bigg)\\
    \end{split}
\end{equation}
Using L'H\^{o}pital's rule, the limiting results for short and long relative epoch lengths (scaled by the eigenvalues) can be determined:
\begin{equation}\label{eq:detAR_mean_lims}
    \begin{split}
        &\lim_{\lambda_n c \rightarrow 0} \E_\vartheta(\vartheta_{j,n})= \lim_{\lambda_n c \rightarrow 0} \frac{b}{a-1}\bigg(1- \frac{b^{a-1}}{a(b+\lambda_n c)^{a-1}}\bigg)=\frac{b}{a-1}\bigg(1- \frac{1}{a}\bigg)
        =\frac{b}{a}\\
        &\lim_{\lambda_n c \rightarrow \infty } \E_\vartheta(\vartheta_{j,n})=\lim_{\lambda_n c \rightarrow \infty} \frac{b}{a-1}\bigg(1- \frac{b^{a-1}}{a(b+\lambda_n c)^{a-1}}\bigg)=\frac{b}{a-1}\,.
    \end{split}
\end{equation}
When epochs are relatively short, the expected value of the long-term temporal coefficients converges towards the harmonic mean of the inverse gamma distribution that was chosen to model the distribution of the overall bias-complemented scaled mutation parameter. When epochs are relatively long, convergence is towards the arithmetic mean of this inverse gamma distribution. Similar results have been obtained for the heterozygosity and the effective size of populations with stochastic cycles of varying size within the framework of the (Wright-Fisher) diffusion \citep{Iizuka01,Iizuka02,Iizuka10}: There, predetermined potential population sizes can easily be modelled as finite states of a Markov Chain. Both the effective size of the population and its heterozygosity generally converge towards the harmonic means when the autocorrelation of transitions between population size states is low. In contrast, high (positive) autocorrelation generates convergence to the arithmetic mean (which is always greater than the harmonic mean), and, when the scenario permits negative autocorrelation, this can lead to results that lie below the harmonic mean. 

\paragraph{Variance}
The method for calculating the variance of the solutions to the long-term temporal coefficients $\vartheta_{j,n}$ is analogous to that for determining the mean. Starting again from Eq.~(\ref{eq:system_detAR}) and using the stationarity of the process for $n\in 2\mathbb{N}$ and $j>0$:
\begin{equation}\label{eq:system_detAR_usqu1}
    \begin{split}
    \E_\vartheta(\vartheta_{j,n}^{2}) \E_{\theta^{*}}\bigg(1-e^{-\lambda_n\frac{c}{\theta^{*}}}\bigg)&=2 \E_{\theta^{*}}\bigg((1-e^{-\frac{c}{\theta^{*}}})e^{-\frac{c}{\theta^{*}}}\theta^{*}\bigg)\E_\vartheta(\vartheta_{j,n})\\
    &\qquad+ \E_{\theta^{*}}({\theta^{*}}^2)\E_{\theta^{*}}\bigg(1-2e^{-\frac{c}{\theta^{*}}}+e^{-\lambda_n\frac{c}{\theta^{*}}}\bigg)\,.
    \end{split}
\end{equation}
Evaluating integrals as before, one obtains for $n\in 2\mathbb{N}$ and $j>0$:
\begin{equation}\label{eq:system_detAR_usqu2}
    \begin{split}
       \E_\vartheta(\vartheta_{j,n}^{2})&=\frac{1}{1-\frac{b^a}{(b+2\lambda_n c)^a}}
        \bigg(\frac{2}{(a-1)}\bigg(\frac{b^a}{(b+\lambda_n c)^{a-1}}-\frac{b^a}{(b+2\lambda_n c)^{a-1}}\bigg)\E_\vartheta(\vartheta_{j,n})\\
        &\qquad+
       \frac{b^a}{(a-1)(a-2)}-\frac{2b^a}{(a-1)(a-2)(b+\lambda_n c)^{a-2}}\\
       &\qquad+\frac{b^a}{(a-1)(a-2)(b+2\lambda_n c)^{a-2}}\bigg)\,.
    \end{split}
\end{equation}
Clearly, the variance can then be determined by the law of total variance; its limiting values for both short and long relative scaled epoch lengths follow directly, for $n\in 2\mathbb{N}$ and $j>0$:
\begin{equation}\label{eq:system_detAR_var}
\begin{split}
        &\Var_\vartheta(\vartheta_{j,n}) = \E_\vartheta(\vartheta_{j,n}^{2}) - \E_\vartheta(\vartheta_{j,n})^2\\
        &\lim_{\lambda_n c \rightarrow 0} \Var_\vartheta(\vartheta_{j,n})=0\\
        &\lim_{\lambda_n c \rightarrow \infty} \Var_\vartheta(\vartheta_{j,n})= \frac{b^2}{(a-1)(a-2)} - \bigg(\frac{b}{(a-1)}\bigg)^2 =\frac{b^2}{(a-1)^2(a-2)}\,.
    \end{split}
\end{equation}

\paragraph{Correlations and Covariances}
Note that we have implicitly assumed that the long-term coefficients $\vartheta_{j+r,n}$ are pairwise-uncorrelated for $r>0$, since there is serial independence between the overall bias-complemented scaled mutation parameters $\theta_j^{*}$ of successive epochs. Using these properties together with the law of total variance, the first order covariances (\ie{} those between $\vartheta_{j+1,n}$ and $\vartheta_{j,n}$) are easily determined, for $n\in 2\mathbb{N}$ and $j>0$:
\begin{equation}\label{eq:system_detAR_cov}
    \begin{split}
        \Phi(r=1)&=\Cov_\vartheta(\vartheta_{j+1,n},\vartheta_{j,n})\\
        &=\Cov_\vartheta(\vartheta_{j,n}e^{-\lambda_n\frac{c}{\theta^{*}_j}}+\theta^{*}_j(1-e^{-\lambda_n\frac{c}{\theta^{*}_j}}),\vartheta_{j,n})\\
        &=\Cov_\vartheta(\vartheta_{j,n}e^{-\lambda_n\frac{c}{\theta^{*}_j}},\vartheta_{j,n})+ \Cov_\vartheta((1-e^{-\lambda_n\frac{c}{\theta^{*}_{j}}})\theta^{*}_j,\vartheta_{j,n})\\
        &=\Cov_\vartheta(\vartheta_{j,n}e^{-\lambda_n\frac{c}{\theta^{*}_j}},\vartheta_{j,n})+ 0\\
        &= \E_\vartheta\E_{\theta^{*}}(e^{-\lambda_n\frac{c}{\theta^{*}}}\vartheta_{j,n}^{2})- \E_\vartheta\E_{\theta^{*}}(e^{-\lambda_n\frac{c}{\theta^{*}}}\vartheta_{j,n}) \E_\vartheta(\vartheta_{j,n})\\
        &=\frac{b^a}{(b+\lambda_n c)^a}\Var_\vartheta(\vartheta_{j,n})\,.
    \end{split}
\end{equation}
Note this is simply the variance times the correlation coefficient. Higher orders of covariance are similarly, for $n\in 2\mathbb{N}$ and $j>0$:
\begin{equation}\label{eq:system_detAR_cov_k}
    \begin{split}
        &\Phi(r>1)=\Cov_\vartheta(\vartheta_{j+r,n},\vartheta_{j,n})=\bigg(\frac{b^a}{(b+\lambda_n c)^a}\bigg)^r \Var_\vartheta(\vartheta_{j,n})\,.
    \end{split}
\end{equation}
Clearly, the limits for short and long scaled relative epoch lengths are:
\begin{equation}\label{eq:detAR_cov_lims}
    \begin{split}
        \lim_{\lambda_n c \rightarrow 0} \Phi(r>0)= 1 \times \lim_{\lambda_n c \rightarrow 0}\Var_\vartheta(\vartheta_{j,n})=0 \\
        \lim_{\lambda_n c \rightarrow \infty } \Phi(r>0)= 0\,.
    \end{split}
\end{equation}

\subsubsection{Results and Discussion for Model~3}
\begin{figure}[!ht]
    \centering
    \includegraphics[width = 13cm, height=9cm]{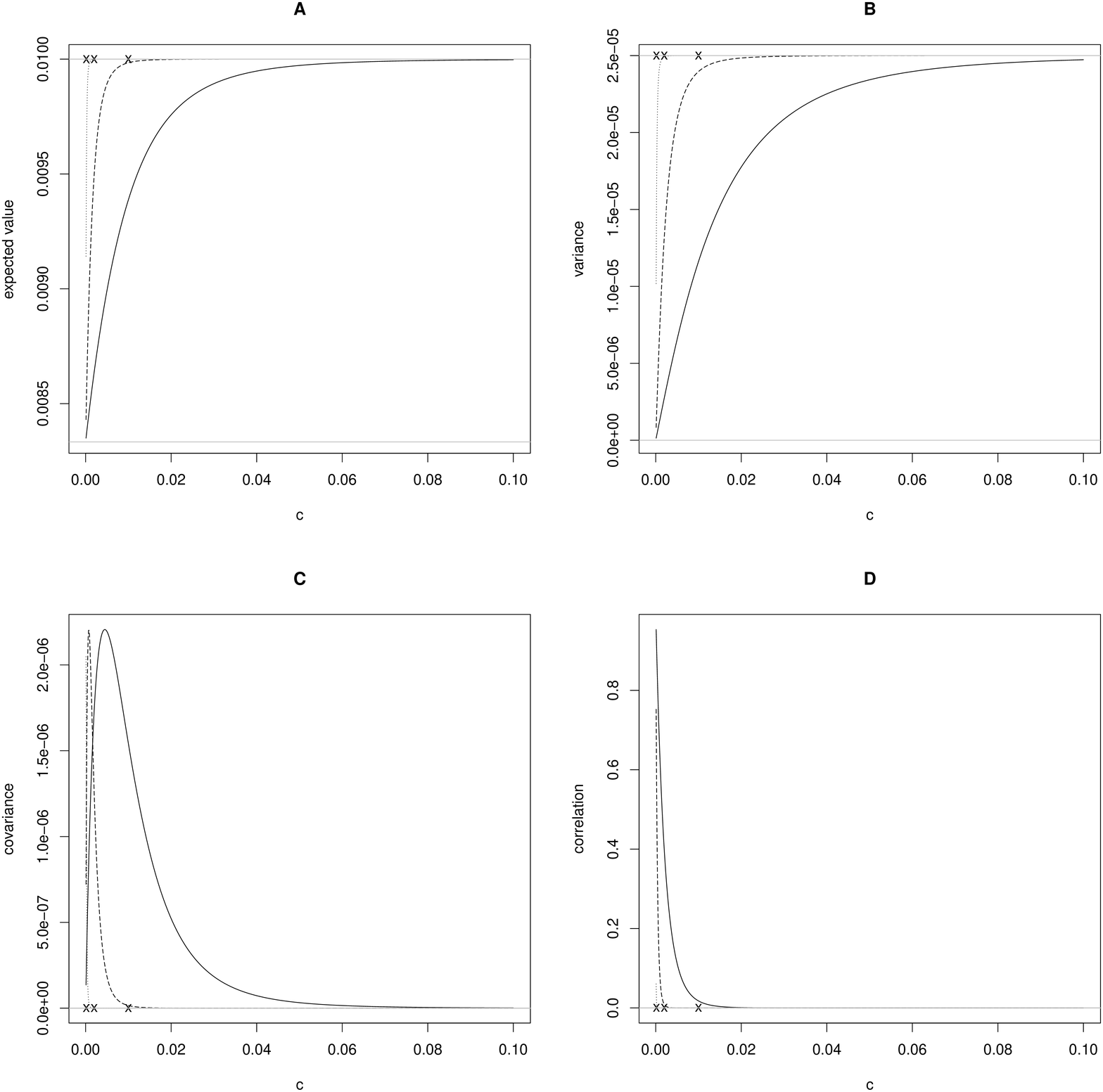}
    \caption{Let the overall bias-complemented scaled mutation rate be distributed according to $\theta_j^{*} \overset{iid}\sim IG(a=6,b=0.01\cdot(a-1))$. The theoretical mean from Eq.~(\ref{eq:detAR_mean2}), variance from Eq.~(\ref{eq:system_detAR_var}), and covariance and correlation Eq.~(\ref{eq:system_detAR_cov}) dependence of the relative scaled epoch length $c$. In each panel, the solutions for expansion orders $n=c(2,4,12)$ are shown in decreasing line strength and the horizontal lines indicate the values of the statistics in the limits $\lambda_n c\rightarrow 0$ and $\lambda_n c\rightarrow \infty$ respectively. The symbols `x' in each panel mark the epoch lengths chosen for Fig.~(\ref{fig:AR_res})(A), (B), and (C), while the epoch length from (D) is beyond the limit of the x-axis.}
    \label{fig:AR_Moments}
\end{figure}

 In Fig.~(\ref{fig:AR_Moments}), the behaviour of the theoretical mean, variance, and first order covariance of the temporal population coefficients is exemplified for varying relative epoch lengths $c$ and select expansion orders $n$. For low orders of expansion $n$, the moments of the temporal mutation coefficients are close to the lower limits (\ie{} the harmonic means) even for rather long relative epoch lengths $c$; convergence towards the lower limits is then very rapid when $c \rightarrow 0$. For the higher expansion orders $n$, the trajectory of the moments of the temporal coefficients shows a more gradual convergence from the higher to lower limits (\ie{} the arithmetic to the harmonic mean) as the relative epoch lengths decrease. Note that the analytical solutions examined here correspond almost precisely to results from simulations; snippets of these comparisons are presented in Appendix A (Section~\ref{sec:TMod3Tables}).

\begin{figure}[!ht]
    \centering
    \includegraphics[width = 11.5cm]{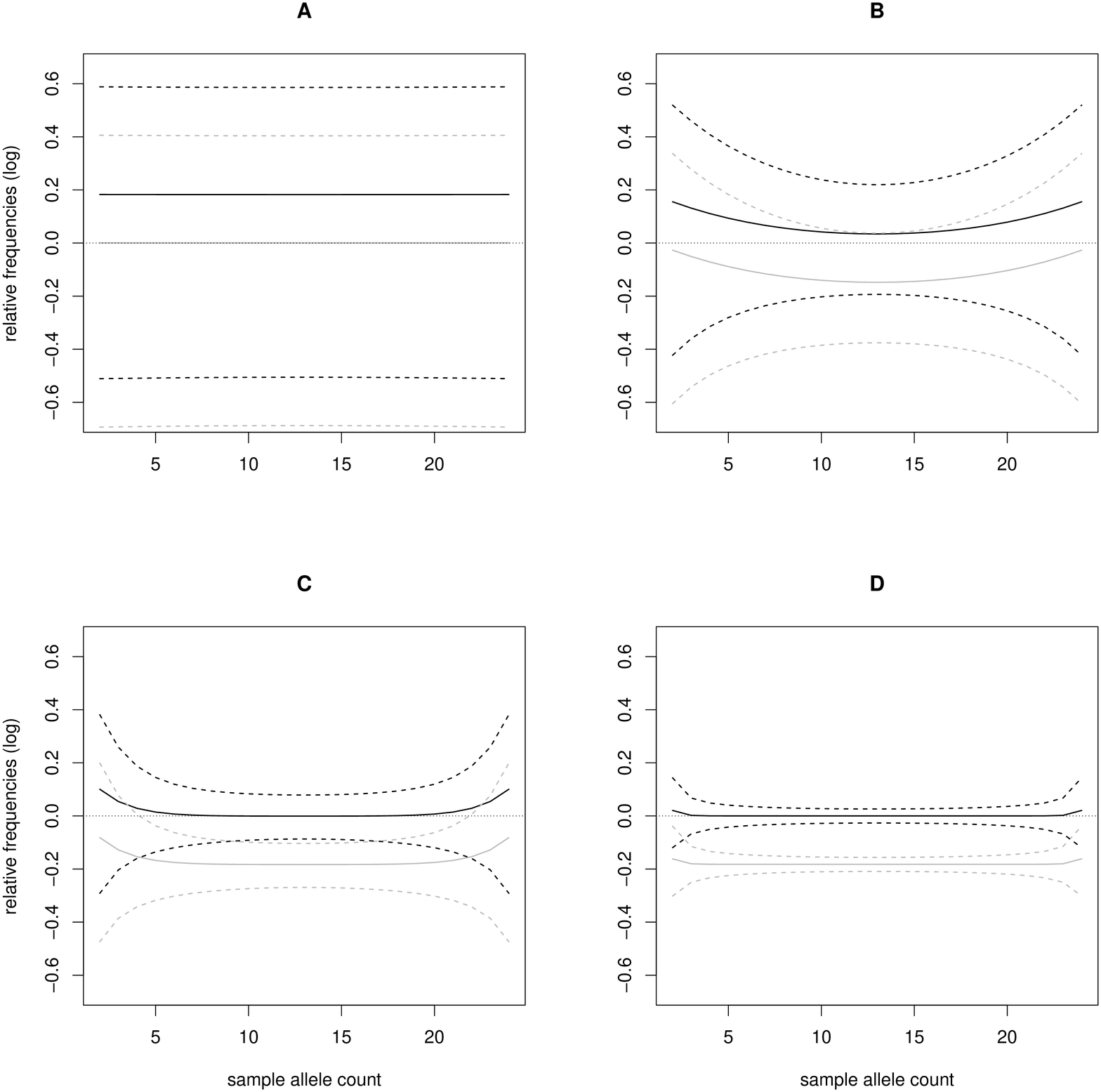}
    \caption{Assuming $\theta_j^{*}\overset{iid}\sim IG(a=6,b=0.01\cdot(a-1))$, $J=1\cdot 10^{6}$ epochs of the autoregression model for the long-term temporal coefficients $\vartheta_{j,n}$ in Eq.~(\ref{eq:system_detAR}) were simulated for relative epoch lengths $c=1e^{-1}$ (A), $c=2\cdot 10^{-2}$ (B), $c=1\cdot 10^{-3}$ (C), and $c=1\cdot 10^{-4}$ (D). Expected sample allele spectra of size $K=24$ were generated for each scenario at the end of epoch $J$. Visualised here are the log-ratios of the expected sample site frequency spectra vs the spectrum generated by the harmonic means of $\theta_j^{*}$ in black, and the log-ratio of the expected sample site frequency spectra vs the spectrum determined by the arithmetic mean of $\theta_j^{*}$ in grey. The dashed lines represent the spectra obtained by adding/subtracting one standard deviation of $\vartheta_{J,n}$ to the mean and calculating the respective expected sample spectra and ratios again.}
    \label{fig:AR_res}
\end{figure}

 Also apparent in each panel of Fig.~(\ref{fig:AR_Moments}) are three little `x' marks: at these epoch lengths, as well as at one additional epoch length that exceeds the limit of the x-axis, we determined the expected polymorphic allele frequencies of a sample of size $K=24$ according to procedure described in Section~\ref{sec:maths_intro_general_framework} and using the expected value of the temporal coefficients from Eq.~(\ref{eq:detAR_mean2}). These are shown relative to both the arithmetic and harmonic equilibrium spectra in Fig.~(\ref{fig:AR_res}). Scenario A corresponds to relatively long epoch lengths; at this point the moments of the temporal mutation coefficients have not yet fully converged to the harmonic mean even for the expansion order $n=2$ (see Fig.~\ref{fig:AR_Moments}) and the polymorphic sample spectrum cannot be distinguished from the arithmetic equilibrium spectrum. Scenarios B and C show epoch lengths between those at which expansion orders $n=2$ and $n=4$ fully converge to the harmonic mean but are close to these respective required lengths (see Fig.~\ref{fig:AR_Moments}); the expected polymorphic sample spectra for both scenarios clearly lie below the arithmetic equilibrium spectrum -- in fact, the intermediate frequency alleles have almost (B) or fully (C) reached harmonic mean frequency while the high/low frequency alleles exceed it. With the very short epoch length of scenario D, the moments have fully converged to the harmonic mean for almost all orders of expansion (note that the highest are not shown for the sake of visibility) and the expected polymorphic allele spectrum is almost identical to the harmonic equilibrium spectrum except for excess singletons. Note that scenario B has epoch lengths of almost the same magnitude as the average overall bias-complemented scaled mutation rate, and scenario D has epoch lengths of the same magnitude as the square of the average overall bias-complemented scaled mutation rate. In general, the variance of the expected polymorphic sample allele spectra visibly decreases with relative epoch length. 

Interestingly, the D-statistic picks up a clear signature of population collapse for scenario A with the relatively long comparative epoch lengths, and equally strong signatures of population expansion for the increasingly short relative epoch lengths of scenarios B-D (see Table~\ref{tab:AR_TajMahal}). Overall, it is apparent that compared to samples of the same size drawn from a population evolving according to a boom-bust model with the same harmonic mean of the overall mutation parameter across epochs, the samples from populations subject to stochastic changes in size generally harbour signatures of more distinct and consistent demographic change. Traces of this are still present even for very short epoch lengths.
In fact, this lingering excess of singletons for very short epoch lengths is particularly interesting: Empirically, excess singletons are often observed and have recently been shown to be well-modelled by multiple-merger coalescents \citep{Freund23}. We will address this further in our Conclusions \ref{sec:discussion}.

After hypergeometrically downsampling all the sample spectra to $K=6$, our agnostic inference approach from Section~\ref{sec:inference} can be applied to infer the current and two historical overall bias-complemented scaled mutation rates (see Table~\ref{tab:AR_TajMahal}). Generally, these vary around the expected means per scenario. 

\newpage
\newpage

\newpage

\begin{table}[!ht]
\centering
\begin{tabular}{llll}
\hline
 (A)  $c=1e^{-1}$  & (B) $c=2e^{-2}$ & (C) $c=1e^{-3}$ & (D) $c=1e^{-4}$ \\ \hline
 0.1323629& -0.01156199 &-0.01160323  & -0.01160785 \\ \hline
\end{tabular}
\caption[]{D-statistic for the expected sample spectra from Fig.~(\ref{fig:AR_res})}
\label{tab:AR_TajMahal}
\end{table}

\begin{table}[!ht]
\centering
\begin{tabular}{l|llll}
\hline
 &(A)  $c=1\cdot 10^{-1}$  & (B) $c=2\cdot 10^{-2}$ & (C) $c=1\cdot 10^{-3}$ & (D) $c=1\cdot 10^{-4}$ \\ \hline
$\widehat{\theta^{*}}_{s_J}$& 0.010009 & 0.008603 & 0.008292 & 0.008295 \\ \hline
$\widehat{\theta^{*}}_{s_{J-}}$ & 0.009938 & 0.010889 & 0.008925 & 0.008681 \\ \hline
$\widehat{\theta^{*}}_{s_{J--}}$& 0.010132 & 0.009646 & 0.009922  & 0.008254 \\ \hline
   \end{tabular}
\caption[]{Inference of temporal coefficients for sampled spectra from Fig.~(\ref{fig:AR_res}). The ``-'' in the naming of epochs indicates one step back in time.}
\label{tab:AR_Inf}
\end{table}

\newpage

\subsection{Model 4: Serially Correlated Population Size and Stochastic Epoch Lengths}\label{sec:model4}

\subsubsection{Rewriting the Autoregression Model}

In Section~\ref{sec:model3}, the relative epoch length $c=s_{j}4\alpha\beta\mu$ is assumed constant. Next, we wish to model the relative epoch lengths as stochastic. Doing so is decently straightforward; for each epoch $j>0$, the distribution of the bias-complemented overall scaled mutation rate $\theta_j^{*}$ is generalised from the inverse gamma distribution with constant parameters in Eq.~(\ref{eq:IG}) to the following:
\begin{equation}
\begin{split}
     &\theta_j^{*} \overset{iid}\sim invgamma(a,b\nu_j) \text{\small{ with $a>2,b\nu_j>0$}}\,,
\end{split}
\end{equation}
where $v_j$ are unit mean exponential random variables:
\begin{equation}\label{eq:EXP1}
\begin{split}
     &\nu_j \overset{iid}\sim exp(1)\\
     &p(\nu)=\Pr(\nu_j=\nu|1)=e^{-\nu}\,.    \end{split}
\end{equation}
Therefore, the constant relative epoch length in the autoregression process for the solution to the long-term temporal dynamics from Eq.~(\ref{eq:system_detAR}) can be replaced with the random relative epoch length $cv_j$; note that this means that we no longer consider the number of generations per epoch to be the same (compare to the reasoning at the beginning of  Section~\ref{sec:model3}). Further setting $\eta_j=\tfrac{1}{\theta_j^{*}}$, the autoregression process for the long-term temporal dynamics becomes:
\begin{equation}\label{eq:system_detAR2}
\begin{split}
    \text{\small{for $n$}}& \text{\small{$\in\{(1,..,N):2n$\} and $j>0$:}}\\
   &\vartheta_{j,n} = \vartheta_{j-1,n} e^{-\lambda_n c \nu_j \eta_j} + \eta_j^{-1} \bigg(1-e^{-\lambda_n c \nu_j \eta_j} \bigg)\,.
\end{split}
\end{equation}

Again, we wish to characterise the distribution of the driving mutation parameters in order to study the dynamics of the long-term temporal coefficients $\vartheta_{j,n}$; we therefore determine the distribution of the $\eta_j$ via: 

\begin{equation}\label{eq:joint_distr2}
\begin{split}
     p(\eta)&=\int_0^\infty  \frac{(b\nu)^a}{\Gamma(a)} \eta^{a-1} e^{-(b\nu)\eta}\,e^{\nu}\,d\nu\\
     &=\frac{b^a}{\Gamma(a)} \eta^{a-1}\int_0^\infty  \nu^a e^{-(b\eta+1)\nu}\,d\nu\\
     &=\frac{b^a}{\Gamma(a)} \eta^{a-1}\frac{\Gamma(a+1)}{(b\eta+1)^{a+1}}\,,
    \end{split}
\end{equation}
which can be re-stated as:
\begin{equation}
\begin{split}
     &\eta_j \overset{iid}\sim genbeta(a,1,1,1/b)\,,
    \end{split}
\end{equation}
where genbeta(.) is the generalised beta prime distribution. Note that the mean of $p(\eta)$ is naturally the inverse of the mean of $p(\theta^{*})$.

\paragraph{Moments}
As in Section~\ref{sec:model3}, the moments of the temporal mutation parameters, which correspond to the moments of the autoregression process in Eq.~(\ref{eq:system_detAR2}), can readily be determined: Once again, stationarity of the process can be invoked and then the expectations can be taken with respect to first $\nu$ and then $\eta$. It is easily seen from there that $\E_\vartheta(\vartheta_{j,n})$ is identical to before. $\E_\vartheta(\vartheta_{j,n})^2$ differs only slightly in that the final three summands of the numerator in Eq.~(\ref{eq:system_detAR_usqu2}) are each multiplied by factor $2$ (observe that this corresponds with the terms multiplied with $E_\eta (\eta^2 p(\eta|\nu))$ from Eq.~\ref{eq:joint_distr2}). The variance, covariance and correlation are therefore inflated compared to the model with fixed relative epoch lengths (see also Appendix~\ref{sec:TMod3Tables}).

\section{Conclusions}\label{sec:discussion}
In this article, a spectral representation of the transition density of the forward diffusion approximation of the biallelic boundary mutation Moran model is utilised to describe population allele trajectories evolving forwards in time. This model separates the dynamics of mutations, which are assumed to occur at a very low overall scaled rate and are thus restricted to the monomorphic boundaries, from those of drift, which determine the polymorphic interior. The eigenfunctions of the approximating diffusion process can be represented using a combination of (i) Gegenbauer polynomials, a class of orthogonal polynomials that describe the spatial component of polymorphic allele frequency trajectories subject to only drift, (ii) temporal coefficients that scale the polymorphic spatial component by the effect induced by the boundary mutations, and (iii) specialised boundary terms that balance the probability flux between the monomorphic boundaries and polymorphic interior, taking into account that mutations enter the polymorphic region at a constant rate during an epoch defined by unchanging (effective) population size. The corresponding eigenvalues of the system are $\lambda_0=0$, which is associated with the equilibrium distribution, $\lambda_1=\theta$, which is associated with the proportion of the focal allele present within the population, and $\lambda_n=n(n-1)$ for $n\geq 2$, which are associated with the polymorphic spectrum. Hence the spectral decomposition comprises both the monomorphic and polymorphic parts of the allele frequency spectrum (compare \citep{Song12, Zivkovic15}); note that classic diffusion and coalescent models only explicitly describe the polymorphic interior. Incorporating piecewise changes in (effective) population size into the forward diffusion of the boundary mutation diffusion model by proxy of changes in the overall bias-complemented scaled mutation rate requires only one full spectral decomposition to determine the spatial component of the process, which remains unchanged over time. One must further solve a system of equations for the temporal coefficients, which differ by time epoch. This is simpler and computationally advantageous to changing the basis function for the spatial component for every piecewise change in (effective) population size, as has been done for spectral representations of general mutation (Wright-Fisher) diffusion models in the past \citep{Lukic11, Zivkovic11, Zivkovic15}.

In this article, the exact form of expected sample site frequency spectra was determined by coupling the spectral representation of the diffusion equation with a forward pass of a forward-backward algorithm \citep{Bergman18a}. We note here that a backward pass of the same forward-backward algorithm could have been employed with the same numerical burden; for a fuller understanding of this, we provided a brief exposition of the backward diffusion equation of the boundary mutation model and its corresponding backward eigensystem in Appendix~(\ref{sec:Appendix_spectral_decomposition}). For now, recall that the spectral representations of the forward and backward processes involving orthogonal polynomials are defined on the space of allelic proportions. In Appendix Section~\ref{sec:appendixB}), we had briefly shown that a dual process to the spectral representation of the backwards diffusion equation can be found in an embedded jump process that runs backwards in time and operates in the space of binomial distributions. From those derivations, it can be seen that the rate of the jump process is intimately related to the rate of Kingman's block counting process (see also \citep{Papaspiliopoulos14}). As of now, it is unclear whether coalescent dual or orthogonal polynomial approaches are numerically preferable within the context of the forward-backwards algorithm. Similarly, the forwards-backwards algorithm approaches have yet to be compared to the methods based on the ``matrix coalescent'' for efficiency \citep{Wooding02,Polanski03,Bhaskar15}. Doing so would also enable an explicit comparison of our results to past literature on the effect of demography on the branch lengths of sample genealogies (see for example \citet{Eriksson2010}). Overall, however, we hope to have at least established an understanding of how the approaches relate to each other as much as the scope of the current article allows. 

Returning to the main part of the article: We examine the temporal coefficients of the forward pass of the spectral forward-backward algorithm as well as the exact sample allele frequency spectra for populations with (i) single deterministic shifts in (effective) population size, (ii) deterministic boom-bust life cycles, and (iii) two models of stochastic changes in (effective) population size. To the best of our knowledge, the stochastic models are novel. Throughout this article, we present conjoint assessment of the temporal population dynamics and the shape of the polymorphic sample spectra (more specifically, the shape of the log ratio of the sample spectrum vs an equilibrium distribution).  Other authors have promoted visual inspection of sample spectra has over summary statistics for a clearer understanding of which regions of sample spectra deviate from equilibrium and to what extent \citep{Nawa08, Achaz09}. We see the advantages of the visual approach ourselves: In our evaluation of the boom-bust model with intermediate epoch lengths in Fig.~(\ref{fig:BoomBust}), the polymorphic sample spectrum from the boom phase has a w-shape that the D-statistic interprets as a collapsing (effective) population size (see Table~\ref{tab:BoomBust_TajMahal}). However, this signal is actually caused by the rapid accrual of intermediate frequency alleles vs high/low frequency alleles after recurrent, intermediately spaced bust phases (\ie{} mild-effect bottlenecks) that fail to eradicate all the population's standing variation \citep[compare][Table 2]{Nawa08}. In our evaluation of sample spectra from populations with stochastic (effective) population size, we see that the D-statistic is blind to the differing effect of the time scale on which the changes occur (Tables~\ref{tab:AR_TajMahal} and \ref{fig:AR_res}). It picks up on the excess high/low frequency alleles vs intermediate as soon as the epochs become shorter and interprets this as population growth, but not on the decrease of excess intermediate alleles for shortening epoch lengths. In both cases, the figures are more informative than the summary statistic.

As informative as the polymorphic sample spectra are per se, we argue that assessing the level of convergence of the temporal coefficients of the source populations leads to a more nuanced understanding of the population dynamics. In practical terms, carrying out forward population simulations and spectral decompositions as part of experimental planning can help inform what sample sizes/orders of expansion are needed to detect departure from equilibrium or, conversely, for what sample size/orders of expansion equilibrium can safely be assumed. Even retrospectively, decomposition of estimated transition rate matrices from population samples can be informative of precisely how far which regions of the sample spectrum are from equilibrium.

Recall that the effective size of a population in time- and space-discrete (pure drift) Wright-Fisher and Moran models without mutation is often defined via the first non-unit eigenvalue $\lambda_{1}$ of the transition probability matrix for the respective neutral models. This is equal to $1-\tfrac{1}{2N}\approx e^{-\frac{1}{2N}}$ or $1-\tfrac{2}{N^2}\approx e^{-\frac{2}{N^2}}$, respectively \citep[][p.~127]{Ewens04}. This eigenvalue determines the rate of decline in population heterozygosity ($H$) over time: In the Wright-Fisher model, heterozygosity at the discrete time point $T$ is $H_{T}=\lambda_{1}^T H_0$, which can be approximated by $H_T=H_0 e^{-\tfrac{T}{2N}}$ in the limit of large population sizes \citep{Ewens82} (but see also \citep{Crow54}). In the Moran model, the equivalent approximation is $H_\iota=H_0 e^{-\tfrac{2\iota}{N^2}}$ per Moran step $\iota$. In the limit of large population sizes, the genealogical tree of a sample drawn from a diploid Wright-Fisher or a haploid Moran model is given by Kingman's coalescent \citep{Kingman82} (with equilibrium coalescent times scaled by $2N$ and $\tfrac{N}{2}$ generations, respectively). Arguably the most prevalent definition of the effective size of a population is through the linear re-scaling of the observed/census population size required for the limiting genealogical tree to still correspond to Kingman's coalescent even when there are temporal or spatial variations in population size \citep{Nordborg02}. In the limit of large $N$, Wright-Fisher/Moran models are approximated by diffusion models; in the case of pure drift, the first non-zero eigenvalue of the corresponding transition rate densities converge towards the coalescent effective population size. If in addition to drift, mutation is also considered, the first non-zero eigenvalue is $\lambda_1=\theta$, and thus only influenced by mutation. As long as $\theta\ll 1$, the second non-zero eigenvalue $\lambda_2$ then corresponds to $\lambda_1$ of the pure drift model.

For populations of short-scale, cyclical variations in size \citep{Wang00a,Wang00b,Pollak02} or for subdivided populations with high migration rates and large demes \citep{Hossjer15}, convergence of the first eigenvalue informative of drift is towards the harmonic mean across time intervals or demes. This eigenvalue is typically $\lambda_1$ in pure drift models, but $\lambda_2$ in the \citet{McKane07} pure drift model and in typical in mutation-drift models.  When deterministic or stochastic changes in population size or migration rates occur on a shorter time scale than the coalescent events, the coalescent effective population size corresponds to the harmonic mean across epochs or demes \citep{Jagers04, Sjodin05, Hossjer11}. However, when changes occur on the same time scale as coalescent events, the limiting ancestral genealogy of the sample is a time-stochastic version of Kingman's coalescent \citep{Kaj03}; and changes on longer time scales than coalescent events are negligible in terms of the sample genealogy. In the latter two cases, the coalescent effective population size is usually considered undefined \citep{Sjodin05} (but see \citet{Sano04}, where non-linear scaling across time is permitted so that the effective population size lies between the harmonic and arithmetic means and converges towards the arithmetic mean in these cases, respectively). Regardless of whether the effective population size is assessed via heterozygosity, leading eigenvalue, or re-scaling to Kingman's coalescent rate, only the variation in a sample of size $K=2$ is typically considered. 

In contrast, the temporal coefficients considered in this article are coefficients of all even orders $n\geq2$ of expansion of the population transition rate density into orthogonal polynomials (as the odd ones are zero when mutation bias is constant). They are of the form $c_1+c_2e^{-\lambda_n f(n,N)}$, where $n$ is the order of the eigenvalue and $N$ the effective population size, and describe the decline of the effect of changes in overall-biased mutation rate on polymorphism over time. Clearly, the temporal coefficient of order $n=2$ converges to its equilibrium value more slowly than the higher order coefficients, which substantiates the reasoning behind using it to estimate the effective population size. However, when past demographic changes are cyclical or fluctuating, the faster-responding higher-order temporal coefficients collect more short-term information and are perturbed relatively further from the harmonic means with every transition to a new epoch. It is therefore not uncommon for the sample allele frequency spectra to deviate from its equilibrium shape even after the temporal coefficient of order $n=2$ has converged, as shown repeatedly in this article (specifically, for the boom-bust model see Fig.~\ref{fig:BoomBust} and for the stochastic model see Fig.~\ref{fig:AR_res}). The deviations from equilibrium in sample spectra drawn from populations of stochastically varying (effective) population size across shorter epoch lengths are of particular interest (\ie{} Fig.~\ref{fig:AR_res}, scenarios B, C, and D): The spectra exhibit excess high and low frequency alleles for epoch lengths of the same magnitude as the average overall bias-complemented scaled mutation rate and below (scenarios B and C), until only excess low frequency alleles/singletons remain when the epochs decrease to roughly the same magnitude as the square of the average overall bias-complemented scaled mutation rate (scenario D). While epoch lengths below the same order of magnitude as the overall-biased mutation rate and especially on the order of magnitude of its square may conventionally be considered ``safely small'' to assume equilibrium, it is apparent in our analyses that this is not the case (see Fig.~\ref{fig:AR_res}). Fluctuations in population size therefore strongly impact the shape of sample spectra, even when limiting results may conventionally be assumed to hold. 

Deviations from neutrality are also commonly observed empirically, when sample allele frequency spectra are polarised into ancestral and derived alleles using outgroup data \citep{Freund23}. In particular, patterns of excess high and low frequency derived alleles as well as excess singletons have recently been noted in genome-wide data of many species. These patterns are often considered more suitably modelled by multiple-merger coalescents than the classic Kingman coalescent \citep{Freund23}. Multiple-merger coalescents reflect skewed offspring distributions across generations, whether these are caused by population substructure, demography, recurrent selection, biased gene conversion, \etc. Note that misleading signals of multiple-merger genealogies can be caused by poor sampling coverage and erroneous allele polarisation. It is well-known that extreme discrete changes in (effective) population size, such as population growth after a bottleneck, can also generate sample spectra with excess singletons; there is some indication that inference procedures can discriminate between these and multiple-merger generated sample spectra since the entire spectrum will not exhibit the same shape between the models for a comparable level of singletons \citep{Eldon15}. In this article, we did not assume polarisation of alleles via an outgroup, but arbitrarily assigned one of two alleles to be the focal allele. Therefore, the shape of any of our sample spectra cannot immediately be related to this body of literature. In particular, information on high frequency derived alleles is not available with our approach. The main outcome from our article in relation to this is that one cannot discount stochastic changes in population size, especially relatively rapid ones, from being responsible for a part of the empirical observations without further investigation. In fact, studies such as that of \citet{Eldon15} could be conducted to address this in the future, alongside more theoretical treatment of the relationship between our model and general coalescent models allowing for simultaneous- and multiple- mergers \citep{Spence16}.

In terms of demographic inference alone, we have shown that inference of population history from sample spectra can be performed using sufficient statistics within the framework of the boundary mutation diffusion model and the spectral forward-backward algorithm. By placing hypothetical demographic events conveniently with respect to the time scales defined by the eigenvalues $\lambda_n=n(n-1)$, we can very efficiently infer population history without specific assumptions or knowledge about past demographic events (including their timing), other than their maximum number \citep[compare to][]{Myers08, Bhaskar14}. This could be a starting point for future computational implementation of demographic inference procedures.

\section*{Declaration of Competing Interest}
The authors declare that they have no known competing financial interests or personal relationships that could have appeared to influence the work reported in this paper.

\section*{Acknowledgements}

The authors would like to thank Bur\c{c}in Yıldırım, Juraj Bergman, Conrad Burden, Joachim Hermisson, and Sandra Peer for conversations and input during this project, and are grateful to Carolin Kosiol for feedback. Further thanks are due to the two anonymous reviewers of this article, whose detailed comments and corrections improved the quality of this article. 

CV's research was supported by the Austrian Science Fund (FWF): DK W1225-B20; LCM's by the School of Biology at the University of St.\ Andrews. 

\newpage

\section{Appendix: Derivation of the Forward Partial Differential Diffusion Equation (FPDE) for Biallelic Boundary Mutations}
\label{sec:FromMoran2Diffusion}

In this Appendix, we derive the forward Kolmogorov forward (Fokker-Planck) diffusion equation for boundary-mutations from the transition rates of the discrete boundary mutation Moran model. 
We do this by first recapitulating how the general biallelic mutation diffusion can be obtained from the transition rates of the general decoupled Moran model \citep{Etheridge09}; this is largely reproduced from Appendix (7.1) of \citep{Bergman18a}. Then we modify this approach for the case of boundary-mutations.

\subsection{General Biallelic FPDE} 
In a discrete time and space Moran model, each reproduction step (which we index by $\iota$) constitutes a randomly chosen individual haploid individual being replaced by the offspring of another randomly chosen individual. The expected lifetime of an individual therefore corresponds to the generation time/population size $N$.

Forward in time, the difference in the probability of observing the focal allele at proportion $i$ at every genomic site per Moran step may be written as \citep[][Eq.~76]{Bergman18a}:
\begin{equation}\label{eq:Moran2}
\begin{split}
    &\Pr(x_{\iota+1}=i)-\Pr(x_{\iota}=i)=\\
    &\qquad\frac{\beta\theta}{N^2}\bigg((N-i+1)\Pr(x_{\iota}=i-1)-(N-i)\Pr(x_{\iota}=i)\bigg)\\
    &\qquad+\frac{\alpha\theta}{N^2}\bigg((i+1)\Pr(x_{\iota}=i+1)-i\Pr(x_{\iota}=i)\bigg)\\
    &\qquad+\frac{1}{N^2}\bigg((i-1)(N-i+1)\Pr(x_{\iota}=i-1)+(i+1)(N-i-1)\Pr(x_{\iota}=i+1)\\
    &\qquad\qquad-2i(N-i)\Pr(x_{\iota}=i)\bigg)\,,
\end{split}
\end{equation}
where the first and second line after the equality sign correspond to mutation events that increase or decrease the proportion of the focal allele respectively, and the final lines corresponds to the effect of drift. 

Let us now introduce infinitesimal steps $\epsilon_t= 1/N^2$ and $\epsilon_x = 1/N$. We use this to re-scale time via $t=\iota\epsilon_t$ and space, \ie{} the allele proportions, via $x = i\epsilon_x$ so that $\phi(x, t)\epsilon_x\epsilon_t=\Pr(x_s = i)$. Then, the infinitesimal transition probabilities can be written, using  Eq.~(\ref{eq:Moran2}), as (\citep[compare][Eq.~77]{Bergman18a}, as well as \citep[][Definition 2.33]{Etheridge12}):
\begin{equation}\label{eq:Moran_infinitesimal}
\begin{split}
    %
    %
    &\frac{\phi(x,t +\epsilon_t)-\phi(x,t))}{\epsilon_t}=\\
    &\qquad\frac{\beta\theta}{\epsilon_x}\bigg((1-x+\epsilon_x)\phi(x-\epsilon_x,t)-(1-x)\phi(x,t)\bigg)\\
    &\qquad+\frac{\alpha\theta}{\epsilon_x}\bigg((x+\epsilon_x)\phi(x+\epsilon_x,t)-x\Pr(\phi(x,t))\bigg)\\
    &\qquad+\frac{1}{\epsilon_x^2}\bigg((x-\epsilon_x)(1-x+\epsilon_x)\phi(x-\epsilon_x,t)+(x+\epsilon_x)(1-x-\epsilon_x)\phi(x+\epsilon_x,t)\\
    &\qquad\qquad-2x(1-x)\phi(x,t)\bigg)   
\end{split}
\end{equation}
Taking the limit $N\to\infty$, which implies $\epsilon_x,\epsilon_t\to0$, and applying the standard definitions for the first and second symmetric derivatives recovers the forward partial differential diffusion equation (FPDE) \citep[][Eq.~78]{Bergman18a}:
\begin{equation}\label{eq:mutation-drift_diffusion2}
\begin{split}
    \frac{\partial\phi(x,t)}{\partial t}&=-\frac{\partial}{\partial x}\bigg(\theta\beta(1-x)\phi(x,t)-\theta\alpha x\phi(x,t)\bigg)+\frac{\partial^2}{\partial x^2} x(1-x)\phi(x,t)\\
    &=-\frac{\partial}{\partial x}\theta(\beta-x)\phi(x,t)+\frac{\partial^2}{\partial x^2} x(1-x)\phi(x,t)\,.
\end{split}
\end{equation}

\subsection{Biallelic Boundary Mutation FPDE}\label{sec:boundaryFPDE}
In the boundary mutation model, we differentiate between the polymorphic interior, \ie{} the interval $[\epsilon_x\leq x\leq 1-\epsilon_x$)], where we denote the infinitesimal transition rate density by $\phi_I(x,t)$, and boundary singularities, where alleles are fixed or lost. Recall that $b_0(t)$ is the proportion of the non-focal allele at time $t$, and $b_1(t)$ that of the focal allele. In equilibrium, $b_0$ and $b_1$ represent the stationary probability that a sample of size one corresponds to the non-focal ($b_0=\alpha$) and focal ($b_1=\beta$) alleles respectively. Convergence towards these equilibrium values occurs exponentially with rate $\theta$, which is low compared to the rate of drift that governs the polymorphic interior (\ie{} the time sites spend in the latter region is negligible). For some starting value $b_0(t_s)$ at time $t_s<0$, we then have $b_0(t)=\alpha-(b_0(t_s)-\alpha) e^{-\theta(t-t_s)}$ and $b_1(t)=1-b_0(t)$. 

At the boundary $x=0$, probability mass is lost to the interior $\phi_I(x,t)$ at a constant  rate due to mutation and gained through drift from the polymorphic region adjoining the boundary at $x=\epsilon_x$. The infinitesimal transition probabilities then are:
\begin{equation}\label{eq:LimitFlow_at_0}
\begin{split}
    \frac{\phi(x=0,t+\epsilon_t)-\phi(x=0,t)}{\epsilon_t}&=\frac{-\beta\theta b_0(t)+0\cdot\alpha\theta}{\epsilon_x}+\frac{\epsilon_x(1-\epsilon_x)\phi_I(\epsilon_x,t)\epsilon_x-0}{\epsilon_x^2}\\
    &=\frac{-\beta\theta b_0(t)+\epsilon_x(1-\epsilon_x)\phi_I(\epsilon_x,t)}{\epsilon_x}\,.
\end{split}
\end{equation}
The zero in the first (mutation) term on the right hand side is due to the absence of mutations in the interior (which is the essence of the boundary mutation model), and the zero in the second (drift) term reflects the absence of drift in the mononorphic state. The other boundary $x=1$ follows analogously:
\begin{equation}\label{eq:LimitFlow_at_1}
\begin{split}
    \frac{\phi(x=1,t+\epsilon_t)-\phi(x=1,t)}{\epsilon_t}&=\frac{-\alpha\theta b_1(t)+(1-\epsilon_x)\epsilon_x\phi_I(1-\epsilon_x,t)}{\epsilon_x}\,.
\end{split}
\end{equation}

Conversely, the same probability masses are gained and lost from and to the boundaries by the polymorphic interior $\phi_I(x,t)$ at the adjacent regions $x=\epsilon_x$ and $x=1-\epsilon_x$. Overall, the infinitesimal transition probabilities in the interior of the boundary mutation model can be written as:
\begin{equation}\label{eq:boundary_inside_whole}
\begin{split}
    &\frac{\phi_I(x,t +\epsilon_t)-\phi_I(x,t)}{\epsilon_t}=\\
&\qquad\begin{cases}
    \displaystyle \frac{1}{\epsilon_x^2}\beta\theta b_0(t)+\frac{1}{\epsilon_x^2}\bigg(0+(2\epsilon_x)(1-2\epsilon_x)\phi_I(2\epsilon_x,t)-2\epsilon_x(1-\epsilon_x)\phi_I(\epsilon_x,t)\bigg)\\
    \qquad \text{for $x=\epsilon_x$;}\\
    \displaystyle     
    \frac{1}{\epsilon_x^2}\bigg((x-\epsilon_x)(1-x+\epsilon_x)\phi_I(x-\epsilon_x,t)+(x+\epsilon_x)(1-x-\epsilon_x)\phi_I(x+\epsilon_x,t)\\
    \qquad-2x(1-x)\phi_I(x,t)\bigg)   
\qquad\text{for $2\epsilon_x\leq x \leq 1-2\epsilon_x$;}\\
    \displaystyle \frac{1}{\epsilon_x^2}\alpha\theta b_1(t)+\frac{1}{\epsilon_x^2}\bigg((1-2\epsilon_x)(2\epsilon_x)\phi_I(1-2\epsilon_x,t)+0\\
    \qquad-2(1-\epsilon_x)\epsilon_x\phi_I(1-\epsilon_x,t)\bigg)\qquad\text{for $x=1-\epsilon_x$.}
\end{cases}
\end{split}
\end{equation}
The zeros in the top and bottom rows above signify that drift does not act on the boundaries to move probability mass into the polymorphic region. Taking the limit $N\to\infty$ yields the forward boundary mutation diffusion equation presented in Eq.~(\ref{eq:boundary-mutation-drift_diffusion}) of the main text, as long as $\phi_I(x,t)$ is twice differentiable and remains finite approaching the boundaries.

\paragraph{Remark} Note that we divide by $\epsilon_x$ in Eqs.~(\ref{eq:LimitFlow_at_0} and \ref{eq:LimitFlow_at_1}) and by $\epsilon_x^2$ in Eq.~(\ref{eq:boundary_inside_whole}) due to the difference in scaling between the interior and the boundary. 

\paragraph{Remark} If $\phi_I(x,t)$ does not remain finite approaching the boundaries, the infinitesimal transition probabilities can still be written as in Eqs.~(\ref{eq:LimitFlow_at_0}, \ref{eq:LimitFlow_at_1}, \ref{eq:boundary_inside_whole}), but the forward boundary mutation diffusion equation (Eq.~\ref{eq:boundary-mutation-drift_diffusion}) does not hold. In particular, substituting $\phi_I(x)=\alpha\beta\theta\,\frac{1}{x(1-x)}$, which approaches infinity towards the boundaries, into Eq.~(\ref{eq:boundary_inside_whole}) shows it to be the stationary solution. On the other hand, substituting $\phi_I(x)$ into into the diffusion equation  Eq.~(\ref{eq:boundary-mutation-drift_diffusion}), we find that the second derivative $\frac{\partial^2}{\partial x^2}x(1-x)\phi_I(x)=0$ for all $x\in\,]0,1[$  and particularly for $x=\epsilon_x$ and $x=1-\epsilon_x$.

\newpage

\section{Appendix: Spectral Representation of Diffusion Models}\label{sec:Appendix_spectral_decomposition}

\subsection{Eigensystem for the Pure Drift Diffusion Equation}
\subsubsection{Forwards Equation}
Consider the familiar forward Komolgorov (Fokker-Planck) diffusion equation for pure drift \citep{Kimura55}: 
\begin{equation}\label{eq:forward_pure_drift2}
    \frac{\partial\phi_I(x,t)}{\partial t}=\frac{\partial^2}{\partial x^2} x(1-x)\phi_I(x,t)\,.
\end{equation}
defined in the open interval $x\in]0,1[$. An analytically tractable form for the transition rate density can be found by decomposing $\phi_I(x,t)$ into eigenfunctions constituted by orthogonal polynomials and corresponding eigenvalues. Specifically, it can be expanded into a series of (modified) Gegenbauer polynomials $U_n(x)$ with temporally dependent coefficients $\tau_n(t)$: $\phi_I(x,t)=\sum_{n=2}^\infty \tau_n(t)U_n(x)$ \citep{Kimura55, Song12,Vogl16}. Substituting this into the above differential equation induces the following system of ordinary homogeneous differential equations:
\begin{equation}\label{eq:DE_homogeneous}
\begin{split}
    \frac{d}{dt}\tau_n(t)U_n(x)&= \frac{\partial^2}{\partial x^2} x(1-x)\tau_n(t)U_n(x)\\
    \frac{d}{dt}\tau_n(t)U_n(x)&=-\lambda_n \tau_n(t)U_n(x)\\
    \frac{d}{dt}\tau_n(t)&=-\lambda_n \tau_n(t)\,.
\end{split}
\end{equation}
In order to solve this system, an initial condition must be defined. At time $t=t_s$ in the past, let a function $\rho(x)$ defined within the polymorphic region represent an ancestral state of the population that can also be represented as a series of (modified) Gegenbauer polynomials $U_n(x)$ with coefficients $\rho_n$:
\begin{equation}
   \rho_n= \frac{1}{\Delta_n}\int_0^1 x(1-x)U_n(x)\rho(x) dx\,.
\end{equation}
Following Kimura \citep{Kimura55}, $\rho(x)$ is commonly taken to be a unit probability mass concentrated at a point $p$ (which is often taken to represent a single mutant allele in the population, $p=1/N$ \citep{Kimura69}) represented as $\delta(x-p)$ \citep[\eg{}][Section~4]{McKane07}. This probability mass is then expanded into (modified) Gegenbauer polynomials $\phi_I(x,t=t_s)=\sum_{n=2}^\infty\rho_n(p)U_n(x)$ with coefficients:
\begin{equation}
\begin{split}
    \rho_n(p)&= \frac{1}{\Delta_n}\int_0^1 x(1-x)U_n(x)\delta(x-p)\,dx\\
    &=\frac{1}{\Delta_n}p(1-p)U_n(p)\,.
\end{split}
\end{equation}
For $t_s\leq t\leq 0$, the temporal component then correspond to $\tau_n(t_s,\rho)=\rho_n(p)e^{-\lambda_n (t-t_s)}$, yielding the overall solution:
$$
\phi_I(x,t_s)=\sum_{n=2}^\infty \rho_n(p)e^{-\lambda_n (t-t_s)}U_n(x).
$$ 
Extending time $t$ into the future, \ie{} to values greater than zero, the system eventually converges to the trivial solution $\tau_n(\infty)=0$, from which it follows that  $\int_0^1\phi_I(x,t)\,dx$ also converges to zero as the entire probability accumulates at boundary singularities. Modelling this accumulation in a way that appropriately conserves probability requires the addition of two eigenfunctions that are linear combinations of delta functions, as in \citet[][Appendix C, last paragraph]{McKane07} and see also \citet{Tran14b}. Given the initial allele frequency $p$, and ensuring that the average frequency of the mutant allele must coincide with it at all times, makes the following full system of forward eigenvalues appropriate:
\begin{equation}\label{eq:forw_McKaneWax_p}
\begin{cases}
    \mathcal{F}_0^{(p)}(x)&=(1-p)\delta(x)+p\delta(x-1)\\
    \mathcal{F}_1^{(p)}(x)&=-\delta(x)+\delta(x-1)\\
    \mathcal{F}_{n\geq2}^{(p)}(x)&=-\frac{(-1)^n}n\delta(x)+U_n(x)-\frac{1}n\delta(x-1)\,.
\end{cases}
\end{equation}
The corresponding eigenvalues are $\lambda_0=\lambda_1=0$ and $\lambda_n=n(n-1)$ for $n\geq 2$ \citep[][Appendix C, last paragraph]{McKane07}. Note that the same spectral representation of the forward pure drift diffusion equation has been obtained from the equivalent representation of the forward biallelic general mutation diffusion equation model via a zeroth order Taylor series expansion in $\theta$ \citep[][Eq.~56]{Bergman18a}.

\subsubsection{Backwards Equation}
Let us now consider the backward Kolmogorov diffusion equation for pure drift:
\begin{equation}\label{eq:backward_pure_drift}
    -\frac{\partial\psi_I(x,t)}{\partial t}=x(1-x)\frac{\partial^2}{\partial x^2}\psi_I(x,t)\,,
\end{equation}
with $t_s \leq t\leq 0$. The complete set of backward eigenfunctions, including appropriate boundary conditions, can be constructed to be orthogonal to the forward eigenfunctions $\mathcal{F}_n^{(p)}$ \citep[compare][Appendix C, last paragraph]{McKane07}:
\begin{equation}\label{eq:back_Bergman18_2}
\begin{cases}
    \mathcal{B}_0^{(p)}(x)&=1\\
    \mathcal{B}_1^{(p)}(x)&=x-p\\
    \mathcal{B}_{n\geq2}^{(p)}(x)&=x(1-x)U_{n}(x)\,,
\end{cases}
\end{equation}
but again compare to the alternative derivation in \citep[][Eq.~56]{Bergman18a}. Note that only $\mathcal{B}_1^{(p)}(x)$ actually depends on $p$. 

The orthogonality relation can be explicitly stated as:
\begin{equation}\label{eq:orthogonality_condition}
    \int_0^1 \mathcal{B}_n^{(p)}(x)\mathcal{F}_m^{(p)}(x)\,dx=\delta_{nm}\Delta_n^{(B)}
\end{equation}
for a nonzero $\Delta_n^{(B)}=\frac{n-1}{(2n-1)n}$ only if $m=n$. 

The initial condition for the backwards equation is generally a pre-specified sampling distribution at the extant time. Recall that in this article, we consider a binomial sample of haploid size $K$ conditional on the population allele frequencies $x$, \ie{} $\Pr(k\given K,x)$, drawn at time $t=0$. Any binomial likelihood $\Pr(k\given K,x)$ can be expressed as a sum of $\mathcal{B}_n^{(p)}(x)$ up to order $n=K$ with coefficients $d_n(k,K)$.
Note that the probabilities of drawing monomorphic samples can only be represented because the spectral representation was intentionally augmented with the boundary terms. Therefore, the backwards diffusion equation can be written as the spectral sum: 
$$
    \psi(x,t)=\sum_{n=0}^K d_n(k,K) e^{\lambda_n t} \mathcal{B}_n^{(p)}(x)\,.
$$

\subsubsection{Sample Probabilities}\label{sec:sample_probs}

Now that explicit analytic representations of the forward and backward diffusion equations have been found, the probability of observing $k$ focal loci in our binomial sample can be determined as the following marginal likelihood:
\begin{equation}\label{eq:marg_like}
 p_k=\Pr(k\given K,\rho,t_s)=\int_0^1 \psi(x,t)\phi(x,t)\,dx\,.
\end{equation}
This corresponds to the forward-backward algorithm introduced in \citep[][Section~2.2]{Bergman18a}, and can be evaluated at any time $t_s\leq t \leq 0$.
Using the results from the previous subsections, particularly that the orthogonality condition simplifies computation, we can analytically determine:
\begin{equation}
\begin{split}
p_k&=\Pr(k\given K,\rho,t_s)\\
    &=\int_0^1 \psi(x,t)\phi(x,t)\,dx\\
    &=\int_0^1 \bigg(\sum_{m=0}^K d_m(k,K)e^{\lambda_n t} \mathcal{B}_m^{(p)}(x)\bigg)\bigg(\sum_{n=0}^\infty \rho_n e^{-\lambda_n(t- t_s)}\mathcal{F}_n^{(p)}(x)\bigg)\,dx\\
    &=\sum_{n=0}^K \int_0^1 d_n(k,K) e^{\lambda_n t}\mathcal{B}_n^{(p)}(x) \rho_n e^{-\lambda_n(t- t_s)}\mathcal{F}_n^{(p)}(x)\,dx\\
    &=\sum_{n=0}^K d_n(k,K) \rho_n(x)\Delta_n^{(B)} e^{\lambda_n t_s}\,.
\end{split}
\end{equation}

Note that, in the main text, we choose to evaluate the probabilities $p_k$ at the extant time $t=0$. In the above equation, $\psi(x,t=0)$ then equates to the expansion of the binomial sampling scheme $\sum_{n=0}^K d_n(k,K) \mathcal{B}_n^{(p)}(x)$, and therefore the calculation as a whole reduces to a forward pass of the forward-backward algorithm.


\subsection{Eigensystem for the Boundary Mutation Diffusion Equation}\label{section:eigensystem_boundary_mutation}

\subsubsection{Derivation of the Forward System}

Formally, the boundary mutation diffusion model resembles a pure drift diffusion model that explicitly takes mutations at the boundaries into account. In this subsection, we will discuss the modifications to the previous results required to represent the transition rate density of the boundary mutation diffusion equation as a spectral sum.

Let us now begin with the forward diffusion equation of the polymorphic interior, which we here rewrite from Eq.~(\ref{eq:boundary-mutation-drift_diffusion}) to the more convenient form:
\begin{equation}\label{eq:boundary-mutation-drift_diffusion3}
\begin{split}
\frac{\partial\phi_I(x,t)}{\partial t}&=\lim_{\epsilon_x\to 0}\bigg(\frac{\beta\theta b_0(t)}{\epsilon_x}\delta(x-\epsilon_x)+\frac{\alpha\theta b_1(t)}{\epsilon_x}\delta(x-1+\epsilon_x)\bigg)\\
    &\qquad+\frac{\partial^2}{\partial x^2} x(1-x)\phi_I(x,t)\,.
\end{split}
\end{equation}

The main novelty here compared to the pure drift model is that there are probability masses coming in from the boundaries due to the input of biased mutation: Importantly, $b_0(t)$ and $b_1(t)$ have an equivalent representation in terms of the eigenfunctions $\mathcal{F}_0^{(\beta)}$ and $\mathcal{F}_1^{(\beta)}$, where the mutation biases $\beta$ and $\alpha$ determine the equilibrium proportion of alleles of the focal and non-focal type in the population, and thus the probability mass concentrated at each boundary in a stationary system. These eigenfunctions are adapted from Eq.~(\ref{eq:forw_McKaneWax_p}) to account for the fact that the average proportion of focal and non-focal alleles may change over time (which $p$ in the pure drift system does not). By comparing to the beginning of Section~\ref{sec:boundaryFPDE}), we state that if the system starts at time $t_s<0$ in the past, convergence towards the equilibrium boundary values must follow according to:
\begin{equation}\label{eq:bs_2_F0_F1}
    b_1(t)\delta(x)+b_1(t)\delta(x-1)=\mathcal{F}_0^{(\beta)}(x)+(b_1(t_s)-\beta)e^{-\lambda_1(t-t_s)}\mathcal{F}_1^{(\beta)}(x)\,,
\end{equation}
where the eigenvalue $\lambda_1$ is equal to the scaled mutation rate $\theta$. 

Furthermore, the Dirac delta function $\delta(x-\epsilon_x)$ in Eq.~(\ref{eq:boundary-mutation-drift_diffusion3}), which defines the point of entry of a mutation into the polymorphic region, can be expressed as the polynomial expansion:
\begin{equation}\label{eq:expansion_of_delta_at_epsilon_x}
    \lim_{\epsilon_x\to 0}\frac{\delta(x-\epsilon_x)}{\epsilon_x}=\sum_{n=2}^{\infty}c_n U_n(x)
\end{equation}
with
\begin{equation}\label{eq:constant_c_n}
\begin{split}
    c_n&= \lim_{\epsilon_x\to 0}\frac{1}{\Delta_n}\int_{\epsilon_x}^{1-\epsilon_x} x(1-x)U_n(x)\frac{\delta(x-\epsilon_x)}{\epsilon_x}\, dx\\
    &= \lim_{\epsilon_x\to 0}\frac{1}{\Delta_n^{(B)}}\epsilon_x(1-\epsilon_x)U_n(\epsilon_x)/\epsilon_x\\
    &=\frac{U_n(0)}{\Delta_n^{(B)}}= -(-1)^n(2n- 1)n\,.
\end{split}
\end{equation}
Analogous operations can of course be performed on $\delta(x-1+\epsilon_x)$ \citep[][Eq.~44]{Vogl16}; these are simplified by noting that $U_n(0)=(-1)^n U_n(1)$.

Observing the above, we can rewrite the diffusion equation for the polymorphic interior from Eq.~(\ref{eq:boundary-mutation-drift_diffusion3}); not that the last step below substitutes in $\phi_I(x,t)=\sum_{n=2}^\infty \tau_n(t)U_n(x)$:
\begin{equation}\label{eq:boundary-mutation-drift_diffusion4}
\begin{split}
\frac{\partial\phi_I(x,t)}{\partial t}&=\lim_{\epsilon_x\to 0}\bigg(\frac{\beta\theta b_0(t)}{\epsilon_x}\delta(x-\epsilon_x)+\frac{\alpha\theta b_1(t)}{\epsilon_x}\delta(x-1+\epsilon_x)\bigg)\\
    &\qquad+\frac{\partial^2}{\partial x^2} x(1-x)\phi_I(x,t)\\
\frac{\partial\phi_I(x,t)}{\partial t}&=\sum_{n=2}^{\infty}U_n(x)\bigg(-\beta\theta b_0(t)\frac{U_n(0)}{\lambda_n\Delta_n^{(B)}}-\alpha\theta b_1(t) \frac{U_n(1)}{\lambda_n\Delta_n^{(B)}}\bigg)\\
    &\qquad+\frac{\partial^2}{\partial x^2} x(1-x)\phi_I(x,t)\\
    \frac{\partial}{\partial t}\sum_{n=2}^{\infty}\tau_n(t)U_n(x)&=
    \sum_{n=2}^{\infty}-\lambda_n U_n(x)\bigg(-\beta\theta b_0(t)\frac{U_n(0)}{\lambda_n\Delta_n^{(B)}}-\alpha\theta b_1(t) \frac{U_n(1)}{\lambda_n\Delta_n^{(B)}}+\tau_n(t)\bigg)\,.
\end{split}
\end{equation}
This induces a system of equations for the temporal coefficients:
\begin{equation}\label{eq:DE_inhomogeneous_Us_1}
\begin{split}
    \frac{d}{dt}\tau_n(t)&=-\lambda_n\bigg(-
    \beta\theta b_0(t)\frac{U_n(0)}{\lambda_n\Delta_n^{(B)}}-\alpha\theta b_1(t) \frac{U_n(1)}{\lambda_n\Delta_n^{(B)}}+\tau_n(t)\bigg)\,.
\end{split}
\end{equation}
The $b_0(t)$ and $b_1(t)$ here can be expressed using $\tau_0(t)$ and $\tau_1(t)$; to this end, define:
\begin{equation}\label{eq:E_n_and_O_n}
\begin{split}
    E_n&=\frac{U_n(0)+U_n(1)}{\Delta_n^{(B)}} =-(n- 1)\frac{(-1)^n+1}{\Delta_n^{(B)}}=-(2n-1)n((-1)^n+1)\\
    O_n&=\frac{U_n(0)\beta+U_n(1)\alpha}{\Delta_n^{(B)}}=-(n- 1)\frac{(-1)^n\beta-\alpha}{\Delta_n^{(B)}}=-(2n-1)n((-1)^n\beta-\alpha)\,.
\end{split}
\end{equation}
Then, we obtain:
\begin{equation}\label{eq:DE_inhomogeneous_Fs_1}
\begin{split}
    \frac{d}{dt}\tau_n(t)\mathcal{F}_n^{(\beta)}(x) &= -\lambda_n\bigg(-\alpha\beta\theta\frac{E_n}{\lambda_n} \tau_0(t)-\theta \frac{O_n}{\lambda_n} \tau_1(t)+\tau_n(t) \bigg)\mathcal{F}_n^{(\beta)}(x)\,.
\end{split}
\end{equation}
This system of forward temporal differential equations for the polymorphic interior is complemented by the following equations of the same form for the boundaries (compare Eq.~\ref{eq:bs_2_F0_F1}): 
\begin{equation}\label{eq:DE_inhomogeneous_Fs_2}
\begin{split}
    \frac{d}{dt}\tau_0(t)\mathcal{F}_0^{(\beta)}(x)&=0\\
    \frac{d}{dt}\tau_1(t)\mathcal{F}_1^{(\beta)}(x)&=-\theta\tau_1(t)\mathcal{F}_1^{(\beta)}(x)
\end{split}
\end{equation}

\subsubsection{Diagonalisation of the Forward System}\label{seq:diagonalisation}

The system of temporal equations we have just derived can also be diagonalised. This will be demonstrated below, in part in order to rectify an erroneous equation in \citet{Bergman18b} and in part as a preparation for the derivations in Appendix Section~\ref{sec:appendixB}, which require a diagonalised eigensystem.

Let us proceed by setting $\tau_0(t)=1$ and observing that $\tau_1(t)=\rho_1 e^{-\theta (t-t_s)}$, where $\rho_1=b_1(t_s)-\beta\,$ (see Eq.~\ref{eq:bs_2_F0_F1}).
Then the solution to the inhomogeneous differential equation from Eq.~(\ref{eq:DE_inhomogeneous_Fs_1}) can be written as:
\begin{equation}\label{eq:DE_inhomogeneous_Fs_1b}
\tau_n(t)=\alpha\beta\theta\frac{E_n}{\lambda_n}+\theta\frac{O_n}{\lambda_n-\theta}\rho_1 e^{-\theta  (t-t_s)}+ \rho_n^{(d)} e^{-\lambda_n  (t-t_s)}\,,
\end{equation}

where $\rho_n^{(d)}$ is set to 
$$
    \rho_n^{(d)}=\rho_n-\alpha\beta\theta\frac{E_n}{\lambda_n}-\theta\frac{O_n}{\lambda_n-\theta}\rho_1
$$
to satisfy the starting the condition $\tau_n(t_s)=\rho_n$. 


Recall that $\phi(x,t)=\sum_{n=2}^\infty\tau_n(t)\mathcal{F}_n^{(\beta)}$. The eigenfunctions $\mathcal{F}_n^{(\beta)}(x)$ in $\phi(x,t)$ associated with $\tau_0(t)=1$, $\tau_1(t)=\rho_1 e^{-\theta t}$ and each $\rho_n e^{-\lambda_n t}$ can therefore be collected to obtain a diagonalised system of forward eigenfunctions, which induces the homogeneous system of temporal differential equations:
\begin{equation}\label{eq:forw_boundary_homogenous_system}
\begin{cases}
    \mathcal{F}_0^{(\beta,\theta)}(x)&=\mathcal{F}_0^{(\beta)}(x)+\alpha\beta\theta\sum_{n=2}^\infty\frac{E_n}{\lambda_n}\mathcal{F}_n^{(\beta)}(x)\\
    \mathcal{F}_1^{(\beta,\theta)}(x)&=\mathcal{F}_1^{(\beta)}(x)+\theta\sum_{n=2}^\infty\frac{O_n}{\lambda_n-\theta}\mathcal{F}_n^{(\beta)}(x)\\
    \mathcal{F}_{n\geq2}^{(\beta,\theta)}(x)&=\mathcal{F}_{n\geq2}^{(\beta)}(x)\,.
\end{cases}
\end{equation}
This corrects an error in the denominator of the summation in the $\mathcal{F}_1^{(\beta,\theta)}$ given by \citet[][Eqs.~63-65]{Bergman18a}. Note that the eigenvalues of the diagonalised system are the same as for the original temporal system, \ie{} correspond to $\lambda_0=0$, $\lambda_1=\theta$, and $\lambda_n=n(n-1)$ for $n\geq 2$ \citep{Vogl14c,Bergman18a}.

This diagonalised system for the eigenfunctions can easily be written in matrix form: Defining
\begin{equation}
    \mathbf{A}=
    \begin{pmatrix}
        1 &0 &\alpha\beta\theta\frac{E_2}{\lambda_2} &\alpha\beta\theta\frac{E_3}{\lambda_3} &\alpha\beta\theta\frac{E_4}{\lambda_4} &\cdots\\
        0 &1 &\theta\frac{O_2}{\lambda_2-\theta} &\theta\frac{O_3}{\lambda_3-\theta}
        &\theta\frac{O_4}{\lambda_4-\theta} &\cdots\\
        0 &0 &1 &0 &0 &\cdots\\
        0 &0 &0 &1 &0 &\cdots\\
        0 &0 &0 &0 &1 &\cdots\\
        \vdots &\vdots &\vdots &\vdots &\vdots &\ddots\\
    \end{pmatrix}\,,
\end{equation}
and the column vectors 
$$
\bs{\mathcal{F}}^{(\beta)'}=(\mathcal{F}_0^{(\beta)}(x),\mathcal{F}_1^{(\beta)}(x),\mathcal{F}_2^{(\beta)}(x),\mathcal{F}_3^{(\beta)}(x),\dots)^{'}
$$ 
as well as an analogous column vector for the diagonalised eigenfunctions $\bs{\mathcal{F}}^{(\beta,\theta)'}$, the homogeneous system is given by:
\begin{equation}
 \bs{\mathcal{F}}^{(\beta,\theta)'}=\mathbf{A}\bs{\mathcal{F}}^{(\beta)'}\,.
\end{equation}
Since the inverse of $\mathbf{A}$ is quite simply:
\begin{equation}
    \mathbf{A}^{-1}=
    \begin{pmatrix}
        1 &0 &-\alpha\beta\theta\frac{E_2}{\lambda_2} &-\alpha\beta\theta\frac{E_3}{\lambda_3} &-\alpha\beta\theta\frac{E_4}{\lambda_4} &\cdots\\
        0 &1 &-\theta\frac{O_2}{\lambda_2-\theta} &-\theta\frac{O_3}{\lambda_3-\theta}
        &-\theta\frac{O_4}{\lambda_4-\theta} &\cdots\\
        0 &0 &1 &0 &0 &\cdots\\
        0 &0 &0 &1 &0 &\cdots\\
        0 &0 &0 &0 &1 &\cdots\\
        \vdots &\vdots &\vdots &\vdots &\vdots &\ddots\\    \end{pmatrix}\,,
        \end{equation}
the diagonalising transformation can be reversed easily. 

The system of inhomogeneous temporal equations for the temporal coefficients alone (\ie{} Eq.~\ref{eq:DE_inhomogeneous_Fs_1b}) can also be expressed in matrix form. To this end, define the row vector of inhomogeneous temporal coefficients
$$
\bs{\tau}=(\tau_0(t),\tau_1(t),\tau_2(t),\dots)
$$ 
and the equivalent row vector of the temporal coefficients of the homogeneous system
$$
\bs{\tau}^{(d)}=(1,\rho_1^{(d)} e^{-\theta (t-t_s)},\rho_2^{(d)} e^{-2 (t-t_s)},\dots)\,.
$$ 
The system of equations from Eq.~(\ref{eq:DE_inhomogeneous_Fs_1b}) can then be stated compactly as:
\begin{equation}
 \bs{\tau}=\bs{\tau}^{(d)}\mathbf{A}\,;
\end{equation}
note that the matrix transformation can once again be reversed easily.

\subsubsection{Derivation of the Backwards System}\label{sec:appendix_diag}

A backwards system of diagonalised eigenfunctions for the boundary mutation model, denoted as $\mathcal{B}_n^{(\beta,\theta)}(x)$ for $n\geq0$, should fulfil the following orthogonality relation for any $n,m>0$:
$$
    \int_0^1 \mathcal{B}_n^{(\beta,\theta)}(x)\mathcal{F}_m^{(\beta,\theta)}(x)\,dx=\delta_{nm}\Delta_n^{(B)},
$$
with $\Delta_0^{(B)}=\Delta_1^{(B)}=1$ and $\Delta_n^{(B)}$ as before. Such a system can be achieved by performing the following diagonalising transformation concurrently to the diagonalisation of the forward eigensystem \citep[][Eq.~68]{Bergman18a}:
\begin{equation}\label{eq:back_boundary_homogenous_system}
\begin{cases}
    \mathcal{B}_0^{(\beta,\theta)}(x)&=\mathcal{B}_0^{(\beta)}(x)\\
    \mathcal{B}_1^{(\beta,\theta)}(x)&=\mathcal{B}_1^{(\beta)}(x)\\
    \mathcal{B}_{n\geq2}^{(\beta,\theta)}(x)&=\mathcal{B}_{n\geq2}^{(\beta)}(x)-\alpha\beta\theta\frac{E_{n}\Delta_{n}^{(B)}}{\lambda_{n}}\mathcal{B}_0^{(\beta)}(x)-\theta\frac{O_{n}\Delta_{n}^{(B)}}{\lambda_{n}-\theta}\mathcal{B}_1^{(\beta)}(x)\,.
\end{cases}
\end{equation}
The above equations correct an error in the $\mathcal{B}_{n\geq2}^{(\beta,\theta)}(x)$ given in \citet[][Eq.~68]{Bergman18a}); aside from this, the orthogonality relationship can be checked precisely as demonstrated in \citep[][Eq.~69]{Bergman18a}.  

The system of diagonalised backwards eigenfunctions can also be written in matrix form. Specifically, defining the matrix
\begin{equation}
    \mathbf{B}=
    \begin{pmatrix}
        1 &0 &0 &0 &0 &\cdots\\
        0 &1 &0 &0
        &0 &\cdots\\
        -\alpha\beta\theta\frac{E_2\Delta_2^{(B)}}{\lambda_2} &-\theta\frac{O_2\Delta_2^{(B)}}{\lambda_2-\theta} &1 &0 &0 &\cdots\\
        -\alpha\beta\theta\frac{E_3\Delta_3^{(B)}}{\lambda_3} &-\theta\frac{O_3\Delta_3^{(B)}}{\lambda_3-\theta} &0 &1 &0 &\cdots\\
        -\alpha\beta\theta\frac{E_4\Delta_4^{(B)}}{\lambda_4} &-\theta\frac{O_4\Delta_4^{(B)}}{\lambda_4-\theta} &0 &0 &1 &\cdots\\
        \vdots &\vdots &\vdots &\vdots &\vdots &\ddots\\
    \end{pmatrix}\,,
\end{equation}
the column vector 
$$
\bs{\mathcal{B}}^{(\beta)'}=(\mathcal{B}_0^{(\beta)}(x),\mathcal{B}_1^{(\beta)}(x),\mathcal{B}_2^{(\beta)}(x),\dots)^{'},
$$ 
and an analogous column vector $\bs{\mathcal{B}}^{(\beta,\theta)'}$, we obtain:
\begin{equation}\label{eq:from_inhomogeneous2homogeneous_B}
 \bs{\mathcal{B}}^{(\beta,\theta)'}=\mathbf{B}\bs{\mathcal{B}}^{(\beta)'}\,.
\end{equation}
The matrix $\mathbf{B}$ can be inverted similarly to the matrix $\mathbf{A}$ above, and therefore the diagonalising transformation can again be reversed as easily.


Expressions for the temporal coefficients can also be found via the backwards system. Let us expressly define the row vector
$$
\bar{\bs{\tau}}^{(d)}=(1,d_1(k,K) e^{-\theta t},d_2(k,K)  e^{-\theta (t-t_s)},\dots)\,.
$$ 
The system of equations from Eq.~(\ref{eq:DE_inhomogeneous_Fs_1b}) can then be stated compactly as:
\begin{equation}
 \bar{\bs{\tau}}=\bar{\bs{\tau}}^{(d)}\mathbf{B}\,.
\end{equation}

We can use these relationships to write the solution of the inhomogeneous temporal equations as:
\begin{equation}\label{eq:DE_inhomogeneous_Bs}
\begin{split}
    -\frac{\partial}{\partial t}\bar{\tau}_0(t)\mathcal{B}_0^{(\beta)}(x)
    &=\alpha\beta\theta\sum_{n=2}^{\infty}E_n\Delta_n \tau_{n}(t)\mathcal{B}_0^{(\beta)}(x)\\
    -\frac{\partial}{\partial t}\bar{\tau}_1(t)\mathcal{B}_1^{(\beta)}(x)
    &=-\theta\bar{\tau}_1(t)\mathcal{B}_1^{(\beta)}(x)+\theta\sum_{n=2}^{\infty} O_n\Delta_n\bar{\tau}_n(t)\mathcal{B}_1^{(\beta)}(x)\\
    -\frac{\partial}{\partial t}\bar{\tau}_n(t)&=-\lambda_n\bar{\tau}_n(t)\\
\end{split}
\end{equation}

Thus a homogeneous temporal system can be obtained for $n\geq 2$ from the last line above: Consider $t_s\leq t \leq 0$, which is solved by $\tau_{2\leq n\leq K}(x)=d_n(k,K)e^{\lambda_n t}$ if one starts  with a binomial sampling scheme at $t=0$. Substituting these temporal functions into the differential equations for $\bar{\tau}_0(t)$ and $\bar{\tau}_1(t)$ results in inhomogeneous equations analogous to those running forward in time. In particular, we have:
\begin{equation}
    -\frac{\partial}{\partial t}\bar{\tau}_0(t)\mathcal{B}_0^{(\beta)}=\alpha\beta\theta\sum_{n=2}^{K}d_n(k,K)E_n\Delta_n e^{\lambda_n t} \mathcal{B}_0^{(\beta)}\,,
\end{equation}
for which
\begin{equation}
    \bar{\tau}_0(t)= d_0(k,K) +\alpha\beta\theta\sum_{n=2}^{K}\frac{E_n\Delta_n}{\lambda_n}d_n(k,K)(1-e^{\lambda_n t})
\end{equation}
is a solution that satisfies the starting condition $\bar{\tau}_0(0)= d_0(k,K)$.


Proceeding similarly for $\tau_1(t)$, we have:
\begin{equation}
    -\frac{\partial}{\partial t}\bar{\tau}_1(t)\mathcal{B}_1^{(\beta)}=-\theta\bigg(\bar{\tau}_1(t)-\sum_{n=2}^{\infty}O_n\Delta_n d_n(k,K)e^{\lambda_n t}\bigg)\mathcal{B}_1^{(\beta)}\,,
\end{equation}
for which the solution satisfying the starting condition is:
\begin{equation}
    \bar{\tau}_1(t)= d_1(k,K)e^{\theta t} +\theta\sum_{n=2}^{K}d_n(k,K)\frac{ O_n\Delta_n}{\lambda_n-\theta}(e^{\theta t}-e^{\lambda_n t})\,.
\end{equation}

This can be repeated again for higher order temporal coefficients.


\paragraph{Backward Partial Differential Equation} 
The results presented above can be used to obtain a system of backwards PDEs corresponding to the forward PDEs of Eq.~\ref{eq:boundary-mutation-drift_diffusion4}. Let us first reflect on the dynamics of mutations in the boundary mutation diffusion model and how these are analytically represented in the spectral representation of the transition rate densities:  Forwards in time, the mutation terms appearing in the first line of Eq.~(\ref{eq:boundary-mutation-drift_diffusion4})
represent the mutational flow from the monomorphic boundaries into the polymorphic interior. The delta functions are then expanded into the orthogonal polynomials (compare Eq.~\ref{eq:expansion_of_delta_at_epsilon_x} and Eq.~\ref{eq:constant_c_n}) and appear in the diffusion operator, \ie{} the polymorphic region. Backwards in time, the mutation terms in the PDE will have to represent the conditioning on the origin of the polymorphism-causing mutation, \ie{} either the non-focal or the focal allele, which will be represented as the backwards eigenfunctions of order $n=0$ and $n=1$. In particular ,the mutation term for the non-focal allele is a product of the mutation rate, $\beta\theta$, the backward function at $\epsilon_x$, $\psi_I(x=\epsilon_x,t)$, and the scaling constant, $1/\epsilon_x$. Taking the limit, we obtain (compare Eq.~\ref{eq:expansion_of_delta_at_epsilon_x} and Eq.~\ref{eq:constant_c_n}):
\begin{equation}\label{eq:derive_mut_backward}
\begin{split}
    \lim_{\epsilon_x\to 0}\beta\theta\cdot\psi_I(x=\epsilon_x,t)\cdot 1/\epsilon_x &=\lim_{\epsilon_x\to 0}\beta\theta\sum_{n=2}^\infty U_n(\epsilon_x)\epsilon_x(1-\epsilon_x)/\epsilon_x\\
    &=\beta\theta\sum_{n=2}^\infty U_n(0)\,.
\end{split}
\end{equation}
The analogous operation can be performed for the focal allele. We must now consider only the spatial dynamics of the system represented by the eigenfunctions. Recall that analogously to the transformation in Eq.~(\ref{eq:bs_2_F0_F1}), $1-x$ and $x$ can be represented by the eigenfunctions $\mathcal{B}_0^{(\beta)}=1$ and $\mathcal{B}_0^{(\beta)}=-\beta+x=-\beta(1-x)+\alpha x$, so that we can overall express the boundary dynamics as:
\begin{equation*}
    \beta U_n(0)(1-x)+\alpha U_n(1) x = 
        \alpha\beta E_n\Delta_n\mathcal{B}_0^{(\beta)}(x)+ O_n\Delta_n\mathcal{B}_1^{(\beta)}(x)\,,
\end{equation*}
with $E_n$ and $O_n$ as in Eq.~\ref{eq:E_n_and_O_n}. Note that for $K=1$, only the first two eigenfunctions need to be considered, yielding a homogeneous temporal system with rates $\lambda_0=0$ and $\lambda_1=1$; for $K\geq 2$, the system is inhomogeneous.
Conveniently, for polynomials of order $n\geq 2$ the operator is identical to the pure drift operator. Dynamics are represented via the eigenfunctions $\mathcal{B}_{n\geq 2}(x)$ resulting in a homogeneous system for $n\geq 2$ (Eq~\ref{eq:DE_inhomogeneous_Bs}). 


\paragraph{Remark} 
Note that in the backward system immediately above, the conditioning on $x$ is taken from $t=0$ to earlier times $t$. This means that although we consider time to be moving backwards since it is the backwards system, the direction of conditioning over time is the same as in the forward system. Thus in Eq.~(\ref{eq:derive_mut_backward}), the scaling constant of the mutational terms enters as $1/\epsilon_x$, \ie{} identical in the forward and backward systems, rather than as the inverse. (Compare the explanation for the direction of the arrows in  Fig.~\ref{fig:e_Betabinomial}.) 

\paragraph{Remark} 
Our diagonalising transformations are similar in spirit to that in \citet{Song12}; however while these authors diagonalise the spatial component of their spectral decompositions via a linear transformation, we diagonalise the temporal component via a linear transformation. 

\subsubsection{Simplifying Assumptions and Convergence}\label{section:simpl_ass}

In the main part of this article, we consider modifications of the forwards system of inhomogeneous differential equations for the temporal coefficients from Eq.~(\ref{eq:DE_inhomogeneous_Fs_1}). In this part, however, we assume that the system has reached an evolutionary equilibrium with respect to the mutation bias at time $t=\infty$, which means that $\tau_1(\infty)=0$, and $b_0(\infty)=\alpha$ and $b_1(\infty)=\beta$. 

Unlike the pure drift system, this system has a non-trivial stationary solution \citep[][Appendix~A2]{Vogl14c}: 
\begin{equation}
    \tau_n(\infty)= \alpha\beta\theta\frac{E_n}{\lambda_n}=-\alpha\beta\theta\frac{(2n-1)n((-1)^n+1)}{\lambda_n}\,,
\end{equation}
whereby one can differentiate between $\tau_{n}(\infty)=-\alpha\beta\theta\frac{4n-2}{n-1}$ for $n\in 2\mathbb{N}$, and $\tau_{n}(\infty)=0$ otherwise. The full forwards diffusion equation therefore has a limiting stationary distribution at $t=\infty$ that can be represented as:
\begin{equation}\label{eq:boundary_diffusion_stationary}
    \phi(x,\infty)=\sum_{n=0}^{\infty}\tau_n(\infty)\mathcal{F}_n^{(\beta)}(x)\,.
\end{equation}
Importantly, this limiting distribution has singularities at the boundaries. We still call it a distribution since it can easily be seen from Eq.~(\ref{eq:forw_McKaneWax_p}) that $\phi(x,\infty)$ integrates to one over the closed interval $[0,1]$: Specifically, only $\mathcal{F}_0^{(\beta)}(x)$ integrates to one while all other $\mathcal{F}_n^{(\beta)}$ integrate to zero. 


The stationary event probabilities $\bar p_k$ of observing exactly $k$ focal alleles in a sample of size $K$ can be obtained by expanding the binomial sampling scheme into the backward boundary polynomials given in Eq.~(\ref{eq:dens_geg}), multiplying the result with $\phi(x,\infty)$, and then integrating:
\begin{equation}\label{eq:boundary_diffusion2discrete_stationary}
\begin{split}
\bar p_k= \Pr(k \mid K,\beta,\theta)&=\int_0^1 \Pr(k\given x,K)\phi(x,\infty)\,dx\\
    &=\sum_{n=0}^K  d_n(k,K) \mathcal{B}_n^{(\beta)}(x)\tau_n(\infty)\mathcal{F}_n^{(\beta)}(x)\\
    &=\begin{cases}
    \displaystyle \alpha(1-\beta\theta H_{K-1}) & k=0 \\
    \displaystyle \alpha\beta\theta\frac{K}{k(K-k)} & 1\leq k \leq K-1\\
    \displaystyle \beta(1-\alpha\theta H_{K-1}) & k=K
    \end{cases}\\
    &=boundbin(k \mid K,\beta,\theta)\,.
\end{split}
\end{equation}
This is precisely the boundary-binomial distribution from Eq.~(\ref{eq:boundary_stationary}), which describes a sample of size $K$ from the stationary distribution of the discrete boundary mutation Moran model with $K\leq N$, where $N$ is the population size
\citep[][Eq.~13]{Vogl15}. Note that while the limiting stationary distribution $\phi(x,\infty)$ has singularities at the boundaries and therefore does not conform to a proper density, the equilibrium distribution of a sample of size $K$ from this limiting distribution conforms to a regular probability distribution as long as the sample size is lower than $K_{max} \approx e^{min(\alpha,\beta)/(\alpha\beta\theta)}$ (since the boundary values otherwise become negative). This restriction on the sample size follows immediately from the fact that the stationary boundary mutation Moran model corresponds to the first order expansion in the overall mutation rate $\theta$ of the beta-binomial sampling distribution of the general mutation drift Moran model from Eq.~(\ref{eq:first_order_stationary}). 

Overall, recovering the stationary distribution of the discrete boundary mutation Moran model as the limiting stationary sampling distribution here validates our representation of the boundary mutation diffusion model as the limiting $\theta\to 0$ approximation of the general mutation diffusion model: Considering this limit in a diffusion setting typically eradicates mutation and leads to a pure drift model on only the polymorphic allele frequency space, but in the boundary mutation diffusion model we then re-instate mutation via constructed boundary terms at the monomorphic states (recall derivations in Appendix~\ref{section:eigensystem_boundary_mutation} and \citep{Vogl16}, as well as the final remark in Section~\ref{sec:boundary_spectral}). While this appears indirect, we can thereby see that this approach yields consistent results. Most importantly, we have therefore also shown that the discrete sampling distribution from the boundary mutation diffusion is not only a proper distribution but also a regular density for even moderately large sample sizes as long as the appropriate assumptions on the upper limits of the mutation rate and sample sizes combined are met. Therefore, it is easily tractable as a starting point for development of inferential procedures in suitable settings.

\paragraph{Remark} While it is beyond the scope of this paper to investigate this matter in detail, the polynomial expansion of the binomial sampling scheme acts as a test function in Eq.~(\ref{eq:boundary_diffusion2discrete_stationary}), against which the limiting stationary distribution integrates to the equilibrium sampling distribution. Although the limiting stationary distribution itself diverges near the boundaries, making the full integral over $\phi(x,\infty)$ on $[0,1]$ improper, it appears that this integral can still be assigned a value: Partitioning the integral into interior and boundary regions as in the examination of the transition rates in Eq.~(\ref{eq:boundary_inside_whole}) and Eqs.~(\ref{eq:LimitFlow_at_0}) and ~(\ref{eq:LimitFlow_at_1}), it can be seen that the 'breakpoints' between the interior and the boundaries can be approached smoothly from either side and that the terms that diverge for $N \to \infty$ near either edge of the interior as well as those that diverge at the boundaries themselves balance precisely. In such cases, analytical considerations can be applied that lead to the divergent terms cancelling out, thus assigning the improper integral a finite value  \citet[see the Cauchy principal value][p.552]{Harris98}. The derivation of the stationary sampling distribution via the orthogonal polynomials in Eq.~(\ref{eq:boundary_diffusion2discrete_stationary}), however, circumvents such analytical difficulties so that we need not expand on them further here.

\paragraph{Remark} \citet[][Eq.~35]{Burden17} provide the analogous discrete distribution to Eqs.~(\ref{eq:boundary_stationary}) or equivalently Eq.~(\ref{eq:boundary_diffusion2discrete_stationary}) for the multiallelic boundary mutation model, however they only show that the equality holds up to order $1/N$ (see their Eq.~41). 

\paragraph{Remark} Substituting $\mathcal{F}_0(x)g(x,t)$, \ie{} the limiting stationary distribution times a test function, into the interior forward PDE (Eq.~\ref{eq:boundary-mutation-drift_diffusion3}) results in the pure drift backward PDE (up to a change in the direction of time) \citep[][Section 3.2]{Bergman18a}, with the exception of the mutation terms, which must be added to first two eigenfunctions in the backward system, as explained in previous sections.

\newpage

\section{Appendix: Embedded Dual Jump Processes}\label{sec:appendixB}

In this section, we aim to provide some understanding of the dual jump processes embedded in the general biallelic reversible mutation and the biallelic boundary mutation diffusion models. Our approach is motivated by both \citet{Etheridge09}, who derive a coalescent dual process for a stationary multiallelic Moran model with selection from the generator of its limiting diffusion approximation, as well as \citet{Papaspiliopoulos14}, who present a comprehensive method for identifying a dual processes for stationary, reversible diffusion models from their diffusion generator. This latter method is particularly relevant, since (similarly to us) the authors view the continuous diffusion process as a hidden process within the framework of a Hidden-Markov-Model, and the sampling distribution as the emission density. In this setting, the identified dual process is the computationally optimal filter for the diffusion. Here, we will present embedded dual jump processes for the general biallelic reversible mutation and the biallelic boundary mutation diffusion models, specifically pure death jump processes that describe the ancestry of population samples. These are comparable to the dual processes of the above articles, but we will focus specifically on their probabilistic interpretation in the context of the forward-backward algorithm.



\subsection{Recap of the Forward-Backward Algorithm}
\label{sec:appendixB.1.new}

Recall that in Appendix Section~\ref{sec:sample_probs}), we determined $p_k$, the probability of observing $k$ focal alleles in a binomial sample of size $K$ drawn from a pure drift diffusion model at time $t=0$, as the marginal likelihood of the product of the forward and backward transition rate densities, both represented as a spectral sum, via a forward-backward algorithm. In the main text, specifically Sections~(\ref{sec:maths_intro_BM}, \ref{sec:maths_intro_general_framework}, and \ref{sec:models}), we use a forward pass of the algorithm to determine the same probability for diffusion equations with mutational input in the form of general biallelic reversible mutations and boundary-mutations respectively; this means that we choose to evaluate the probability of $p_k$ at $t=0$. In this first subsection, our main aim is to establish a notation for $p_k$ that holds for both of these patterns of mutation, allowing us to evaluate it at any time $t_s \leq t \leq 0$ via a combination of forward and backward passes of the forward-backward algorithm.

In order to do so, we will again assume a population allele proportion that evolves forward in time via a diffusion equation; specifically, it evolves from an ancestral distribution $\rho(x)$ at time $t_s<0$ up to the extant time $t=0$, at which point we draw a binomial sample of haploid size $K$. We will treat this as a single time epoch, assuming constancy of mutation parameters for its duration. Note that this could be because we do not wish to model changes in these parameters, or because we assume that we are considering the last of a series of epochs with piecewise constant mutation parameters as in the main text.

The core principle of the forward-backward algorithm is that the marginal probabilities of the observations, here the event probabilities $p_k$, depend on the length of the interval between $t=t_s$ and $t=0$ and can be evaluated to the same result at any point $t$ within this interval. This is achieved by a combination of forward probabilities of population allele proportions covering the events from $t_s$ up to $t$, and backward probabilities that evaluate the likelihood of the sampling distribution from $t$ to $t=0$. Implicitly, this requires an adjoint relationship between the forward and backward diffusion generators \citep[][Section 3.1., specifically Eq.~32]{Bergman18a}. It is more straightforward to see that this holds in time homogeneous systems; therefore, we will from here on use the spectral decomposition of the general mutation diffusion model from Section~\ref{sec:maths_intro_BM} \citep[adjoint relationship verified in][Section 3.4]{Bergman18a} as well as the diagonalised eigensystem of the boundary mutation diffusion from Appendix (\ref{sec:appendix_diag}) \citep[][Section 4.1]{Bergman18a}.

Following on from this, we expand the distribution that describes the distribution of ancestral population allele proportions at time $t_s$ into a series of polynomials $\rho(x)= \phi(x,t=t_s)= \sum_{m=0}^\infty \rho_m^{(\beta,\theta)} \mathfrak{F}_m^{(\beta,\theta)}(x)$, and then similarly expand the binomial sampling distribution into a complementary series of orthogonal polynomials $\Pr(k\given K,x)= \psi(x,t=0) =\sum_{n=0}^K d_n(k,K)\mathfrak{B}_n^{(\beta,\theta)}(x)$. The functions of the indeterminate $x$ in these spectral sums are constituted by the orthogonal polynomials themselves, which here represent the eigenfunctions of \textit{either} time homogeneous eigensystem with mutational input, and the coefficients of these function in the spectral sums therefore depend on both the mutation rate and bias.

Within either homogeneous eigensystem, the probability of observing $k$ focal alleles in the sample drawn at the extant time $t=0$ can then be determined by evaluating the following marginal likelihood at any time $t$ within $t_s\leq t \leq 0$ \citep[][Section~3.1]{Bergman18a}: 
\begin{equation}\label{eq:marginal_any_time}
\begin{split}
    p_k &=\Pr(k\given K,\beta,\theta,\rho,t_s)\\ &=\int_0^1 \psi(x,t)\phi(x,t)\,dx\\
    &=\int_0^1 \bigg(\sum_{n=0}^K d_n(k,K) e^{\lambda_n t} \mathfrak{B}_n^{(\beta,\theta)}(x)\bigg)\\
    &\qquad\times\bigg(\sum_{m=0}^\infty \rho_m^{(\beta,\theta)} e^{-\lambda_m(t-t_s)}\mathfrak{F}_m^{(\beta,\theta)}(x)\bigg)\,dx\\
    &=\sum_{n=0}^K \int_0^1 d_n(k,K) e^{\lambda_n t} \mathfrak{B}_n^{(\beta,\theta)}(x) \rho_n^{(\beta,\theta)}e^{-\lambda_n(t-t_s)}\mathfrak{F}_n^{(\beta,\theta)}(x)\,dx\\
    &=\sum_{n=0}^K d_n(k,K) \rho_n^{(\beta,\theta)}\Delta_n^{(\beta,\theta)} e^{\lambda_n t_s}\,.
    \end{split}
\end{equation}

Note that the above distribution is discrete, \ie{} $0\leq p_k\leq 1$ for $0\leq k\leq K$, whereby $\sum_k p_k=1$. In the remainder of this Appendix, we will show that the transition rates $\psi(x,t)$ of the backwards process can be interpreted probabilistically for $t_s\leq t\leq 0$. Specifically, we will rewrite its representation as a spectral sum into temporally weighted mixtures of binomial samples across time, thereby tracing the genealogy of the observed sample back through the population history via an embedded time-dependent jump process.

\paragraph{Remark} In this Section, we considered a single interval between $t=t_s$ and $t=0$; however, multiple epochs can easily be incorporated, \ie{} the backward process can be extended to additional epochs going further back in time. At the time where the mutation parameters change, which we denote as $t_j$ (which may coincide with $t_s$), $\phi(x,t_j)$ is represented by a weighted sum of binomial distributions. This can be used as the initial distribution, \ie{}  $\phi(x,t)$, in determining the backwards process for epochs further back in time; note that previously, $\phi(x,t)$ was set to a single binomial distribution $\Pr(k\given K, x)$ at time $t=0$.

\subsection{General Description of Dual Embedded Pure Death Jump Processes}

Let us begin by providing an outline of the embedded pure death jump processes we will derive: To start, denote the observed sample allele configuration at the extant time $t=0$ as $(k,K)$, which simply means that $k$ of the total $K$ sampled alleles are of the focal type. In the history of this sample, we can ``lose'' alleles of both the focal or non-focal type as they coalesce with a common ancestor as well as through parent-independent mutation. Looking back in time, any reduction in sample size due to these events will be considered a jump step. After $K-n$ of these, we will denote our allele configuration as $(i,n)$, meaning that we have $i$ (where $\max(0,k+n-K)\leq i \leq \min(k,n)$) focal alleles among our remaining total of $n\leq K$ from the original sample. The transition probabilities between configurations of different historical sample sizes will be denoted as $\Pr(n\given K,t)$, and they determine the temporal aspect of the sample genealogy. Subordinated to this is a spatial process that governs the number of focal alleles $i$ across this genealogy, whereby we take "spatial" to mean "dependent on the population allele proportion $x$": This process is constituted by a sequence of pure death jump steps on the space of binomial distributions $\Pr(i\given n,x)=\binom{n}{i}x^i(1-x)^{n-i}$ with $0\leq i \leq k$ and $0\leq (n-i)\leq (K-k)$, with coefficients $e_{i,n}$ recording the probability of observing $k$ focal alleles in the sample at the extant time across all possible sample paths from configuration $(i,n)$ to $(k,K)$: 
$$
e_{i,n}={\Pr}(k\given K, i, n, \beta,\theta)\,.
$$ 
Putting these two processes together, the the transition rates $\psi(x,t)$ of the backwards process can be alternatively expressed as the following time-dependent pure death jump process:
\begin{equation}\label{eq:B_to_sum_of_binomials}
\begin{split}
        \psi(x,t)&=\Pr(k\given K,\beta,\theta,\rho,x,t)\\
        &=\sum_{n=0}^K \Pr(n\given K,t) \sum_{i=0}^n e_{i,n}\Pr(i\given n,x)\,.
    \end{split}
\end{equation}

We will now proceed to derive this representation of the backwards transition rate density for the general biallelic reversible mutation diffusion model, and explicitly show that it is an embedded dual process with direct connections to the spectral sum we have typically expanded the transition rate density into in this article. Then, we will appropriately modify the derivation to obtain corresponding results for the boundary mutation model. 

\paragraph{Remark}
Note that terms analogous to our Eq.~(\ref{eq:B_to_sum_of_binomials}) appear in
an expression for the stationary transition distribution of the backward multiallelic Moran diffusion equation in \citet[][Eq.~2]{Ethe09}.

\subsection{Derivation of Dual Embedded Pure Death Jump Processes}\label{sec:appendixB.2.new}

\subsubsection{General Biallelic Reversible Mutation Model}

In order to separate the temporal and spatial processes, we take an approach comparable to \citep[][Eq.~7]{Ethe09}. Specifically, we substitute the binomial likelihood $\Pr(i\given n,x)= \binom{n}{i}\,x^i(1-x)^{n-i}$ into the backwards diffusion equation of the general biallelic reversible mutation model (Eq.~\ref{eq:mutation-drift_diffusion}):
\begin{equation}\label{eq:death_general_mutation_rec_backward}
\begin{split}
    \mathcal{L}^{*}\Pr(i\given n,x)&=\mathcal{ L}^{*}\binom{n}{i}x^i(1-x)^{n-i}\\
    &=i(i-1+\beta\theta)\binom{n}{i}x^{i-1}(1-x)^{n-i+1}\\
    &\qquad+(n-i)(n-i-1+\alpha\theta)\binom{n}{i}x^{i}(1-x)^{n-i-1}\\
    &\qquad-(i(n-i+\alpha\theta)+(i+\beta\theta)(n-i))\binom{n}{i}x^i(1-x)^{n-i}\\
    &=\lambda_n\bigg(\tfrac{i-1+\beta\theta}{n-1+\theta}\Pr(i-1\given n-1,x)\\
    &\qquad
    +\tfrac{n-i-1+\alpha\theta}{n-1+\theta}\Pr(i\given n-1,x)-\Pr(i\given n,x)\bigg)\,.
\end{split}
\end{equation}
Recall that the eigenvalues of this system, which govern its temporal aspect, are given by $\lambda_n=n(n-1+\theta)$. Regarding the number of focal alleles rather than the population allele proportion $x$ as the dependent variable, the last line of the above equation can be seen as an infinitesimal generator of the spatial process within the large parentheses \citep[compare][Eq.~7]{Ethe09}. Setting this spatial process zero yields a recursive set of equations on the probability of observing $i$ focal alleles within historical samples of size $n$:
\begin{equation}\label{eq:recursion}
    \Pr(i\given n,x)=\tfrac{i-1+\beta\theta}{n-1+\theta}\Pr(i-1\given n-1,x)+\tfrac{n-i-1+\alpha\theta}{n-1+\theta}\Pr(i\given n-1,x)\,,
\end{equation}
which can also be written as:
\begin{equation}\label{eq:recursion_2}
\begin{split}
    \Pr(i\given n,x)&=\frac{i}{n}\frac{betabin(i\given n,\beta,\theta)}{betabin(i-1\given n-1,\beta,\theta)}\Pr(i-1\given n-1,x)\\
    &\qquad+\frac{n-i}{n}\frac{betabin(i\given n,\beta,\theta)}{betabin(i\given n-1,\beta,\theta)}\Pr(i\given n-1,x)\,.
\end{split}
\end{equation}

This recursion defines a time-independent, pure death jump process: Since it goes backwards in time, it starts from the allele configuration $(i=k,n=K)$ at the extant time $t=0$, proceeds via configurations $(i=k-1,n=K-1)$ and $(i=k,n=K-1)$ at the next lower possible sample size $n$, and down to the trivial empty sample $(i=0,n=0)$, which has $\Pr(i=0\given n=0,x)=1$, since the only possible value of $i$ after $K$ steps is zero. Let $e_{i,n}\geq 0$ be the solution of this recursive system; it is given by:
\begin{equation}\label{eq:e_in_back_const}
\begin{split}
    e_{i,n}={\Pr}(k\given K, i, n, \beta,\theta)\,
    &=\binom{K-n}{k-i}\frac{\Gamma(\theta+n)}{\Gamma(\theta+K)}\frac{\Gamma(\beta\theta+k)}{\Gamma(\beta\theta+i)}\frac{\Gamma(\alpha\theta+K-k)}{\Gamma(\alpha\theta+n-i)}\\
    &=hyper(i\given K,n,k)\,\frac{betabin(k\given K,\beta,\theta)}{betabin(i\given n,\beta,\theta)}\,.
\end{split}
\end{equation}

Note that $e_{0,0}$, which is the solution for the trivial historic sample configuration $(i=0,n=0)$, corresponds to the stationary beta-binomial sampling distribution for the configuration $(k, K)$ from Eq.~(\ref{eq:betabin}). From this, it is easy to see that the $e_{i,n}$ are indeed the probabilities of observing $k$ focal alleles in a sample configuration $(k,K)$ at the extant time $t=0$ given the historic sample configuration $(i,n)$ at a time $K-n$ jump steps in the past; note also that this process is stationary for constant $\beta$ and $\theta$.  

\begin{figure}[!ht]
    \centering
    \includegraphics[width = 12cm]{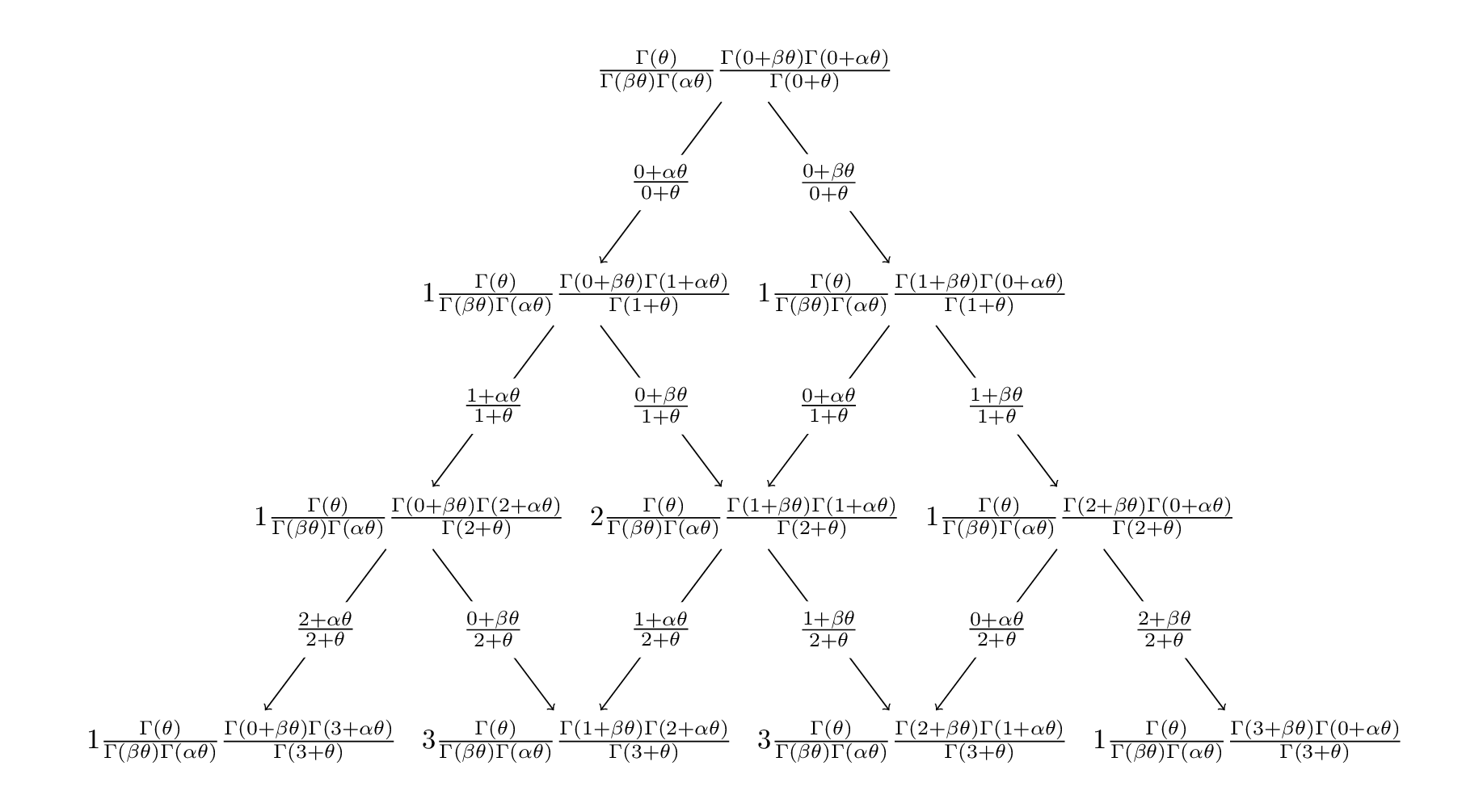}
    \caption{This is a schematic for the Polya-type urn scheme that is reminiscent of Pascal's triangle. Specifically, we trace the history of a sample allele configuration from an ancestral state to the observed configuration (\ie{} forwards in time) under the assumption of an underlying general reversible biallelic mutation model. The entry at the tip of the triangle represents the trivial sample $(i=0,n=0)$ (which has a probability of $1$), and as we move down the triangle we proceed forwards in time in the urn process. The edges of the arrows in this scheme denote the transition probabilities from Eq.~(\ref{eq:ind_betabin}). The entries in the $n$th row from the top correspond to the increasing values of the historic sample size $n$, while the rows represent the possible values $i$ of the focal allele within each row, \ie{} a historic sample. Note that the expressions at each node are the stationary beta-binomial sampling distributions from Eq.~(\ref{eq:betabin}) for the appropriate sample configuration.}   
   \label{fig:Betabinomial_Pascal}
\end{figure}
\begin{figure}[!ht]
    \centering
    \includegraphics[width = 12cm]{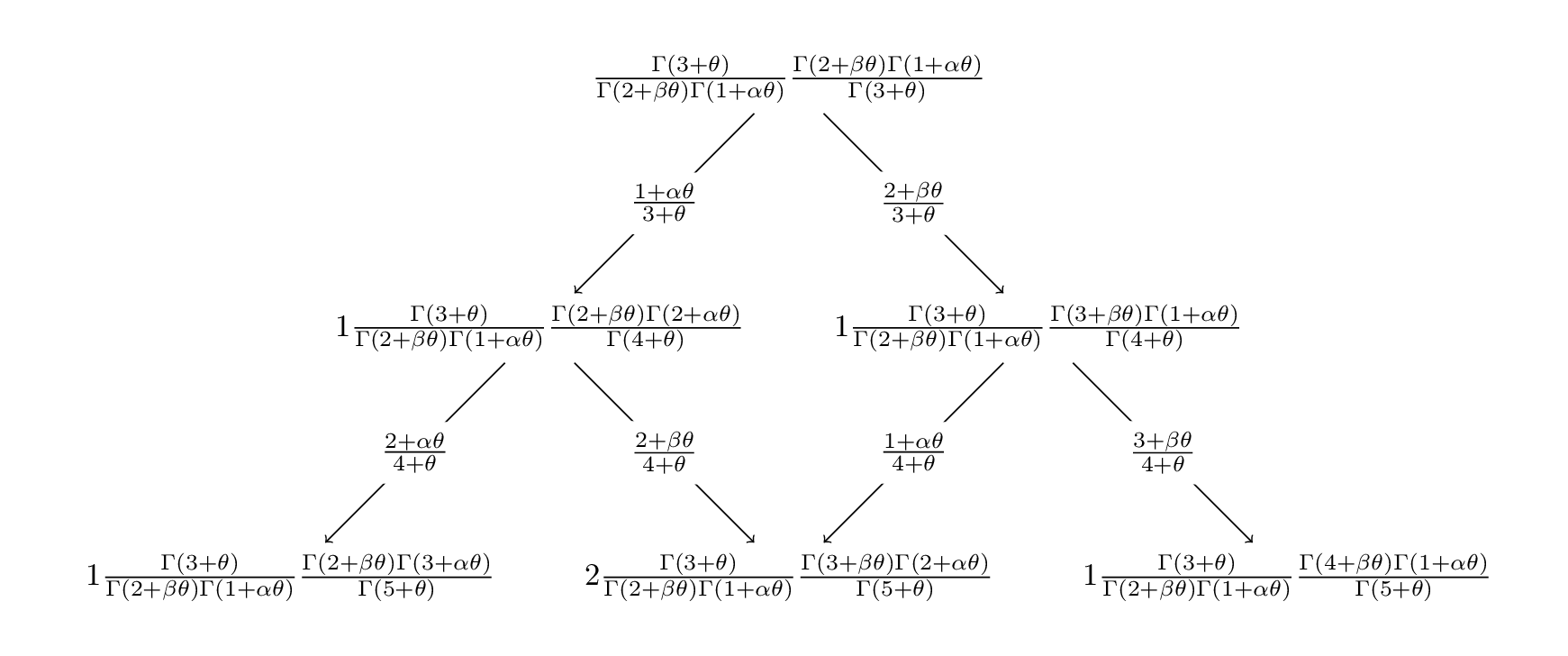}
    \caption{This illustration elucidates the jump process determined by the probabilities $e_{i,n}$ from Eq.~(\ref{eq:e_in_back_const}). We assume that a sample of size $K=5$ is drawn at the extant time $t=0$. The bottom row in the triangle shows the probabilities $e_{k,K}$ for the number of focal alleles $k=(2,3,4)$, assuming that this configuration has evolved from the historical sample configuration $(i=2,n=3)$ at the top of the triangle two jump steps previously. While this process runs backwards in time, the conditional probabilities are constructed forwards in time and this is indicated by the direction of the arrows that represent the same transition probabilities as described by the Polya-type urn process in Eq.~(\ref{eq:ind_betabin}). Note that the node at the tip of the triangle corresponds to a probability of $1$.}   
   \label{fig:e_Betabinomial}
\end{figure}

 Connections to well-established models can be made by reversing the dependency relationships in this jump process: By beginning from the trivial historic sample configuration $\Pr(i=0\given n=0)=1$, we can construct a trivially modified Polya-type urn process for the probabilities of past sample configurations as though we were moving forwards in time \citep[][compare Eqs. (1),(16)]{Hoppe87} and \citep[][Algorithm 2.1]{Step00}. The process starts from an urn containing balls of two colours, one focal and the other non-focal, with respective weights of $\beta\theta$ and $\alpha\theta$. At each jump step, we randomly draw a ball from the urn and then return it with a ball of the same type that has unit weight. Expressly, let the indicator variable $z$ record whether the next ball drawn is of the focal color. The transition probabilities from the sample configuration $(i,n)$ to the sample configuration $(i+z,n+1)$ are proportional to the weight of the focal and non-focal balls, and can be written as: 
\begin{equation}\label{eq:ind_betabin}
\Pr(z\given i, n,\beta,\theta)=p^z(1-p)^{1-z}\end{equation}
with 
$$
    p=\frac{i+\beta\theta}{n+\theta}=\frac{betabin(i+1\given n+1,\beta,\theta)}{betabin(i\given n,\beta,\theta)}\,.
$$
For further clarification of the spatial process, we provide a visualisation for both the urn model (see Fig.~\ref{fig:Betabinomial_Pascal}) and the jump process determining the $e_{i,n}$ (see Fig.~(\ref{fig:e_Betabinomial}).

\clearpage 

\clearpage 

We have now characterised the spatial component of the backwards diffusion equation operating on a binomial sampling function given in Eq.~(\ref{eq:death_general_mutation_rec_backward}). This leaves the temporal component:
Looking backwards in time, a jump step in the sample genealogy corresponds to a reduction in the historical sample size $n$ by either a coalescent or a parent-independent mutation event. Let $\Pr(n\given K,\theta,t)$, with $0\leq n \leq K$, denote the probability of a realised historic sample of size $n$ at time $t$. Considering this as the dependent variable in the backwards diffusion equation from Eq.~(\ref{eq:death_general_mutation_rec_backward}), we can re-write the latter as the following system of temporal linear differential equations:
\begin{equation}\label{eq:temp_system_back}
\begin{split}
    -\frac{d}{dt}\Pr(n\given  K,\theta,t)&=\lambda_K\,\Pr(n\given  K,\theta,t)\,, \quad\text{for $n=K$, and}\\ 
    -\frac{d}{dt}\Pr(n\given  K,\theta,t)&=-\lambda_{n+1}\,\Pr(n+1\given  K,\theta,t)\\
    &\qquad+\lambda_{n}\,\Pr(n\given  K,\theta,t)\,, \quad\text{for $K>n\geq 0$.}
\end{split}
\end{equation}

Imposing the starting conditions $\Pr(n=K\given K,\theta, t=0)=1$, and $\Pr(K>n\geq 0\given K,\theta,t=0)=0$, the solution of this system is given by the standard solution for a pure death process (except that our historical time $t$ is negative) \citep[][chapter 6, Eq.~2.2]{Taylor98} \citep[and compare][Eq.~4.7]{Tavare84}:
\begin{equation}\label{eq:temp_system_back_solution}
\begin{split}
    \Pr(n\given K,\theta,t)&=e^{\lambda_K t}\, \text{, for $n=K$}\,\\
    \Pr(n\given K,\theta,t)&=\bigg(\prod_{j=n+1}^K \lambda_j\bigg) \sum_{i=n}^{K}\frac{e^{\lambda_i t}}{\prod_{j=n,j\neq i}^K(\lambda_j-\lambda_i)}\, \text{, for $K-1\geq n\geq 0$}\,.
\end{split}
\end{equation}
Clearly, this solution requires the eigenvalues $\lambda_n=n(n-1+\theta)$ to be distinct, which they are in this model for realistic values of $\theta$, \ie{} values distinct from zero and infinity. Further, all $\Pr(n\given K,\theta,t)$ are positive and it is easily seen that $\sum_{n=0}^K\Pr(n\given K,\theta,t)= 1$ for $t_s\leq t\leq 0$, \ie{} probability must be conserved across the sample lineages at all times.

\vspace{5mm}

Combining the temporal pure death jump process and the subordinated spatial pure death jump process yields a time-dependent pure death jump process representation of the transition rates $\psi(x,t)$ of the backwards process:
\begin{equation}\label{eq:jump_full_backward}
    \begin{split}
        \psi(x,t)
        &=\sum_{n=0}^K \Pr(n\given K,t) \sum_{i=0}^n e_{i,n}\Pr(i\given n,x)\,.
    \end{split}
\end{equation}

If we rewrite this equation to make the exponential function visible, we obtain the following:
\begin{equation}
    \begin{split}
        \psi(x,t)
        &=\sum_{n=0}^K \Pr(n\given K,t) \sum_{i=0}^n e_{i,n}\Pr(i\given n,x)\\
        &=\sum_{n=0}^K \bigg(\sum_{j=n}^K f_{n,j}e^{\lambda_j t}\bigg)\sum_{i=0}^n hyper(i\given K,n,k)\,g_{i,n}x^i(1-x)^{n-i}\,
    \end{split}
\end{equation}
where we use $f_{j,n}$ and $g_{i,n}$ as placeholders for the remaining coefficients.

Recall that the expansion of the transition rates of the backwards process into a sum of orthogonal polynomials yields uniquely:
$$\psi(x,t)=\sum_{n=0}^K e^{\lambda_n t} d_n(k,K) R_n^{(\beta,\theta)}(x)$$
Collecting the polynomials multiplied by each of the $e^{\lambda_n t}$ for $n=0,...,K$ in both the time-dependent jump process and the spectral sum, the coefficients for every order of expansion $n$ can be equated between the two representations as in the Appendix SubSection~\ref{seq:diagonalisation}). 

We will not pursue this further here. Nevertheless, we hope to have established that the jump process on the space of binomial distributions and the spectral expansion on the space of orthogonal polynomials are equivalent processes for evaluating the probability of sample genealogies at each point in time. In other words, multiplication of either representations of $\psi(x,t)$ with the probability measure $\phi(x,t)$ at any time $t$ and integration over the population allele proportion will yield the same result for the probability of the observed sample configuration; \ie{} these processes are both dual to the backwards diffusion equation with respect to $\phi(x,t)$. Specifically, instead of determining the probability of observing $k$ focal in a total of $K$ alleles in a sample drawn at the extant time via the spectral expansion in the main text, we could utilise the time-dependent jump process:
\begin{equation}\label{eq:marg_like_equivalence}
\begin{split}
    p_k&=\Pr(k\given K,\beta,\theta,\rho, t)\\
    &=\int_0^1 \psi(x,t)\phi(x,t)\,dx\\
    &=\int_0^1 \sum_{n=0}^K \Pr(n\given K,t) \sum_{i=0}^n e_{i,n} \Pr(i\given n,x)\,\phi(x,t)\,dx\\
    &=\sum_{n=0}^K \Pr(n\given K,t) \sum_{i=0}^n e_{i,n} \sum_{m=0}^{n} d_m(i,n)\,\rho_m^{(\beta,\theta)}\Delta_m^{(G;\beta,\theta)} e^{-\lambda_m (t-t_s)}\,.
\end{split}
\end{equation}
While this expression may be easier to interpret probabilistically in the light of this past subsection than the equivalent calculation involving the spectral sum, its evaluation clearly involves more intermediate summation steps. Note however, that this is specifically true for non-equilibrium scenarios since the final sum reduces to a single coefficient in equilibrium. 

\paragraph{Remark}
Our embedded time-dependent jump process bears the most similarity to \citet[][Section 3.3]{Papaspiliopoulos14}, where the same general result is presented as the dual process for a reversible multiallelic Wright-Fisher model; incidentally, our process corresponds to the special case of theirs with only two alleles, and thus also to the dual process derived by \citep{Chaleyat-Maurel09}. Both of these articles concern filtering the Wright-Fisher model on the basis of binomially distributed data, so that all tasks performed by a forward-backward algorithm involve finite mixture expansions as in the above Eq.~(\ref{eq:marg_like_equivalence}). However, whilst we arrive at the same time-dependent pure death jump process as a dual process for the general biallelic reversible diffusion model as \citep{Papaspiliopoulos14} with a superficially analogous method, we additionally elaborate on the probabilistic interpretation of its component processes and explicitly allow for non-equilibrium scenarios. We do so because the partitioning into spatial and temporal components reflects the logic of the orthogonal polynomial expansion. Note that the start of the derivation of a coalescent dual process for the Wright-Fisher diffusion requires a similar separation \citep{Etheridge09}. However, these authors consider a model with selection and differ methodologically in their approach in that they assume an infinite entrance boundary rather than starting from an observed sample at a specified time point.

\paragraph{Remark}

In terms of determining the expected sample allele configuration of a population evolving according to Moran and Wright-Fisher diffusion models, a dual processes is generally found in Kingman's coalescent \citep{Kingman82} or a variant thereof \citep{Tavare84}. As pointed out in \citet{Papaspiliopoulos14}, the expression $\sum_{i=n}^{K}\frac{e^{\lambda_i t}}{\prod_{j=n,j\neq i}^K(\lambda_j-\lambda_i)}$ with $K-1\geq n\geq 0$, which occurs within our temporal process, is precisely the transition probability of Kingman's block-counting process with mutation. The duality to Kingman's coalescent is more explicitly shown in \citet[][Eq.~7.25]{Tavare84}, who recovers an ancestral process on the number of distinct ancestors in a population immediately from the diffusion equation (this is cited as Kingman's own diffusion time scale approximation to the original discrete time and space coalescent). Further, \citep{Griffiths10} has shown that the Wright-Fisher diffusion with transition rate densities expressed via expansion into Jacobi polynomials is dual to Kingman's coalescent by showing that these transition rate densities correspond to the mixture distribution tracing the non-mutant lineages in the population arising from the coalescent (whereby the latter also appear in \citep{Etheridge09}); note that these are population level considerations involving infinite sums.

\paragraph{Remark}
 Computational methods related to the ``matrix coalescent'' \citep{Wooding02} employ a spectral decomposition of the transition rate matrix of the ancestral processes of a sample \citep[][compare Section 5.1]{Tavare84}, \ie{} the rate matrix describing the decrease in distinct lineages looking back in time. For Kingman's coalescent, this would be exactly the corresponding rate matrix to the transition probabilities recovered within our time-homogeneous pure death process (see previous remark). Within this framework, time-inhomogeneous coalescent processes, which include non-constant demographic scenarios as well as multiple-merger coalescents, can be modelled by realising that such inhomogeneity affects the observed site-frequency spectrum only via the first coalescent time within each (historical) sample size and explicitly incorporating the dynamics between it and the other coalescent times into the system \citep{Spence16}. Separating out the impact of temporal changes in this way is similar in spirit to our approach; it is simply the genealogy that is considered rather than allele proportions.

\subsubsection{Boundary Mutation Model}

\begin{figure}[!ht]
    \centering
    \includegraphics[width = 12cm]{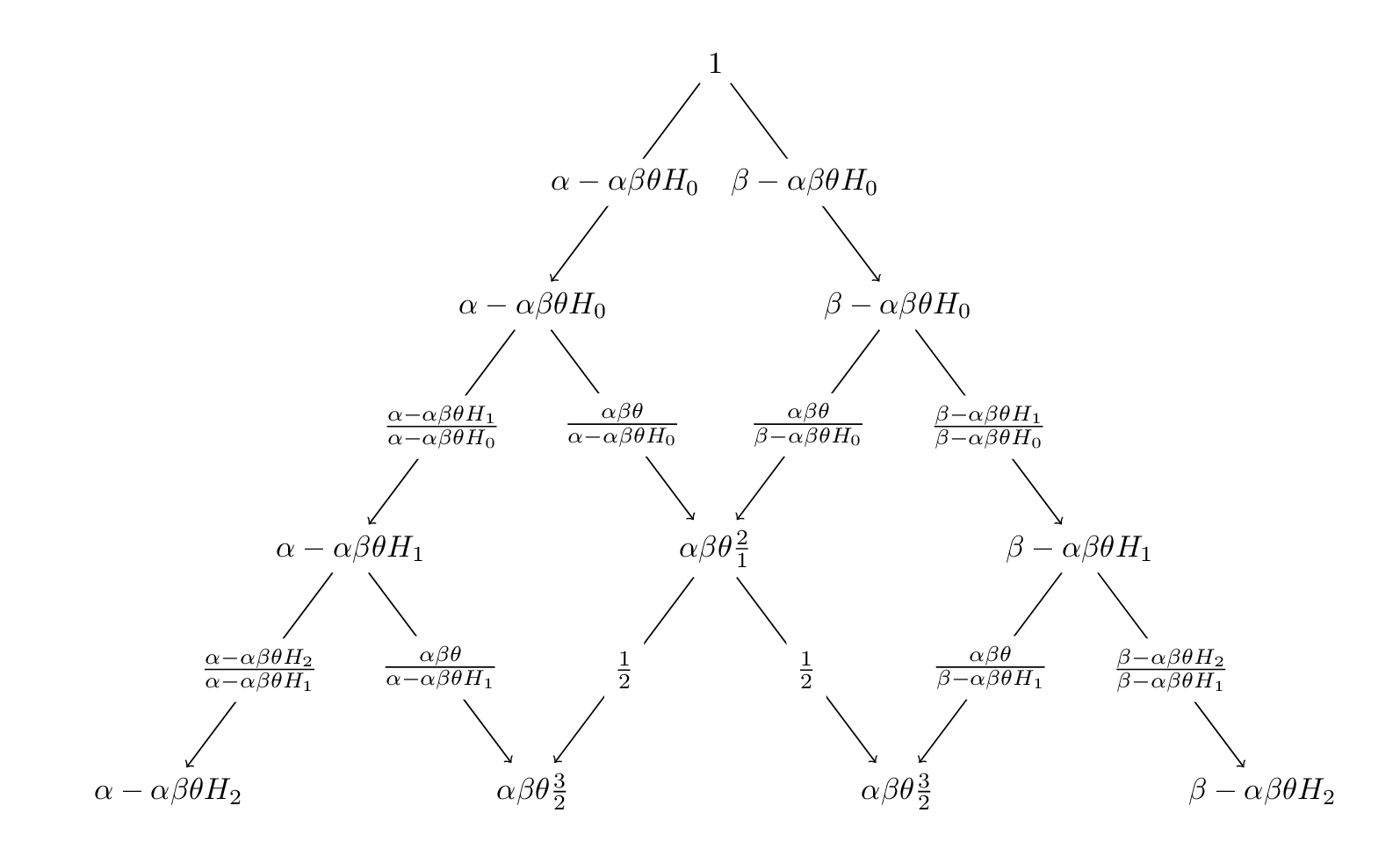}
    \caption{This is a schematic for the Polya-type urn scheme that is reminiscent of Pascal's triangle. Specifically, we trace the history of a sample allele configuration from an ancestral state to the observed configuration (\ie{} forwards in time) under the assumption of an underlying boundary mutation model. The entry at the tip of the triangle represents the trivial sample $(i=0,n=0)$ (which has a probability of $1$), and as we move down the triangle we proceed forwards in time in the urn process. The edges of the arrows in this scheme denote the transition probabilities from Eq.~(\ref{eq:ind_boundary}). The entries in the $n$th row from the top correspond to the increasing values of the historic sample size $n$, while the rows represent the possible values $i$ of the focal allele within each row, \ie{} a historic sample. Note that the expressions at each node are the stationary beta-binomial sampling distributions from Eq.~(\ref{eq:boundary_stationary}) for the appropriate sample configuration. For the visualisation scheme to work, we defined the trivial harmonic number $H_0=0$ (so that the recurrence relation $H_{n+1}=H_n+\frac{1}{n+1}$ still holds).}
   \label{fig:Boundary_Pascal}
\end{figure}

Because of the unique dynamics at the boundaries, substituting the binomial likelihood representing the sampling scheme into the appropriate backwards diffusion equation for the boundary mutation model in order to derive an embedded dual process is not immediately possible. However, the recursive system for the spatial process analogous to Eq.~(\ref{eq:recursion_2}) in the general mutation model can be directly given as:
\begin{equation}\label{eq:recursion_3}
\begin{split}
    \Pr(i\given n,x)&=\frac{i}{n}\frac{boundbin(i\given n,\beta,\theta)}{boundbin(i-1\given n-1,\beta,\theta)}\Pr(i-1\given n-1,x)\\
    &\qquad+\frac{n-i}{n}\frac{boundbin(i\given n,\beta,\theta)}{boundbin(i\given n-1,\beta,\theta)}\Pr(i\given n-1,x)\,.
\end{split}
\end{equation}
Recall that the factored out eigenvalues must be $\lambda_0=0$, $\lambda_1=\theta$, and $\lambda_n=n(n-1)$ for $n\geq 2$ for the boundary mutation model. The above recursive relationship then describes a spatial process for the history of the number of focal alleles in a sample (compare Eq.~\ref{eq:recursion}); its result is given by:
\begin{equation}\label{eq:e_in_back_const_bmm}
\begin{split}
    e_{i,n}&=hyper(i\given K,n,k)\,\frac{boundbin(k\given K,\beta,\theta)}{boundbin(i\given n,\beta,\theta)}\,.\\
\end{split}
\end{equation}
Reversing time in this spatial process, a Polya-type urn model can again be formulated to describe the system: The transition rates are determined by a Bernoulli distributed indicator variable as in Eq.~(\ref{eq:ind_betabin}), but with the following success probabilities: 
\begin{equation}\label{eq:ind_boundary}
p=\frac{boundbin(i+1\given n+1,\beta,\theta)}{boundbin(i\given n,\beta,\theta)}\,.
\end{equation}
For clarity, this urn process is visualised in Fig.~(\ref{fig:Boundary_Pascal}).

Importantly, we must draw attention to the fact that this dual embedded pure death jump process is actually not equivalent to the process considered in the rest of this article. Instead, it is analogous to the considerations made by \citet{Burden18a} in their proofs of their results for the limits of low scaled mutation rates; however, to our knowledge the spectral representation of their backwards process for low scaled mutation rates is not known. Recall that the backwards system considered in the rest of this article specifically contrasts the monomorphic boundary region, consisting of the first two eigenfunctions associated with eigenvalues $\lambda_0$ and $\lambda_1$ in the spectral expansion, with the polymorphic interior. The interior eigenfunctions associated with the eigenvalues $\lambda_{n\geq 2}$ in the spectral expansion are assumed to follow pure drift dynamics. Hence, the corresponding embedded dual pure death process for $\Pr(i\given n, x)$ from any starting values of $n$ down to $n=2$ can be readily obtained from Eq.~(\ref{eq:death_general_mutation_rec_backward}) by setting $\theta=0$ and following the pattern in the general mutation model. In contrast, to recover the dynamics of $\Pr(i\given n=1, x)$ in terms of this death process, the system from Eq.~(\ref{eq:DE_inhomogeneous_Bs}) would need to be expressed as a function of $\Pr(i\given n, x)$ for $n=0$ and $n=1$ (see Section~\ref{sec:appendix_diag}). A fuller treatment of this approach, or also an in depth comparison of the two alternative approaches, would over-extend the scope of our manuscript. 

\newpage


\paragraph{Remark}
Note that while the general structure of the temporal pure death process for the boundary mutation model is identical to that of the general reversible biallelic mutation model, the transition probabilities will differ slightly: Instead of the  stationary distribution being a Beta distribution, the stationary distribution of the boundary mutation diffusion from Eq.~(\ref{eq:boundary_diffusion_stationary}) must be employed; thus, instead of the Beta-binomially distributed terms in Eq.~(\ref{eq:recursion_2}), boundary-binomially distributed terms (see Eq.~\ref{eq:boundary_stationary}) must be utilised as in Eq.~(\ref{eq:recursion_3}). Note also that the eigenvalues must also be those of the boundary mutation model.

\paragraph{Remark}
The methodology formalised by \citep{Papaspiliopoulos14} for identifying a dual process for diffusion models that allows for performing the calculations involved in a forward-backward algorithm via finite sums appears to require diffusion models that are reversible with respect to their stationary distribution. This is indeed the case for the general biallelic reversible mutation model (compare \citep[][p.~3]{Song12} and \citep[][Eq.~36]{Bergman18a}) where the stationary distribution is Beta distributed (or Dirichlet distributed when multiallelic models are considered, as in \citet{Papaspiliopoulos14}). However, they note that the methodology can be extended to non-stationary diffusion processes that are reversible with respect to a reversible measure (which is not necessarily a well-defined distribution) as long their sampling distributions are well-defined (see Discussion in \citet{Papaspiliopoulos14}). While we were not able to adapt their methodology to the boundary mutation diffusion process without workarounds and additional reasoning,  we would like to note that it does fall into this latter category: The process is reversible with respect to the stationary distribution with boundary singularities from Eq.~(\ref{eq:boundary_diffusion_stationary}) (see Section 4.2 in \citep{Papaspiliopoulos14})), and we have shown that, for $K$ small enough, the sampling distribution is well-defined in the sense that it is a regular density both in equilibrium (Appendix Section~\ref{section:simpl_ass}) and non-equilibrium (as in this section).


\clearpage

\section{Appendix: Miscellaneous Additional Results for the Demographic Models}\label{sec:app_misc}

\subsection{Population Explosion}\label{sec:PopExpl}

Note that if $\theta_1^{*}\ll\theta_2^{*}$, the expression in the exponent of the solution to Eq.~(\ref{eq:TauExpansion_AC_solution}) tends to zero. A first order Taylor expansion around it yields for $t_1<t\leq 0$: 
\begin{equation}\label{eq:TauExpansion_AC_approx}
\begin{split}
 \tau_n(t)\, \Xi_n ^{-1} & \approx \theta_2^{*} + \big( \theta_1^{*} - \theta_2^{*}\big)\bigg(1-\lambda_n \frac{\theta_1^{*}}{\theta_2^{*}} t\bigg)\\
 &= \theta_1^{*} + \frac{\theta_2^{*}-\theta_1^{*}}{\theta_2}\theta_1^{*} \lambda_n t\\
 &\approx \theta_1^{*}(1+\lambda_n t)\,.
\end{split}
\end{equation}


\subsection{D-statistic}\label{sec:TajMahal}

Neutrality tests based on the infinite sites model are pervasive in population genetics. Formulated for site frequency data, they take the general form \citep{Korneliussen13}:
$$D=\frac{\hat{\theta}_W-\hat{\theta}_Z}{\sqrt{\Var(\hat{\theta}_W-\hat{\theta}_Z)}},$$
where $\hat{\theta}_W$ and $\hat{\theta}_Z$ are unbiased estimators for the overall scaled mutation rate that differ in their weighting of polymorphic sites assuming neutral equilibrium. In the case of Tajima's~D \citep{Tajima89a}, $\hat{\theta}_W$ is Watterson's (or Ewens' and Watterson's) estimator \citep{Ewens72,Ewens74,Watterson75}, which counts the expected number of segregating sites, and $\hat{\theta}_Z$ is a statistic that counts the number of pairwise differences between ancestral and derived nucleotides \citep{Tajima83}. The variance in the denominator of the statistic $D$ is conventionally calculated assuming complete linkage between sites, \ie{} sites are not independent. Note that $\hat{\theta}_Z$ weights intermediate frequency alleles higher than $\hat{\theta}_W$. Therefore recent positive selection or population expansion, which increase the frequency of rare alleles, lead to values $D<0$; balancing selection and population collapse elevate the relative numbers of intermediate allele frequencies and therefore lead to values $D>0$. 

Here, we reformulate Tajima's D for the context of the biallelic boundary mutation Moran model:
Let us assume a sample site frequency spectrum of size $K$ comprised of $L$ loci, and recall that $y_k$ denoted the realised count of focal alleles. We further assume 
recombination rates high enough for linkage to be negligible and sites therefore independent, which we in fact do implicitly throughout this article. For low overall scaled mutation rates the distribution of the counts of focal alleles $y_k$ of the polymorphic spectrum ($k\in \{1,\dots,K-1\}$) can be approximated by a Poisson distribution with :
\begin{equation*}
\begin{split}
     &y_{k} \overset{iid}\sim Pois\bigg(L \bar p_k\bigg)\,.
\end{split}
\end{equation*}
In this case, the stationary distribution $\bar p_k= \alpha\beta\theta\frac{K}{k(K-k)}$ is given by the boundary binomial (Eq.~\ref{eq:boundary_stationary}). Then,
$$\hat{\theta}_W=\frac{\sum_{k=1}^{K}y_k}{2L \sum_{k=1}^{K-1}\tfrac{1}{k}}$$
is the minimum variance, unbiased maximum likelihood estimator for the bias-complemented overall scaled mutation rate $\theta^{*}=\alpha\beta\theta$ \citep{Vogl20}. Since we do not distinguish between ancestral and derived alleles, the average pairwise distance between alleles of a sample can also be used as an estimator of the same parameter:
$$\hat{\theta}_Z=\frac{\sum_{k=1}^{K} y_k\,k(K-k)}{L K(K-1)}\,.$$
Substituting the estimator $\hat\theta_W$ for the parameter $\theta^{*}$, the estimator for the variance of the difference between the two estimators, recalling that mean and variance of the Poisson distribution are equal, is:
\begin{equation*}
\begin{split}
\widehat\Var(\hat{\theta}_W-\hat{\theta}_Z)
    &=\sum_{k=1}^{K-1}\E\big(\tfrac{y_k}{L}\big)\bigg(\frac{k(K-k)}{K(K-1)} - \frac{1}{2\sum_{k=1}^{K-1}\tfrac{1}{k}}\bigg)^2\\ 
    &=\hat\theta_W\sum_{k=1}^{K-1}\frac{K}{k(K-k)}\bigg(\frac{k(K-k)}{K(K-1)} - \frac{1}{2\sum_{k=1}^{K-1}\tfrac{1}{k}}\bigg)^2.
\end{split}
\end{equation*}


\newpage


\subsection{Tabulated Comparisons of Analytical and Simulated Results for Models~3,~4}\label{sec:TMod3Tables}

\begin{table}[!htbp] \centering 

\begin{tabular}{@{\extracolsep{5pt}} c|cccc} 
 
\hline \\[-1.8ex] 
\textbf{Additional Example 1} & Mean & Variance & Covariance & Correlation \\ 
\hline \\[-1.8ex] 
\textbf{Model 3} &  &  &  &  \\ 
\hline 
\text{$c=1e^{-4}$} &  &  &  &  \\ 
\hline 
 $n=2$ - Theoretical  & $0.008350$ & $0$ & $0$ & $0.976332$ \\ 
  \hspace{9mm}  Simulated  & $0.008351$ & $0$ & $0$ & $0.976349$ \\ 
 $n=12$ - Theoretical  & $0.009142$ & $0.000008$ & $0.000002$ & $0.245198$ \\ 
 \hspace{9mm}  Simulated  & $0.009142$ & $0.000008$ & $0.000002$ & $0.245154$ \\ 
 \hline 
\text{$c=1e^{-2}$} &  &  &  &  \\ 
\hline 
 $n=2$ - Theoretical  &  $0.009387$ & $0.000012$ & $0.000002$ & $0.132810$ \\ 
 \hspace{9mm}  Simulated  & $0.009387$ & $0.000012$ & $0.000002$ & $0.133339$ \\ 
$n=12$ - Theoretical & $0.01$ & $0.000025$ & $0$ & $0$ \\ 
  \hspace{9mm}  Simulated  & $0.009999$ & $0.000025$ & $0$ & $$-$0.000345$ \\
  \hline 
\text{$c=1$} &  &  &  &  \\ 
\hline 
$n=2$ - Theoretical & $0.01$ & $0.000025$ & $0$ & $0$ \\ 
 \hspace{9mm}  Simulated  & $0.009998$ & $0.000025$ & $0$ & $$-$0.000095$ \\ 
$n=12$ - Theoretical & $0.01$ & $0.000025$ & $0$ & $0$ \\ 
 \hspace{9mm}  Simulated & $0.009998$ & $0.000025$ & $0$ & $0.000071$ \\ 
 \hline 
\textbf{Model 4}  &  &  &  &  \\
\hline 
\hline 
\text{$c=1$} &  &  &  &  \\ 
\hline 
$n=2$ - Theoretical  & $0.01$ & $0.000150$ & $0$ & $0$ \\ 
\hspace{9mm}  Simulated & $0.01$ & $0.000150$ & $0$ & $$-$0.000205$ \\ 
$n=12$ - Theoretical  & $0.01$ & $0.000150$ & $0$ & $0$ \\ 
\hspace{9mm}  Simulated & $0.009996$ & $0.000150$ & $0$ & $$-$0.000332$ \\ 
\hline \\[-1.8ex] 
\end{tabular} 
\caption{A population is assumed evolve forwards in time according to the autoregression model from Eq.~(\ref{eq:system_detAR}), with the overall bias-complemented scaled mutation rate $\theta_j^{*}$ randomly fluctuating across epochs according to $\theta_j^{*} \overset{iid}\sim IG(a=6,b=0.01\cdot(a-1))$. The arithmetic mean of $\theta_j^{*}$ is then $0.01$ and the harmonic mean $0.008333$. After $J=1e^{6}$ epochs, the mean, variance, covariance, and correlations for the temporal coefficients $\vartheta_{j,n}$ for $j$ between $1$ and $J$ and select $n$ are evaluated and compared with their theoretical counterparts (rounded to 6 post-decimal digits).} 
  \label{Table:AR-res1} 
  \end{table} 
\newpage


\begin{table}[!htbp] \centering 
\begin{tabular}{@{\extracolsep{5pt}} c|cccc} 
\hline \\[-1.8ex] 
\textbf{Additional Example 2} & Mean & Variance & Covariance & Correlation \\ 
\hline \\[-1.8ex] 
\textbf{Model 3} &  &  &  &  \\ 
\hline 
\text{$c=1e^{-4}$} &  &  &  &  \\ 
\hline 
$n=2$ - Theoretical  & $0.006700$ & $0$ & $0$ & $0.970590$ \\ 
  \hspace{9mm}  Simulated  & $0.006700$ & $0$ & $0$ & $0.970490$ \\ 
$n=12$ - Theoretical  & $0.008153$ & $0.000015$ & $0.000003$ & $0.218613$ \\ 
 Simulated & $0.008153$ & $0.000014$ & $0.000003$ & $0.218262$ \\ 
 \hline 
\text{$c=1e^{-2}$} &  &  &  &  \\ 
\hline 
$n=2$ - Theoretical  & $0.008571$ & $0.000020$ & $0.000003$ & $0.125$ \\ 
 \hspace{9mm}  Simulated  & $0.008571$ & $0.000020$ & $0.000003$ & $0.124526$ \\ 
 $n=12$ - Theoretical & $0.009998$ & $0.000096$ & $0$ & $0.000003$ \\ 
 \hspace{9mm}  Simulated  & $0.009998$ & $0.000096$ & $0$ & $$-$0.000071$ \\ 
 \hline 
\text{$c=1$} &  &  &  &  \\ 
\hline 
 $n=2$ - Theoretical & $0.009999$ & $0.000097$ & $0$ & $0.000001$ \\ 
 \hspace{9mm} Simulated  & $0.009999$ & $0.000097$ & $0$ & $$-$0.000172$ \\ 
 $n=12$ - Theoretical & $0.01$ & $0.0001$ & $0$ & $0$ \\ 
 \hspace{9mm}  Simulated & $0.01$ & $0.0001$ & $0$ & $0.000629$ \\ 
 \hline \\[-1.8ex] 
\textbf{Model 4} &  &  &  &  \\ 
\hline 
\hline 
\text{$c=1$} &  &  &  &  \\ 
\hline 
$n=2$ - Theoretical & $0.009999$ & $0.000294$ & $0$ & $0.000001$ \\ 
\hspace{9mm}  Simulated & $0.009995$ & $0.000288$ & $0$ & $0.000387$ \\ 
$n=12$ - Theoretical  & $0.01$ & $0.0003$ & $0$ & $0$ \\ 
\hspace{9mm}  Simulated & $0.009998$ & $0.000299$ & $0$ & $0.000074$ \\ 
\hline \\[-1.8ex] 
\end{tabular} 
\caption{A population is assumed to be evolving forwards in time according to the autoregression model from Eq.~(\ref{eq:system_detAR}), with the overall bias-complemented scaled mutation rate $\theta_j^{*}$ randomly fluctuating across epochs according to $\theta_j^{*} \overset{iid}\sim IG(a=3,b=0.01\cdot(a-1))$. The arithmetic mean of $\theta_j^{*}$ is then $0.01$ and the harmonic mean $0.006667$. After $J=1e^{6}$ epochs, the mean, variance, covariance, and correlations for the temporal coefficients $\vartheta_{j,n}$ for $j$ between $1$ and $J$ and select $n$ are evaluated and compared with their theoretical counterparts (rounded to 6 post-decimal digits).} 
\label{Table:AR-res2} 
\end{table} 

\newpage


\begin{table}[!htbp] \centering 
\begin{tabular}{@{\extracolsep{5pt}} c|cccc} 
\hline \\[-1.8ex] 
\textbf{Additional Example 3}& Mean & Variance & Covariance & Correlation \\ 
\hline \\[-1.8ex] 
\textbf{Model 3} &  &  &  &  \\ 
\hline 
\hline 
\text{$c=1e^{-4}$} &  &  &  &  \\ 
\hline 
$n=2$ - Theoretical  & $0.004183$ & $0$ & $0$ & $0.953316$ \\ 
 \hspace{9mm}  Simulated  & $0.004184$ & $0$ & $0$ & $0.953433$ \\ 
 $n=12$ - Theoretical  & $0.004775$ & $0.000004$ & $0$ & $0.078571$ \\ 
 \hspace{9mm}  Simulated  & $0.004774$ & $0.000004$ & $0$ & $0.077895$ \\ 
\hline 
\text{$c=1e^{-2}$} &  &  &  &  \\ 
\hline
 $n=2$ - Theoretical  & $0.004879$ & $0.000004$ & $0$ & $0.029401$ \\ 
 \hspace{9mm}  Simulated  & $0.004880$ & $0.000004$ & $0$ & $0.029057$ \\ 
 $n=12$ - Theoretical & $0.0050$ & $0.000006$ & $0$ & $0$ \\ 
  \hspace{25mm}  Simulated  & $0.004999$ & $0.000006$ & $0$ & $0.000088$ \\ 
  \hline 
  \hline 
\text{$c=1$} &  &  &  &  \\ 
\hline
 $n=2$ - Theoretical & $0.005$ & $0.000006$ & $0$ & $0$ \\ 
 \hspace{9mm}  Simulated  & $0.005001$ & $0.000006$ & $0$ & $0.000084$ \\ 
 $n=12$ - Theoretical & $0.005$ & $0.000006$ & $0$ & $0$ \\ 
 \hspace{9mm}  Simulated & $0.005001$ & $0.000006$ & $0$ & $0.0000273$ \\ 
 \hline 
\textbf{Model 4}  &  &  &  &  \\
\hline 
\hline 
\text{$c=1$} &  &  &  &  \\ 
\hline 
$n=2$ - Theoretical  & $0.005$ & $0.000038$ & $0$ & $0$ \\
\hspace{9mm}  Simulated & $0.004999$ & $0.0000375$ & $0$ & $0.000510$ \\
$n=12$ - Theoretical  & $0.005$ & $0.000038$ & $0$ & $0$ \\ 
\hspace{9mm}  Simulated & $0.005$ & $0.000037$ & $0$ & $$-$0.000396$ \\ 
\hline \\[-1.8ex] 
\end{tabular} 
\caption{A population is assumed to be evolving forwards in time according to the autoregression model from Eq.~(\ref{eq:system_detAR}), with the overall bias-complemented scaled mutation rate $\theta_j^{*}$ randomly fluctuating across epochs according to $\theta_j^{*} \overset{iid}\sim IG(a=6,b=0.005\cdot(a-1))$. The arithmetic mean of $\theta_j^{*}$ is then $0.005$ and the harmonic mean $0.004167$. After $J=1e^{6}$ epochs, the mean, variance, covariance, and correlations for the temporal coefficients $\vartheta_{j,n}$ for $j$ between $1$ and $J$ and select $n$ are evaluated and compared with their theoretical counterparts (rounded to 6 post-decimal digits).} 
\label{Table:AR-res3} 
\end{table}


\newpage

\bibliography{coal}
\end{document}